\documentclass{aa}
\usepackage{graphicx}
\usepackage{txfonts}

\def\etal{{et\,al. }}
\def\EE#1{\times 10^{#1}}

\def\cm3{\rm ~cm^{-3}}
\def\kms{\rm ~km~s^{-1}}

\def\wl{~\lambda}
\def\wll{~\lambda\lambda}

\def\Msun{~{\rm M}_\odot}
\def\Ti44{M(^{44}{\rm Ti})}

\def\snr{SNR~0540-69.3}

\def\lsim{\!\!\!\phantom{\le}\smash{\buildrel{}\over
  {\lower2.5dd\hbox{$\buildrel{\lower2dd\hbox{$\displaystyle<$}}\over
                               \sim$}}}\,\,}
\def\gsim{\!\!\!\phantom{\ge}\smash{\buildrel{}\over
  {\lower2.5dd\hbox{$\buildrel{\lower2dd\hbox{$\displaystyle>$}}\over
                               \sim$}}}\,\,}

\begin{document}
\title{Kinematics, structure and abundances of supernova remnant 0540-69.3\thanks{Based on 
observations performed at the European Southern Observatory, La Silla and 
Paranal, Chile (ESO Programmes 56.C-0731 and 68.D-0394).}}
 \subtitle{}
   
\author {P. Lundqvist\inst{1,2}
\and N. Lundqvist\inst{1,2}
%\and C. Sandin\inst{2}
\and Yu. A. Shibanov\inst{3,4}
}

\institute{
Department of Astronomy, AlbaNova University Center, Stockholm University, SE-10691 Stockholm, Sweden\\
\email{peter@astro.su.se}
 \and
The Oskar Klein Centre, AlbaNova, SE-10691 Stockholm, Sweden\\
\and 
Ioffe Institute, Politekhnicheskaya 26, St. Petersburg, 194021, Russia\\
\and
Peter the Great St. Petersburg Polytechnic University, Politekhnicheskaya 29, St. Petersburg, 195251, Russia
}

\date{Received ?, ?, 2021; accepted ?, ?, 2021}
%\mail{peter@astro.su.se}

\titlerunning{Supernova remnant 0540-69.3}
\authorrunning{Lundqvist et al.}
%\offprints{P.~Lundqvist;\hfill\\
%e-mail: peter@astro.su.se}

  \abstract
   {}
   {To investigate the structure, elemental abundances, physical conditions and the immediate surroundings of supernova remnant 0540-69.3 in the Large Magellanic Cloud.}
   {Imaging in [\ion{O}{iIi}] and spectroscopic studies through various slits were utilized using European Souther Observatory's Very Large and New Technology Telescopes. Densities, temperatures and abundances were estimated applying nebular analysis for various parts of the remnant.} 
   {Several new spectral lines are identified, both in the pulsar-wind nebula part of the remnant, and in interstellar clouds shocked 
   by the supernova blast wave. In the pulsar-wind nebula, all lines are redshifted by $440\pm80 \kms$ with respect to the rest 
   frame of the host galaxy, and a 3D-representation of the [\ion{O}{iii}] emission displays a symmetry axis of 
   ring-like structures which could indicate that the pulsar shares the same general redshift as the central supernova ejecta. 
   [\ion{O}{ii}], [\ion{S}{ii}], [\ion{Ar}{iii}] and H$\beta$ share a common more compact structure than [\ion{O}{iii}], and possibly 
   [\ion{Ne}{iii}]. The average [\ion{O}{iii}] temperature in the pulsar-wind 
   nebula is $23\,500 \pm 1\,800$~K, and the electron density derived from [\ion{S}{ii}] is typically $\sim 10^3 \cm3$. By mass, 
   the relative elemental abundances of the shocked ejecta in the pulsar-wind nebula are ${\rm O:Ne:S:Ar} \approx 1:0.07:0.10:0.02$, 
   consistent with explosion models of $13-20 \Msun$ progenitors, and similar to that of SN~1987A, as is also the explosive mixing of 
   hydrogen and helium into the center. From H$\beta$ and \ion{He}{i}~$\lambda$5876, the mass ratio of He/H in the center is estimated to
   be in excess of $\sim 0.8$. The rapid cooling of the shocked ejecta could potentially cause variations in the relative abundances 
   if the ejecta are not fully microscopically mixed, and this is highlighted for S/O for the period 1989--2006. [\ion{O}{iii}] is also seen in 
   presumably freely coasting photoionized ejecta outside the pulsar-wind nebula at inferred velocities out to well above $2\,000 \kms$, 
   and in projection [\ion{O}{iii}] is seen out to $\sim 10$\arcsec\ from the pulsar. 
   This is used to estimate that the pulsar age is $\approx 1\,200$ years. The freely coasting [\ion{O}{iii}]-emitting
   ejecta have a strictly non-spherical distribution, and their mass is estimated to be $\sim 0.12 \Msun$. A possible outer boundary of 
   oxygen-rich ejecta is seen in [\ion{O}{ii}]~$\lambda\lambda$3726,3729 at $\sim 2\,000-2\,100 \kms$. 
 Four filaments of shocked interstellar medium are identified, and there is a wide range in degree of ionization of iron, from Fe$^+$ to
  Fe$^{13+}$. One filament belongs to a region also observed in X-rays, and another one has a redshift of $85\pm30 \kms$ relative to the
    host. From this we estimate that the electron density of the [\ion{O}{iii}]-emitting  gas is $\sim 10^3 \cm3$, and that the line of the most 
    highly ionized ion, [\ion{Fe}{xiv}]~$\lambda$5303, come from an evaporation zone in connection with the radiatively cooled gas 
    emitting, e.g., [\ion{O}{iii}], and not  from immediately behind the blast wave. We do not find evidence for nitrogen-enriched ejecta in 
   the south-western part of the remnant, as was previously suggested. Emission in this region is instead from a severely reddened 
    \ion{H}{ii}-region.
}
 {}

\maketitle

\section{Introduction}

Supernova remnant (SNR) 0540-69.3 (henceforth simply 0540) has been 
observed at wavelengths ranging from X-rays to the radio. Both in the radio 
and in X-rays. the remnant is bounded by an outer shell, which has a radius 
of $\sim 20-35\arcsec$  \citep{man93b,GW00}. In the optical, the outer 
shell coincides with two bright filaments to the west and southwest \citep{Mat80}, 
one emitting [\ion{O}{iIi}] and the other [\ion{N}{ii}], respectively. As cautioned by 
these authors, the [\ion{N}{ii}] filament may not belong to the remnant, as it coincides
with two bright stars. 

Inside the outer shell, the emission from the remnant is concentrated to a
substantially smaller nebula \citep{Mat80,Kirshner89,Car92,Mo03,Ser05,nlun11}.
In [\ion{O}{iii}] the diameter of the main emission is $\sim 8\arcsec$ \citep{Car92}.
Henceforth we will refer to this as the `central part of the SNR' (SNRC),
and in [\ion{S}{ii}] and H$\alpha$ the observed structures are even smaller and 
weaker, decreasing in size and strength in this order. 

The SNRC also contains an optical continuum source; this is synchrotron emission 
from the pulsar wind nebula (PWN) \citep{Cha84,Ser04,nlun11}. The synchrotron
emission and its spectral characteristics have been studied throughout the energy
range from radio to X-rays \citep{Ser04,nlun11,Mignani12,bra14,plun20}  A PWN is 
expected since the remnant harbors the young pulsar (PSR) B0540-69 whose pulsed emission 
has been documented in X-rays \citep{sew84}, in the UV \citep{Mignani19}, in the optical 
\citep{Middleditch85}, and in the radio \citep{man93a}. 

The SNRC with its pulsar bears many similarities to the Crab Nebula, and a detailed 
discussion and comparison of 0540 with the Crab pulsar and its PWN can be found in 
\citet{Ser04}. However, as revealed by spectral studies in the optical 
\citep{Mat80,DT84,Kirshner89,Morse06}, the elemental 
abundances in 0540 are very different from those in the Crab. While the Crab is 
helium-rich, but not conspicuous in other respects \citep{M96}, 
0540 is dominated by forbidden oxygen and sulphur lines, and is 
classified as an ``oxygen-rich SNR'' (OSNR). The most studied object in this class 
is undoubtedly Cas A, which, however, is not pulsar-powered, although it contains
a neutron star \citep{tan99,Pavlov00}. SN~1987A is another case that is now
entering its remnant stage, with an oxygen-mass of $\sim 1.8-1.9  \Msun$ within 
its innermost ejecta, i.e., out to $\sim 2\,000 \kms$ \citep{KF98,Jerk11}.
0540 makes an interesting link between pulsar-powered remnants in 
general and OSNRs.

In all the emission lines, projected images on the sky of the SNRC appear to be 
concentrated to a few blobs \citep[e.g.,][]{nlun11}. In addition, spectroscopy of 0540 by 
\citet{Kirshner89} showed that the weighted  emission of the SNRC is redshifted with respect to the 
LMC rest velocity. This was further studied by \citet{Morse06} who, however, also revealed 
a faint  [\ion{O}{iIi}] ``halo" between $-1400$ and $+1900 \kms$, with the center of expansion 
close to the systemic velocity of the surrounding H~II region. \citet{Sand13} used the
the integral field unit (IFU) ESO/VLT/VIMOS to probe the 3D structure of the whole SNRC
and found that [\ion{O}{iIi}] the full velocity range between $-1650 \leq v_{[\ion{O}{iIi}]} \leq +1700 \kms$, 
where $v_{[\ion{O}{iIi}]}$ is relative to the systemic velocity of local LMC redshift. In the plane of
the sky, an [\ion{O}{iIi}] halo can be traced out to $\sim 8  \arcsec$ from the pulsar
\citep{Morse06} . \citet[][who also included infrared observations with the Spitzer Space telescope]{Williams08} 
interpreted the halo emission to come
from photoionized unshocked supernova ejecta outside the PWN, but inside the
reverse shock of the SNR. At a distance of 50 kpc, $\sim 8 \arcsec$ and 
$v_{[\ion{O}{iIi}]} \approx 1700 \kms$ would imply an age of $\sim 1.1\times10^3$ years for 0540,
which is less than the spin down age of the pulsar \citep[$1.6\times10^3$ years,][]{rey85}. 

Spectroscopy and imaging show that there are clear differences, not only in the
size of the SNRC, but also in the structure of the emitting gas depending on the emission 
line traced. In particular, the structures of [\ion{O}{iIi}] and [\ion{S}{Ii}]  emission have a very low 
correlation in the bright southwest part of the SNRC, while the [\ion{S}{Ii}] emission correlates 
well with continuum emission \citep{nlun11}. This could be due to differences in
the ionization structure as a function of position angle, or it could be difference in
elemental structure. 

%\input{table_0.tex}
%--------------- Table 1 --------------
\begin{table*}[htb]
\begin{center}
\caption{Log of observations of \snr.}
\label{t:obs} 
\begin{tabular}{lccccccccc}
\hline
\hline
\multicolumn{10}{c}{{\it Imaging of 0540}} \\ %\cline{5-5}
Telescope & Instrument & & Filter & & UT & Exp. time & & sec~z & Seeing$^a$ \\ 
 & & & & & & (s) & & & (arcsec) \\
\hline
NTT & EMMI & & {[\ion{O}{iii}]$/$0}    & & 04:24:54$^b$& 900 & & 1.37& 1.20 \\
NTT & EMMI & & {[\ion{O}{iii}]$/$6000} & & 05:05:00	& 900 & & 1.43& 1.47\\
NTT & EMMI & & {[\ion{O}{iii}]$/$0}    & & 06:27:17	& 900 & & 1.62& 1.13 \\ \\
\hline
\hline
\multicolumn{10}{c}{{\it Spectroscopy of 0540}} \\
Telescope & Instrument & & Range & & UT & Exp. time & & sec~z & Seeing$^a$ \\ 
 & & & (nm) & & & (s) & & & (arcsec) \\
\hline
NTT & EMMI  & & 385-845 & & 07:26:45$^c$ & 2200 & & 1.89&1.35\\
VLT & FORS1 & & 360-606 & & 01:59:32$^d$ & 1320 & & 1.43&1.25 \\
& & & & & 02:23:25&1320&& 1.42&1.30\\
& & & & & 02:52:38&1320&& 1.41&1.18\\
& & & & & 03:16:31&1320&& 1.41&1.12\\
& & & & & 03:42:32&1320&& 1.41&1.24\\
& & & & & 04:06:25&1320&& 1.43&1.09\\
& & & & & 04:32:42&1320&& 1.45&0.88 \\
VLT&FORS1 & & 360-606 & & 02:07:57$^e$&1320&&1.42&1.48\\
& & & & & 02:31:50&1320&& 1.41&1.25\\
& & & & & 03:39:58&1320&& 1.42&0.92\\
& & & & & 04:03:52&1320&& 1.43&0.99\\
& & & & & 04:30:26&1320&& 1.46&1.03\\
& & & & & 04:54:20&1320&& 1.49&1.03\\  \\   
\hline
\hline
\multicolumn{10}{c}{{\it Spectroscopy of standard star LTT 3864}} \\
Telescope & Instrument & & Range & & UT & Exp. time & & sec~z & Seeing$^a$ \\ 
 & & & (nm) & & & (s) & & & (arcsec) \\
\hline
NTT & EMMI  & & 385-845 & & 08:14:55$^b$ & 900 & & 1.91 & 1.02\\
VLT & FORS1 & & 360-606 & & 08:51:03$^f$ &  20 & & 1.03 & 0.67\\
\hline
\end{tabular} \\
\end{center}
\begin{tabular}{ll} \hspace{20mm}
$^a$~FWHM from stellar profiles & $^d$~2002 Jan 9, slit ``1" in Fig.~\ref{f:OIIIim}, PA=88\degr\ \\ \hspace{20mm}
$^b$~1996 Jan 17 & $^e$~2002 Jan 10, slit ``3" in Fig.~\ref{f:OIIIim}, PA=51\degr\  \\ \hspace{20mm}
$^c$~1996 Jan 17, slit ``2" in Fig.~\ref{f:OIIIim}, PA=22\degr\ & $^f$~2002 Jan 10  \\ 
\end{tabular}
\end{table*}

%-----------------------figure1------------------------------
\begin{figure*}[t]
\begin{center}
\includegraphics[height=175mm,angle=270, clip]{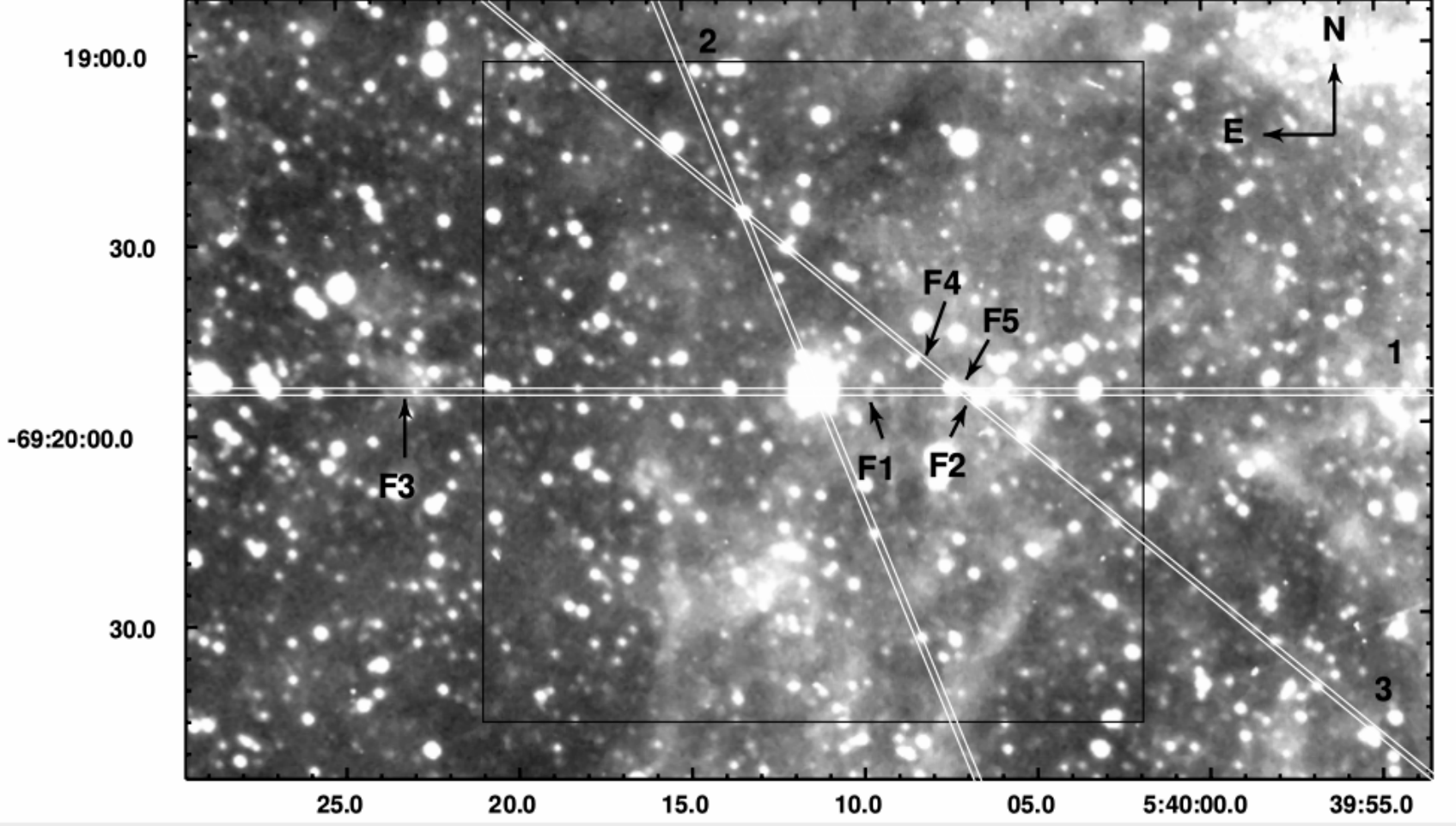}
\end{center}  
\caption{
3.2\arcmin$\times$2.0\arcmin\ [\ion{O}{iii}] image of \snr~obtained with 
NTT/EMMI in 1996 using the narrow  [\ion{O}{iii}]$/$0 filter centered at  
the rest wavelength of [\ion{O}{iii}]~$\lambda5007$. The bright extended object in the 
center of the image is the central part of the remnant containing the pulsar. 
The white lines marked by numbers show the slit positions used  at the 
spectral observations of the SNR with NTT/EMMI (2) and VLT/FORS1  
(1 and 3). The areas marked ``F1-F5'' show the positions of five H~II regions 
identified from the spectral observations and described in the text. 
A 100\arcsec$\times$100\arcsec\ black box outline the region shown in Fig.~\ref{f:OIIIim2}.
}
\label{f:OIIIim}
\end{figure*}
%-----------------------figure2------------------------------

\begin{figure*}[tbh]
\begin{center}
\includegraphics[width=60mm, clip]{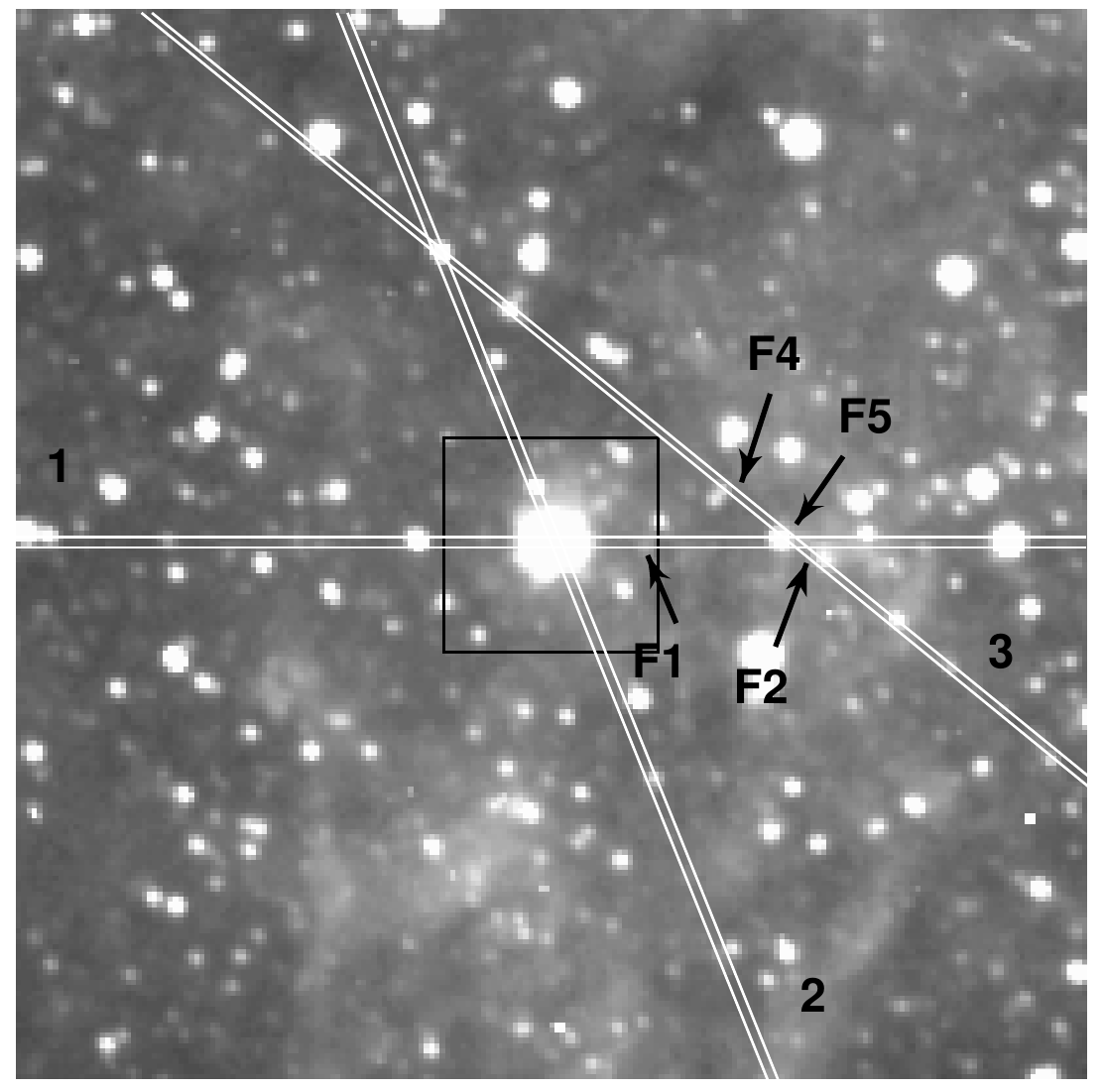}\includegraphics[width=60mm, 
clip]{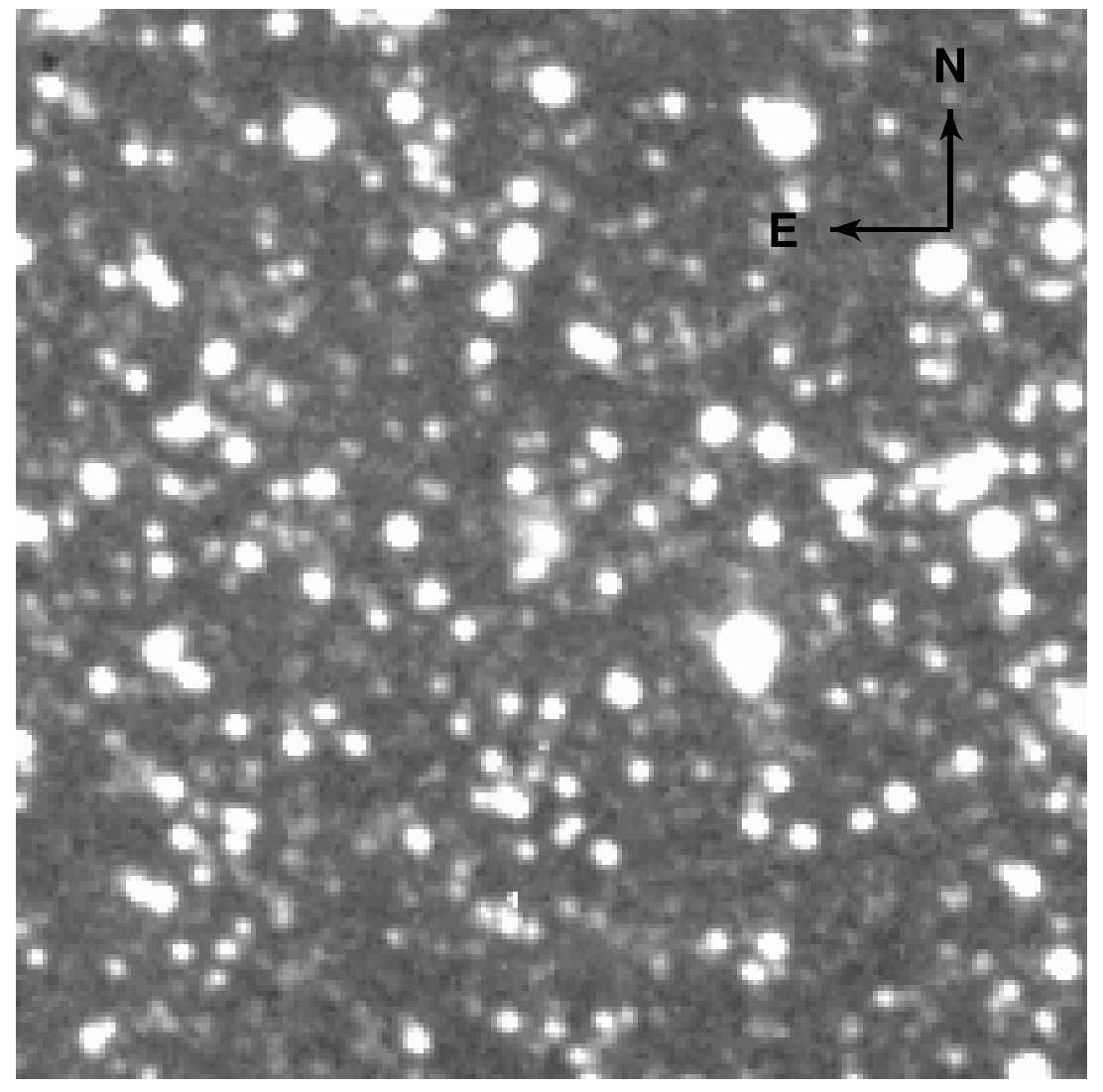}\includegraphics[width=60mm, clip]{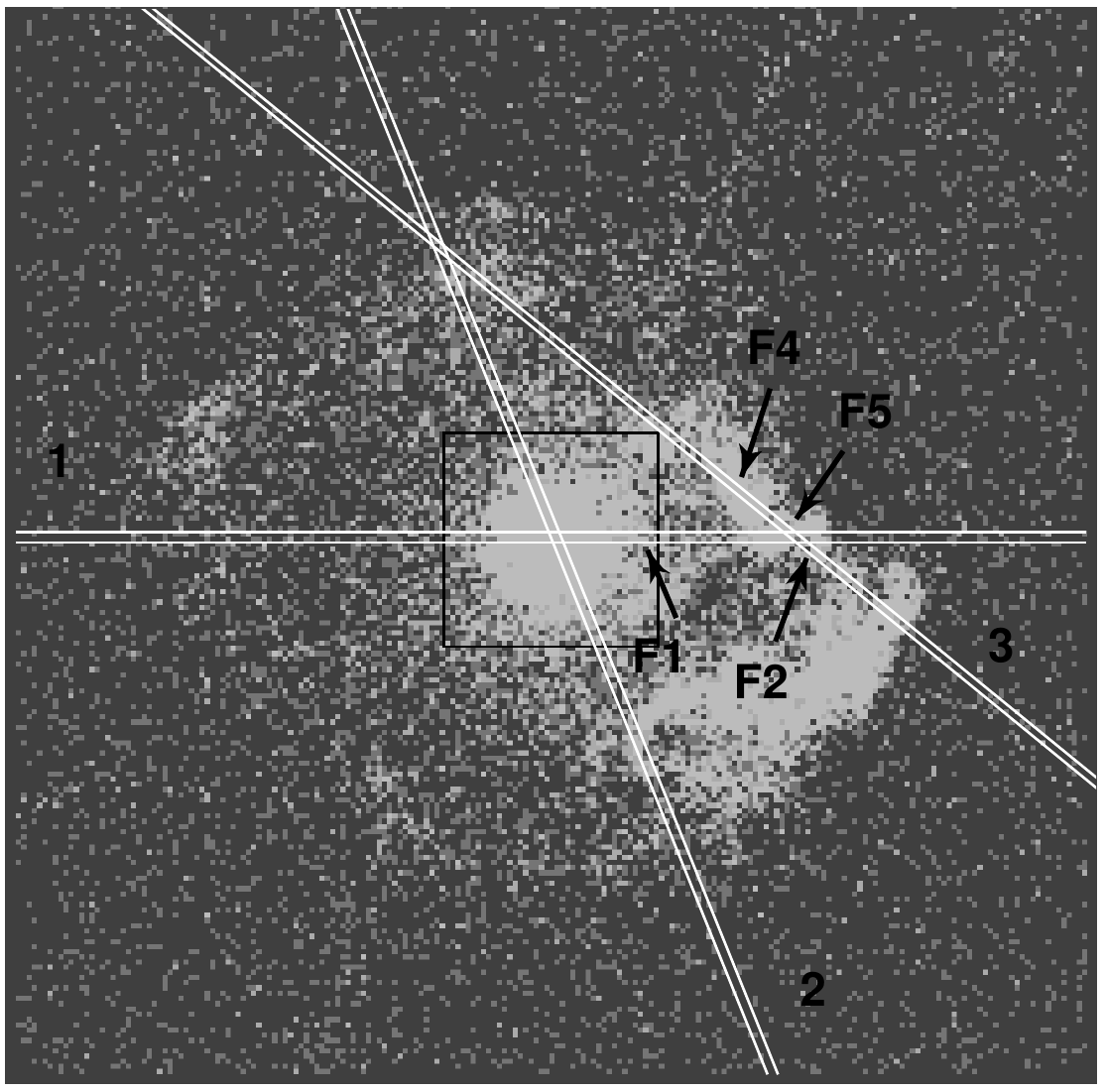}
\end{center}
\caption{The 100\arcsec$\times$100\arcsec\ region of the field around 0540 
marked with a box in Fig.~\ref{f:OIIIim}. The images are  obtained in 
the [\ion{O}{iii}]$/$0 ({\it left}) and [\ion{O}{iii}]$/$6000 ({\it middle}) bands with 
NTT/EMMI, and in the 1.5$-$6.4 keV X-ray range ({\it right}) with Chandra/ACIS 
(Hwang et al. 2001). The positions of all the slits used in the spectral 
observations, and of the detected H~II regions are marked as in 
Fig.~\ref{f:OIIIim}. A 20\arcsec$\times$20\arcsec\ box marks the inner region
of 0540 blown up in Fig.~\ref{f:SII_Chan}. Note that the H~II structures named 
as ``F1", ``F2", ``F4" and ``F5'' are projected on regions with strong X-ray 
emission from the SNR.} 
\label{f:OIIIim2}
\end{figure*}

The 3D structure probed by \citet{Sand13} shows that [\ion{S}{Ii}] comes from a more central 
part, $-1200 \leq v_{[\ion{S}{ii}]} \leq +1200 \kms$, than [\ion{O}{Iii}] with very weak emission 
in the $-1200 \leq v_{[\ion{S}{ii}]} \leq -800 \kms$ range. \citet{Sand13} used the [\ion{S}{Ii}] line
to make a map of the electron density, $n_{\rm e}$, and found that the [\ion{S}{Ii}] emission
mainly comes from regions with $n_{\rm e} \lsim 750 \cm3$, although there could be
``pockets" with $n_{\rm e} \sim 2\times10^{-3}\cm3$. In particular, [\ion{S}{Ii}] emission
is strong from a specific feature called ``the blob" in our previous papers 
\citep{Ser04,Ser05,nlun11,Sand13}, located $\sim 1\farcs3$ south-west of the pulsar.
This blob is conspicuous in continuum emission from radio to X-rays \citep{nlun11,plun20}, and
\citet{DeLuca07} argued for that the blob appears to change position and continuum brightness 
between 1995 and 2005. \citet{nlun11} offered the alternative explanation, also including 
optical polarimetry and X-ray data in their analysis, that a local energy deposition may have 
occurred around 1999, and that the emission from that faded until later epochs. 
\citet{Sand13} suggested that the energy deposition in the blob region could be due to 
interaction of the blob with the pulsar-wind torus. 

There are also more recent evidence of time-varying emission from the PWN. \citet{Ge19}
showed that the sudden change in the spin-down rate in December 2011 could be linked to a gradual
brightening of the PWN in X-rays by $\sim 30\%$ in about $\sim 2$ years. \citet{nlun11} suggested
that spatial changes in polarization angle along an axis in the northeast-southwest direction,
and crossing the blob, point to past changes in activity along this axis. It is intriguing that the
presumed pulsar jet has been suggested to point nearly orthogonal to this axis \citep[e.g.,][]{GW00}.

Here we report on imaging and spectroscopic observations of 0540 and its
immediate surroundings using NTT/EMMI from 1996 and VLT/FORS from 2002 (Sect. 2 \& 3).
We also include results from \citet{Sand13}.
Since our data are deeper than previously reported spectral studies, our aim is to improve the
knowledge about the spatial velocity distribution of various elements, and on the 
temperature and ionization of the different parts of the remnant and its neighborhood (Sect. 4). 
%As the data were taken at periods close to the brightening of the southwestern region, we
%discuss in Sect. 4 possible constraints on that from our data as well.
We summarize our conclusions in Sect.5. A preliminary version of parts of
our work was presented in \citet{Ser05}.

\section{Observations and data analysis}

%%%%%%%%%%%%%%%%%%%%%%%%%%%%%%%%%%%%%%%%%%%%%%%%%%%
\subsection{NTT Observations}
%%%%%%%%%%%%%%%%%%%%%%%%%%%%%%%%%%%%%%%%%%%%%%%%%%%
Observations of 0540 were performed on 1996 January 17, 
using the 3.58m ESO/NTT equipped with the
ESO Multi-Mode Instrument~(EMMI). Narrow-band images were obtained 
in [\ion{O}{iii}]~$\lambda=5007$~\AA\ at zero velocity using the 
[\ion{O}{iii}]$/$0 filter\footnotemark \footnotetext{http://www.ls.eso.org/lasilla/Telescopes/NEWNTT/
emmi/emmiFilters.html}, and through the [\ion{O}{iii}]$/$6000 
filter which is shifted to +6000 km s$^{-1}$ from the rest wavelength 
of [\ion{O}{iii}]. %~$\lambda=5007$\AA. 
The pixel size was 0\farcs268$\times$0\farcs268.
On the same night we also carried out low-resolution (2.23 \AA$/$pixel)
long-slit spectroscopy of 0540 in the 3850-8450~\AA\ range using 
the EMMI RILD mode with 
grism\#3\footnotemark \footnotetext{http://www.ls.eso.org/lasilla/Telescopes/NEWNTT/
emmi/emmiRild.html}.
The log of the observations is presented in 
Table~\ref{t:obs}. The EMMI [\ion{O}{iii}]$/$0 image of the 0540 neighborhood is 
shown in Fig.~\ref{f:SII_Chan}, where we also show the slit position of the 
NTT spectral observation (marked by ``2''). The position angle of the slit 
is PA=22\degr, and the slit crosses the SNRC containing the pulsar and the
PWN.  

%-----------------------figure3------------------------------
\begin{figure}[t]
\center{\includegraphics[width=85mm, angle=0, clip]{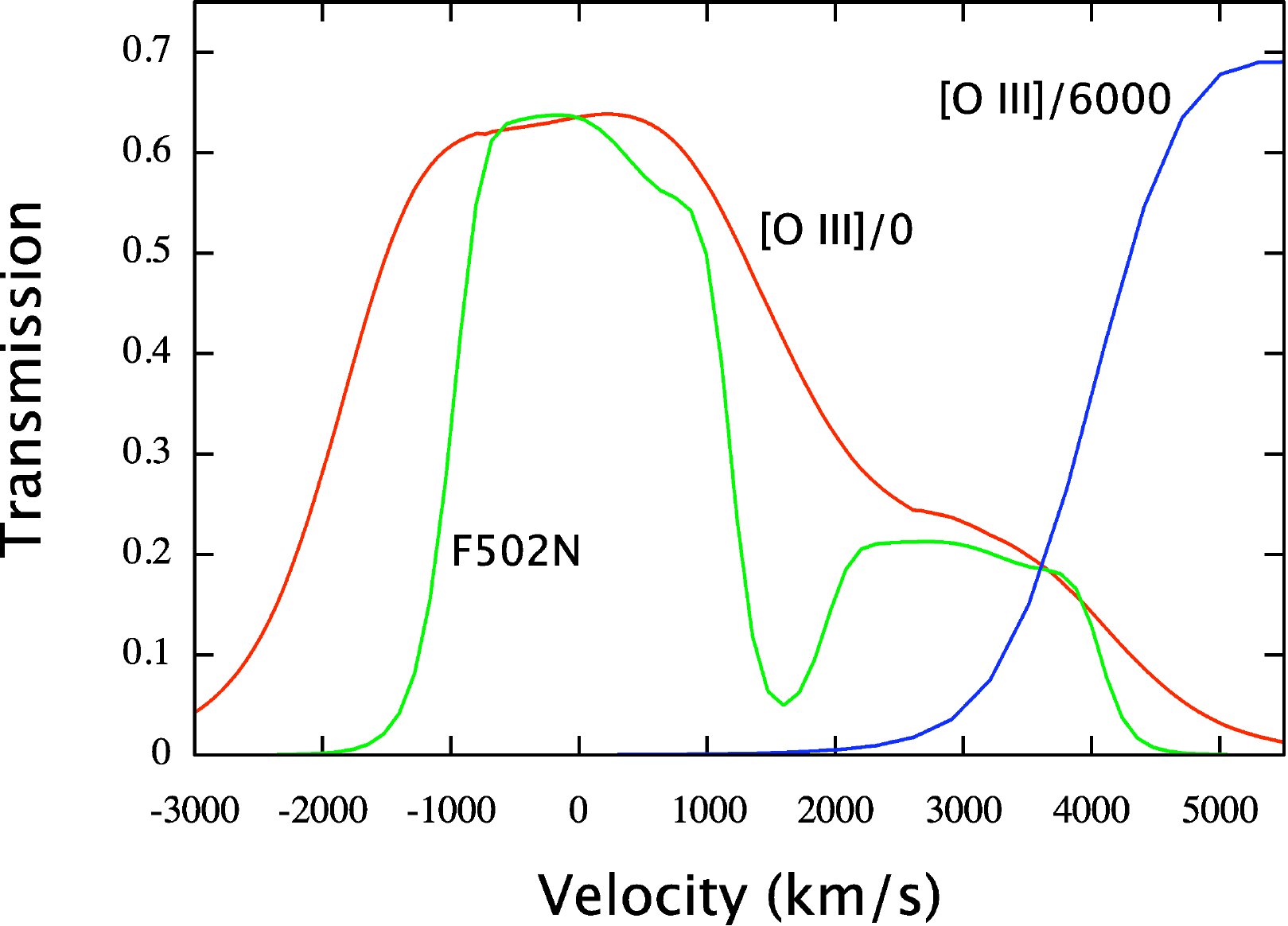}}
\caption{
Effective transmission curves for the ESO/[\ion{O}{iii}]/0 (red), ESO[\ion{O}{iii}]/6000 (blue) and 
HST/F502N (green) filters. The velocity is centered on the rest velocity of
[\ion{O}{iii}]~$\lambda$5007 in the local gas in the LMC, which we assume has a systemic 
shift of $+273 \kms$ compared to Earth \citep[cf.][]{Morse06}. The red wings of the curves 
ESO/[\ion{O}{iii}]/0 and HST/F502N are due to emission from [\ion{O}{iii}]~$\lambda$4959 that enters 
the filter with a shift of $\sim 2\,870 \kms$. We have assumed that this line is a factor of three 
weaker than [\ion{O}{iii}]~$\lambda$5007.
} \label{f:Filters}
\end{figure} 

%-----------------------figure4----------------------------
\begin{figure*}[tbh]
\begin{center}
\end{center}  
\begin{center}
\includegraphics[width=83mm, clip]{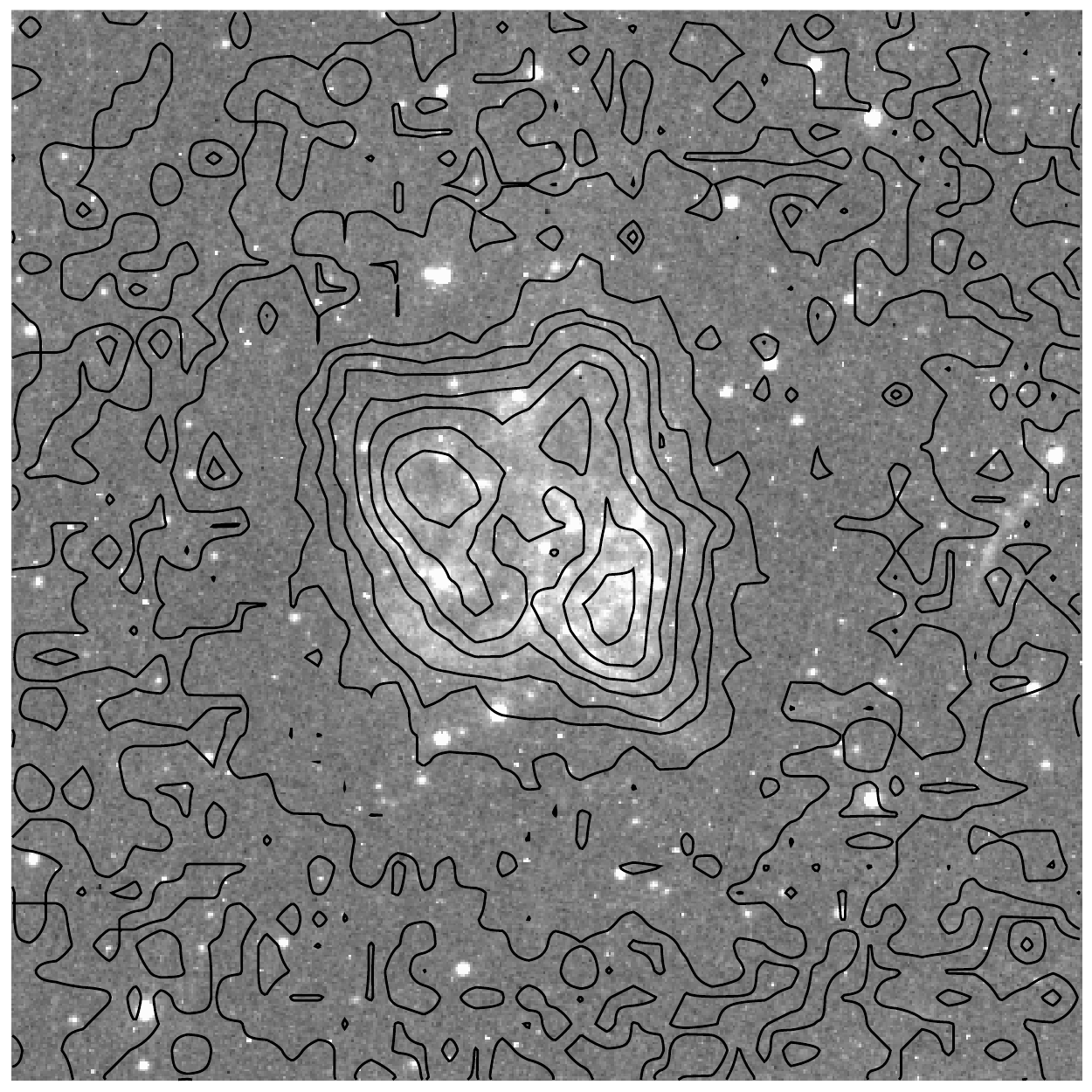}\includegraphics[width=83mm, clip]{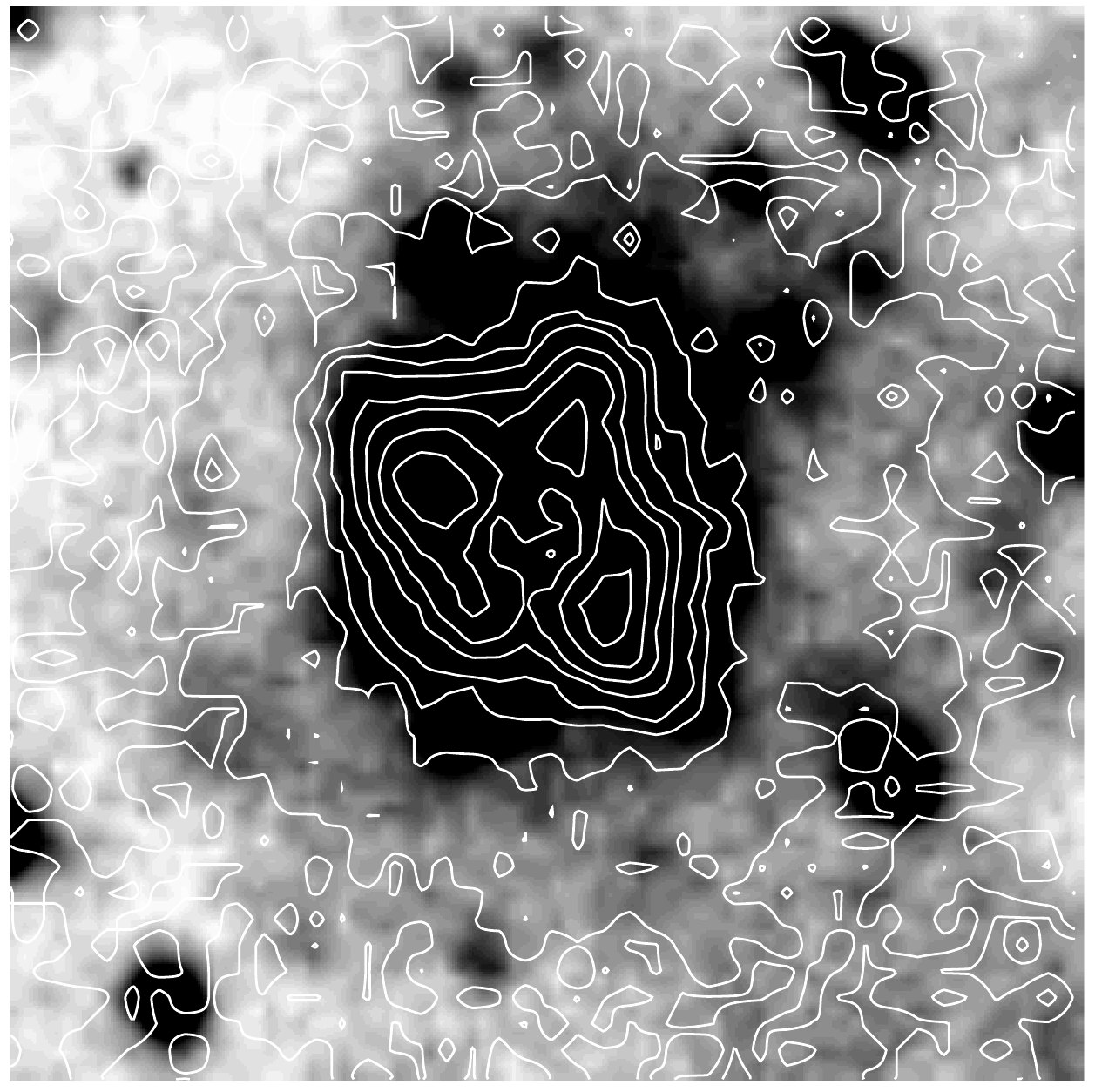}
\end{center}
\caption{ 
Central 20\arcsec$\times$20\arcsec\ field of 0540 as viewed by HST/WFPC2/F673N 
({\it left, positive}) and NTT/EMMI/[\ion{O}{iii}]/0 ({\it right, negative}). 
North is up and east to the left.
The contours of the Chandra/ACIS X-ray fluxes in Fig.~\ref{f:OIIIim2} are overlaid in 
both images. The spatial distribution of the X-ray emission 
shows an elongated structure with NW-SE jets, which is associated with the 
PWN \citep[cf.][]{Ser04,nlun11}. As seen, the [\ion{S}{ii}] emission dominating 
in the HST image does not extend outside the PWN, whereas the [\ion{O}{iii}] glow 
in the NTT image extends far outside the PWN.
} \label{f:SII_Chan}
\end{figure*}

%-----------------------figure5----------------------------
\begin{figure*}[tbh]
%\begin{center}
%\end{center}  
\begin{center}
\center{\includegraphics[width=180mm, angle=0, clip]{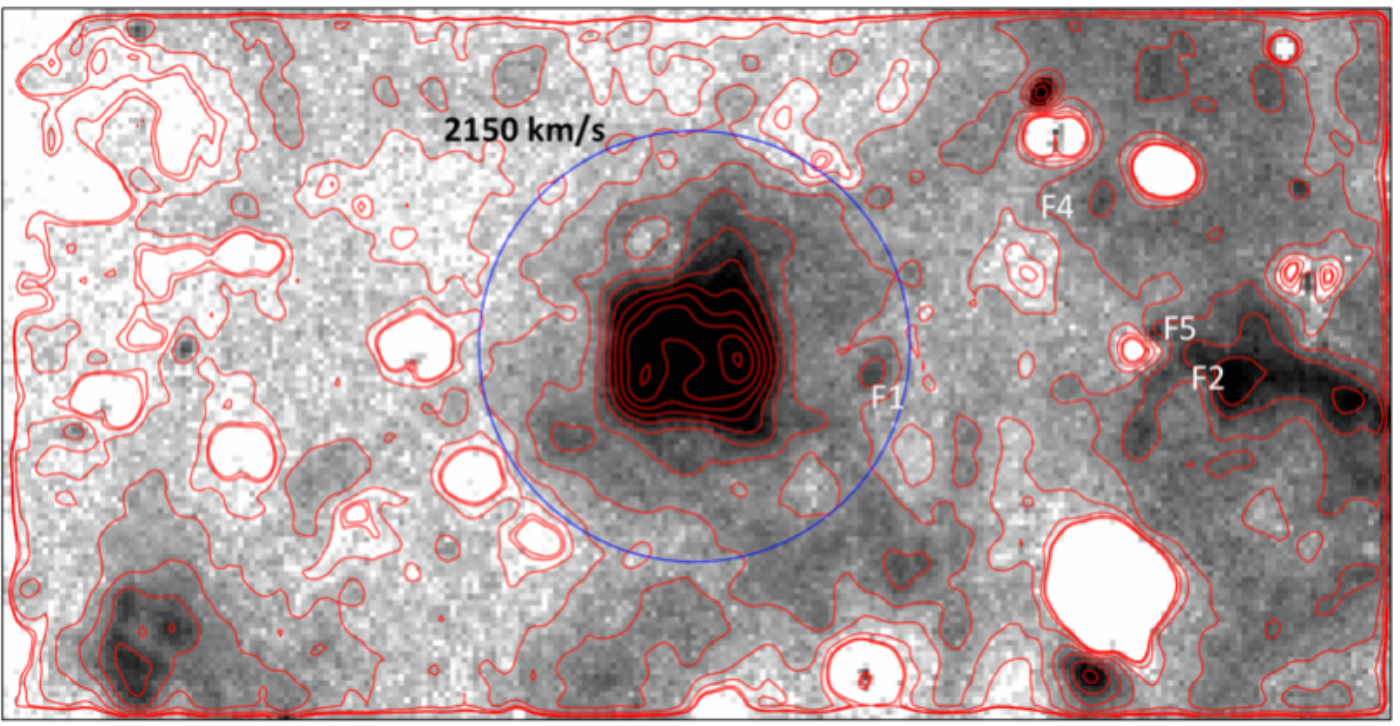}}
\end{center}
\caption{ 
{\it Right:} 65\arcsec\ $\times$ 33\arcsec\ difference image between NTT/EMMI/[\ion{O}{iii}]/0 and NTT/EMMI/[\ion{O}{iii}]/6000 with 
square root intensity scaling to bring out faint details. The [\ion{O}{iii}]/0 image was smoothed to match the slightly worse seeing
of the [\ion{O}{iii}]/6000 image before subtraction. North is up and east is to the left.  The image can trace faint [\ion{O}{iii}] emission
 to large distances from the pulsar. A blue circle is drawn with radius 10.0\arcsec, corresponding to $\sim 2\,150 \kms$ for ejecta coasting 
 freely for 1\,100 years (or $\sim 1\,980 \kms$ for 1\,200 years) at at distance of 50 kpc. The filaments F1, F2, F4 and F5, discussed 
 in the text, are marked. Note PWN protrusions 
 to the  north, presumably in the pulsar jet direction, and to the southwest, which could be a structure similar to the Crab chimney.
} \label{f:O3_diff}
\end{figure*}

%-----------------------figure6----------------------------
\begin{figure*}[tbh]
%\begin{center}
%\end{center}  
\begin{center}
\center{\includegraphics[width=180mm, angle=0, clip]{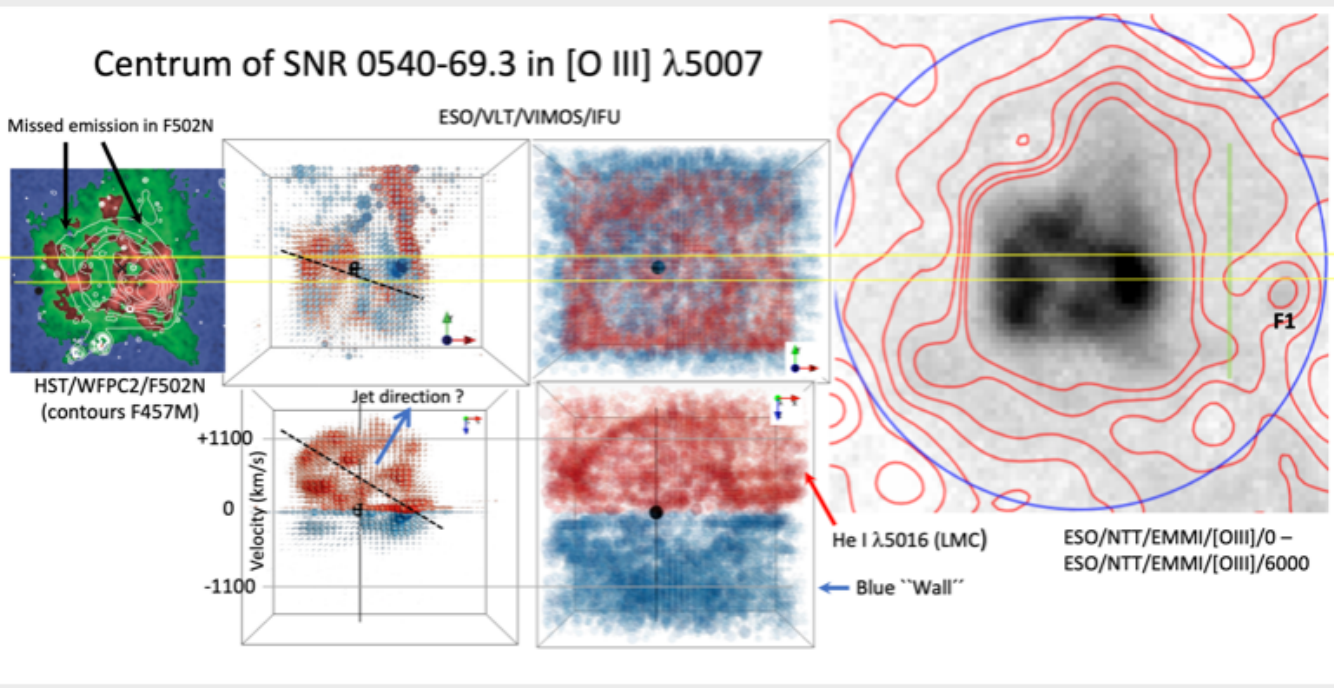}}
\end{center}
\caption{ 
{\it Right:} 20\arcsec$\times$20\arcsec\ difference image between the NTT/EMMI/[\ion{O}{iii}]/0 and  NTT/EMMI/[\ion{O}{iii}]/6000 images.
The blue circle has the same meaning as in Fig.\ref{f:O3_diff}. The flux scaling is linear. To highlight the [\ion{O}{iii}] glow, red 
contours are inserted for intensities up to $23\%$ of the peak surface intensity of the PWN. Filament F1 has been marked. North is up 
and east to the left. {\it Middle two top panels:} [\ion{O}{iii}]~$\lambda5007$ as viewed by VIMOS/IFU \citet{Sand13}. The left of the 
two middle panels shows the central part, where blue is for approaching ejecta, and red for receding. The right of the two panels brings 
out  fainter halo emission. {\it Middle two bottom panels:} Same as the two top panels, but in velocity space. A symmetry axis 
(also shown in the top panel) is marked that goes through rings of [\ion{O}{iii}] emitting ejecta, and a possible jet axis is highlighted 
for the pulsar jet. For the lower right panel, likely contamination from LMC H~II region \ion{He}{i}~$\lambda5016$ is marked, as is also
a region with emission on the approaching side ($\leq -750 \kms$) named the blue ``Wall" (also seen for [\ion{O}{iii}]~$\lambda5007$ in
 Fig.~\ref{f:OIIIim_slit1}). {\it Left:} Wavelet filtered HST/WFPC2/F502N map and contours of a wavelet filtered HST/WFPC2/F457M 
 map \citep{nlun11}. Areas where the F502N fails to detect [\ion{O}{iii}] emission are highlighted. To guide the eye, a
 1$\arcsec$ slit with PA~$=90$\degr\ has been drawn across all top panels. A green line in the rightmost panel marks how far west the 
 VIMOS field reaches. All images are to scale.
} \label{f:O3_new}
\end{figure*}

%-----------------------figure7----------------------------
\begin{figure}[t]
\center{\includegraphics[width=175mm, angle=270, clip]{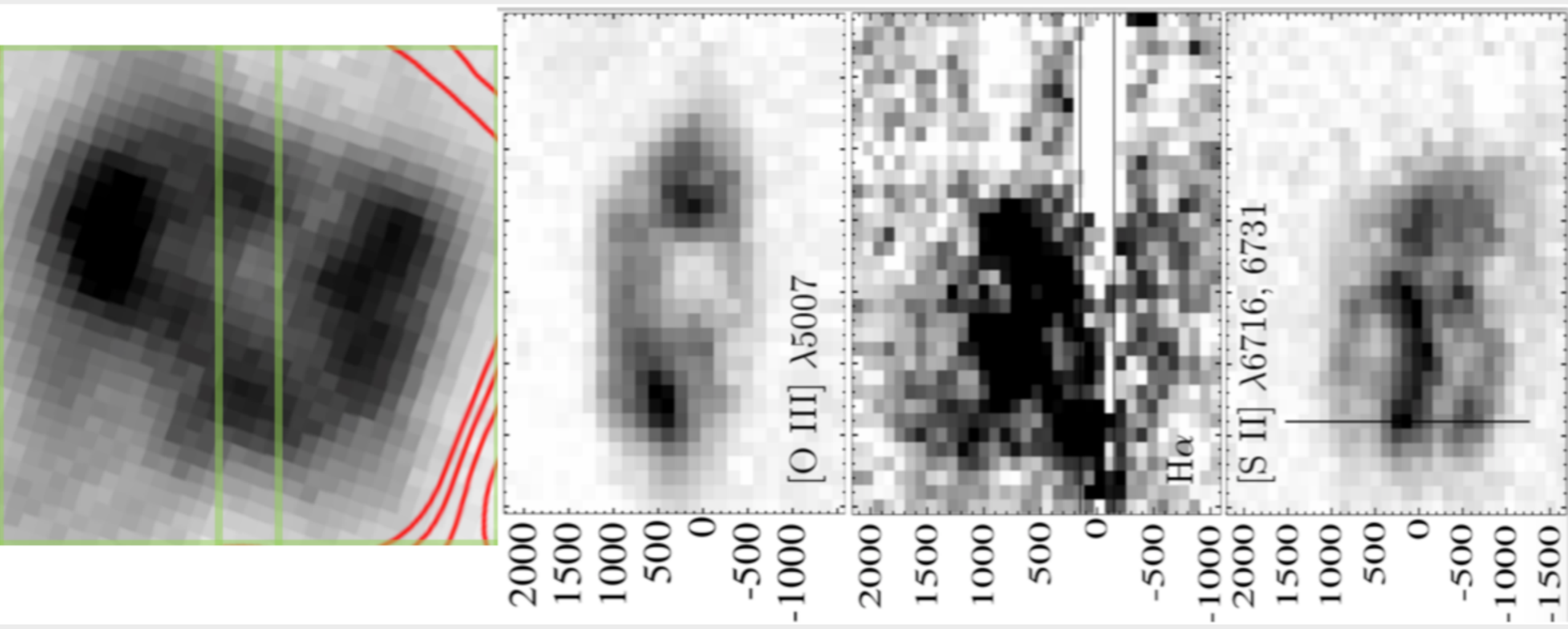}}
\caption{{\it Top panel:} Central 8\arcsec$\times$8\arcsec\ of \snr\ using the NTT/EMMI difference image 
with linear intensity scaling in  Fig.~\ref{f:O3_new}, rotated $68\degr$ to match the horizontally marked slit 
2. {\it Lower panels:} Space-velocity images along slit 2 for [\ion{O}{iii}]~$\lambda$5007, H$\alpha$, and 
[\ion{S}{ii}]~$\lambda\lambda$6716,6731. The vertical and horizontal axes show 
the velocity (in $\kms$, corrected for the LMC redshift) and the spatial 
coordinate (in arcseconds) along the slit, respectively.
The horizontal black lines (for H$\alpha$) mark the velocity spread of H~II 
regions in LMC ($\sim \pm 125 \kms$). The velocity interval affected 
by the subtraction of the uneven LMC background is somewhat larger 
because of the finite spectral resolution. H$\alpha$ is more affected by the background subtraction 
than other lines, and blended with [\ion{N}{ii}]~$\lambda\lambda$6548,6583. The vertical line in the 
[\ion{S}{ii}] image is described in Sect. 3.2.1.
} \label{f:velom1}
\end{figure}

%-----------------------figure8------------------------------
\begin{figure}[t]
\center{\includegraphics[width=135mm, angle=270, clip]{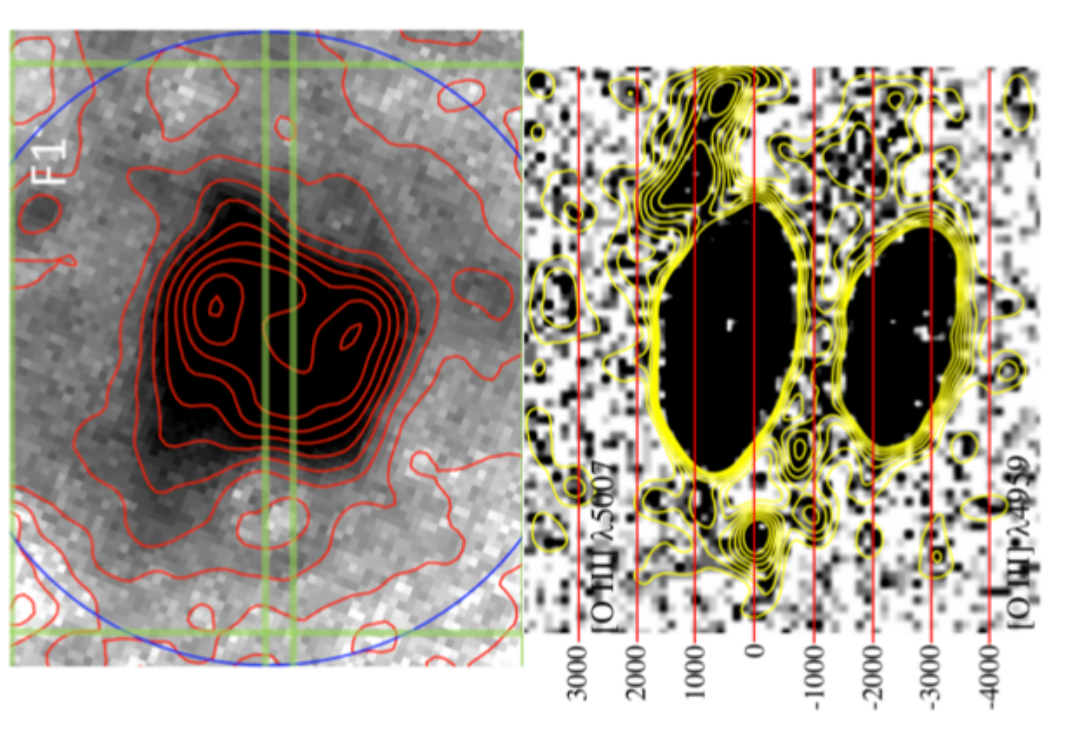}}
\caption{{\it Top panel:} Central $20\arcsec\times$17\arcsec\ of \snr\ using the 
NTT/EMMI difference image with square root intensity scaling in Fig.~\ref{f:O3_diff}, rotated $68\degr$ to 
match the horizontally marked slit 2 {\it Bottom panel} Space-velocity image along slit 2 for 
[\ion{O}{iii}]~$\lambda\lambda$4959,5007. Velocity, corrected for the LMC redshift, is for  [\ion{O}{iii}]~$\lambda$5007.
The large dynamic range of the images reveals the faint glow emission outside the SNRC of 0540. 
} \label{f:OIIIim_slit2}
\end{figure}  

%-----------------------figure9------------------------------
\begin{figure}[t]
\center{\includegraphics[width=180mm, angle=270, clip]{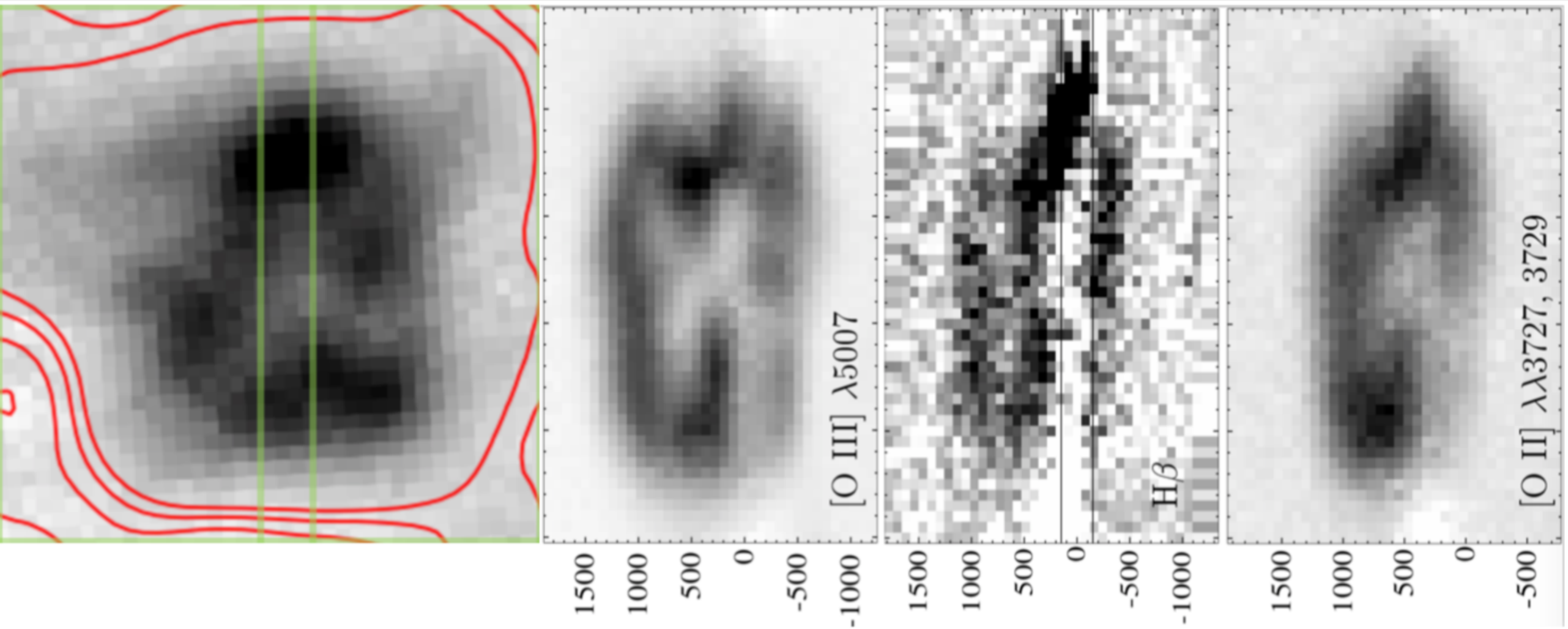}}
\caption{Same as in Fig.~\ref{f:velom1}, but for slit 1 (cf. Fig.~\ref{f:OIIIim2}), 
and for a different set of emission lines, as marked in the panels. 
The spatial extent of the images along the slit is 10\arcsec. 
As for H$\alpha$ in Fig.~\ref{f:velom1}, the H$\beta$ image is corrupted 
by subtraction of the uneven LMC background. Note the very different 
structures in [\ion{O}{ii}] and  [\ion{O}{iii}].
} \label{f:velom2}
\end{figure}

%-----------------------figure10------------------------------
\begin{figure}[t]
\center{\includegraphics[width=145mm, angle=270, clip]{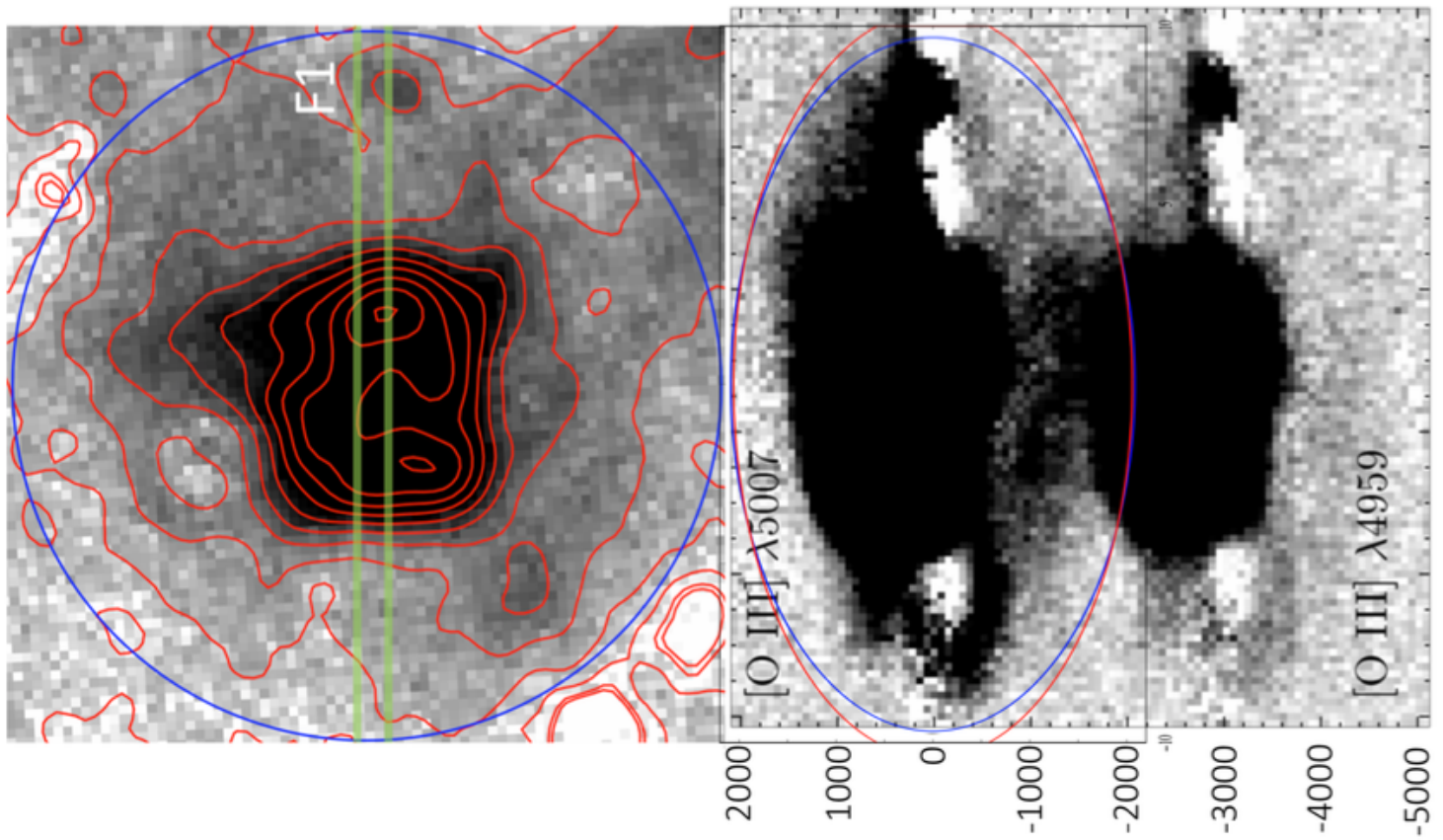}}
\caption{Same as in Fig.~\ref{f:OIIIim_slit2}, but for slit 1.  
The spatial extent of the images along the slit is 20\arcsec. Filament F1 is
highlighted. Ellipses are drawn around [\ion{O}{iii}]$\wl$5007 marking an ejecta velocity of $+1\,900 \kms$ 
at 4\arcsec\ west of the pulsar, assuming freely coasting ejecta and an age of 1\,100 years (blue) and 1\,200 years (red) 
for 0540. See text for further details.
} \label{f:OIIIim_slit1}
\end{figure}

%%%%%%%%%%%%%%%%%%%%%%%%%%%%%%%%%%%%%%%%%%%%%%%%%%%
\subsection{VLT observations}
%%%%%%%%%%%%%%%%%%%%%%%%%%%%%%%%%%%%%%%%%%%%%%%%%%%

Further spectral observations were carried  out in the 3600--6060~\AA\ range  
on 2002 January 9 and 10 with the FOcal Reducer/low dispersion Spectrograph 
(FORS1) on the 8.2m UT3 (MELIPAL) of ESO/VLT, using the grism 
GRIS\_600B\footnotemark \footnotetext{http://www.eso.org/instruments/fors1/grisms.html}. 
This has a dispersion of 50~\AA$/$mm, or 1.18~\AA$/$pixel. 
The optical path also includes a Linear Atmospheric Dispersion Corrector that 
compensates for the effects of atmospheric dispersion \citep{av97}. 
The spatial scale of the FORS1 CCD is 0\farcs2 per pixel. 
On January 9 and 10 we obtained seven and six exposures of 1320 s each, 
respectively (see Table~\ref{t:obs}), i.e., a total exposure time of 154 and 
132 minutes. The position angles, PA=88\degr\ and PA=51\degr, 
were the same in all these exposures, and the slit positions are marked 
by ``1'' and  ``3'' in Fig.~\ref{f:OIIIim}, respectively. 

Slits 1 (VLT) and  2 (NTT) were chosen to include the SNRC and the pulsar. 
Slit 3 (VLT) does not cross the SNRC, but was placed west of it 
to probe the emission from the outer shell that is most clearly identified 
in the Chandra X-ray images \citep[][see Fig.~\ref{f:OIIIim2}]{Hw01}.  
All spectroscopic observations (both NTT and VLT) were performed using a 
slit width of~1\arcsec~and the seeing was generally about 1\arcsec. 

\subsection{NTT and VLT data reduction}

The spatial and spectral images were bias-subtracted and flat-fielded 
using standard procedures
and utilities from the NOAO {\sf IRAF} package. We used the averaged sigma 
clipping algorithm {\sf avsigclip} with the 
{\sf scale} parameter set equal to {\sf none} to combine the images.
Wavelength calibration for the 2D spectral images were done using arc 
frames obtained with a He/Ar lamp and procedure {\sf transform}. 
The spectra of the objects were then extracted from the 2D image
using the {\sf apall} and {\sf background} tasks.
Flux calibration of the spectra was accomplished by comparison to the 
spectrophotometric standard star LTT 3864 \citep{ham94} for both the NTT and 
VLT spectra. Atmospheric extinction corrections were performed using 
a spectroscopic extinction table provided 
by ESO\footnote{http://www.eso.org/observing/dfo/quality/UVES/files/ \\ 
/atmoexan.tfits}. 

%%%%%%%%%%%%%%%%%%%%%%%%%%%%%%%%%%%%%%%%%%%%%%%%%%%
\subsection{HST and Chandra observations}
%%%%%%%%%%%%%%%%%%%%%%%%%%%%%%%%%%%%%%%%%%%%%
The \snr\ field has been imaged with the HST several times 
in various bands.  In particular, narrow and medium band filters are useful to 
compare with our spectral information. The HST data highlighted here are described 
and discussed in \citet{Mo03}, \citet{Morse06} and \citet{nlun11}.

To compare the optical and X-ray data  we also discuss data
retrieved from the Chandra archive. In \citet{nlun11} we discuss these
data in greater detail and how they were reduced. 

\section{Results and Discussion}

\subsection{Imaging}

Narrow-band [\ion{O}{iii}]  %~$\wl$5007 
images of 0540 obtained with NTT/EMMI through the zero velocity and 
+6\,000 $\kms$ filters ([\ion{O}{iii}]/0 and [\ion{O}{iii}]/6000, respectively) are shown 
in the left and middle panels of Fig.~\ref{f:OIIIim2}, 
respectively. The FWHM of the transmission curve for both filters corresponds 
to 3\,300 $\kms$ (cf. Fig.~\ref{f:Filters}). The +6\,000 $\kms$ image shows no obvious sign of
[\ion{O}{iii}] emission in any part of the remnant, confirming the spectroscopic
observations by \citet{Kirshner89} who did not find any line emission 
with redshifts higher than $\sim 3\,000 \kms$. The only emission seen from the 
remnant in the +6\,000 $\kms$ image is the continuum radiation from the PWN
powered by the pulsar B0540-69.3 in SNRC \citep[see][for 
details on the continuum part]{Ser04,nlun11}.

The zero-velocity [\ion{O}{iii}] image of Fig.~\ref{f:OIIIim2} reveals more extended emission
in accordance with \citet{Car92}. Faint patchy nebulosity from
the LMC background is also seen. A comparison between the [\ion{O}{iii}]/0 image and the 
X-ray image obtained with Chandra/ACIS (right panel of Fig.~\ref{f:OIIIim2}), 
where the shell-like structure associated with the outer SNR shock activity 
is clearly detected \citep{Hw01},     
shows no obvious morphology that connects the optical to the X-ray emission.  
As seen in Fig.~\ref{f:SII_Chan}, the slits used for 
our spectroscopic observations encapsulate X-ray structures of the SNR. 
This enables us to test (see Sect.~3.2) whether or not
optical structures projected on the regions with strong X-ray emission 
and named ``F1", ``F2", ``F4'' and ``F5'' in Figs.~\ref{f:OIIIim} 
and \ref{f:OIIIim2} belong to, or are affected by the SNR.        

The high spatial resolution of HST/WFPC2 allows us to resolve in great detail  
the filamentary structure of the SNRC. In Fig.~\ref{f:SII_Chan} we present 
an image of this field obtained in the narrow band 
F673N, centered on [\ion{S}{ii}]~$\lambda\lambda$6716,6731.
The F637N data, and data for the F502N band, centered on [\ion{O}{iii}]~$\lambda$5007,
were shown and discussed by \citet{Morse06} and \citet{nlun11}. 
%See also Fig.~\ref{f:SII_Chan}.    

Although detailed, the HST [\ion{S}{ii}] and [\ion{O}{iii}] narrow-band images  may not fully 
reliably reveal the real structure and size of the SNRC. The reason is that the narrow band 
filters F502N and F673N may not fully cover the whole line profiles of the [\ion{O}{iii}] and 
[\ion{S}{ii}] emission, respectively. As shown in \citet{Morse06}, the F502N filter cuts 
around 5035~\AA, which corresponds to $\sim +1\,415 \kms$ from the rest wavelength 
of [\ion{O}{iii}]~$\lambda$5007 (if we include a systemic shift of $+273 \kms$ for
the local gas in the LMC), but spectroscopic studies show 
\citep[][Sect.~3.2 of this paper]{Kirshner89} that the faint [\ion{O}{iii}] emission 
extends up to at least +1\,700 $\kms$ (in the rest frame of  LMC). This can, as pointed out
by \citet{Morse06}, mean that F502N image misses out some high-velocity
features of the SNRC (see also below). 

The F673N filter cuts around 6770~\AA, which corresponds to $\sim +1\,475 \kms$ from the rest wavelength 
of [\ion{S}{ii}]~$\lambda$6731 (for the same systemic redshift). Since the [\ion{S}{ii}] emitting region
is truly more compact than that emitting [\ion{O}{iii}] also along the line of sight \citep{Sand13}, 
the F673N filter (cf. Fig.~\ref{f:SII_Chan}) should encapsulate most of the [\ion{S}{ii}]  emission. 

Figure~\ref{f:SII_Chan} demonstrates the difference between the
HST/WFPC2/F673N (left) and NTT/EMMI/[\ion{O}{iii}]/0 (right) images.  
The images are presented for the same spatial scale and the contours 
of the Chandra/ACIS X-ray image are overlaid on each optical image 
to help us to better compare the sizes and shapes of the SNRC in both images. 
In the F673N image the emission from the SNRC in the optical traces the X-ray
emission, and we see no [\ion{S}{ii}] emission outside the X-ray PWN
\citep[see also][]{nlun11}.

One effect not discussed in \citet{Morse06} is the influence of the [\ion{O}{iii}]~$\lambda$4959
line on filter observations centered on [\ion{O}{iii}]~$\lambda$5007. Since the velocities of the 
remnant are high enough for the $\lambda$4959 component to be probed by the F502N filter
at velocities in the range $1\,900-4\,000 \kms$ (including LMC redshift), and since the expected intensity ratio 
$I_{\lambda5007}/I_{\lambda4959}$ is 3, the effective F502N filter transmission of 
[\ion{O}{iii}]~$\lambda\lambda4959,5007$ looks like the green curve in Fig.~\ref{f:Filters}. From this
we can clearly see that [\ion{O}{iii}]~$\lambda\lambda4959,5007$ is not probed well by F502N for 
$v_{[\ion{O}{iii}]} \leq -1\,200 \kms$ and at a redshift between $1\,300-1\,900 \kms$, bottoming at around 5\% 
effective transmission at $\sim +1\,600 \kms$, while emission from gas at higher velocities has about 20\% 
effective transmission. If we compare with the velocity maps of \citet{Sand13} for the VIMOS/IFU field, 
the F502N filter mainly misses emission from gas moving away from us at projected positions $\sim 2\arcsec\ - 3\arcsec$
northeast and northwest of the pulsar, close to the pulsar and 
$\sim 3\arcsec$ southwest of the pulsar. This is shown in Fig.~\ref{f:O3_new}. 
As we discuss in Sect. 3.2, there are also regions outside
the VIMOS/IFU field that are not probed well by the F502N filter.

The [\ion{O}{iii}]/0 filter used in our NTT/EMMI observations has its peak transmission at 
5009~\AA\ and has 25\% transmission of the peak value at 5042~\AA. Including
the emission in [\ion{O}{iii}]~$\lambda$4959, Fig.~\ref{f:Filters} shows that 20\% effective
transmission extends out to $\sim +3500 \kms$ (including LMC redshift), without any dropouts in transmission
like those of the F502N filter. The 20\% transmission cutoff on the blue side of [\ion{O}{iii}]~$\lambda$5007
is at $\sim -2\,200 \kms$. The [\ion{O}{iii}]/0 filter is therefore likely to probe gas with
all likely velocities, and with less bias than the F502N filter. Fig.~\ref{f:Filters} also shows
that the effective transmission is the same (19 \%) for [\ion{O}{iii}]/0 and [\ion{O}{iii}]/6000 filters at 
$\sim 3\,600 \kms$, and that the [\ion{O}{iii}]/6000 filter only kicks in at velocities $\gsim 3\,000 \kms$,
which is higher than seen in spectra for [\ion{O}{iii}]. 

A difference image between [\ion{O}{iii}]/0 and [\ion{O}{iii}]/6000 filter observations should therefore remove 
stars and the synchrotron continuum from the PWN, and in principle provide a cleaner image than using HST/F502N
to probe [\ion{O}{iii}]-emitting gas moving at velocities between $\sim -2\,200 \kms$ and $\sim +3\,000 \kms$,
%We show such an image in Fig.~\ref{f:Diffimage}. This figure enables us to reliably estimate,
albeit at lower spatial resolution. We show such an image in Fig.~\ref{f:O3_diff}. The [\ion{O}{iii}]/0 image was smoothed 
to match the slightly worse seeing of the [\ion{O}{iii}]/6000 image before subtraction.  However, some residuals from the 
image subtraction remain. The surface intensity has a square root scaling to bring out faint emission. To guide the eye, we have 
included a circle, which corresponds to the distance freely expanding SN ejecta would reach in 1\,100 years if moving at 
$2\,150 \kms$ (or in 1\,200 years if moving at $1\,980 \kms$) and for a distance to the LMC of 50 kpc \citep[e.g.,][]{Pietr19}.
Although the nebulous  [\ion{O}{iii}] emission is complex, there is a hint of 
emission associated with the SNRC out to this radius. Fig.~\ref{f:O3_diff} also highlights filaments F1, F2, F4 and F5. 

For the PWN part of the SNRC, the ground-based VIMOS/IFU [\ion{O}{iii}] images in \citet{Sand13},
are superior to the combination of [\ion{O}{iii}]/0 and[\ion{O}{iii}]/6000 as it also adds velocity information, but those images 
only cover $13\arcsec \times 13\arcsec$. A problem with all the mentioned methods to study the SNRC, even with the good 
spectral resolution of the VIMOS/IFU data, is to clean the images from [\ion{O}{iii}], and to a minor extent, 
\ion{He}{i}~$\lambda5016$ (cf. Fig.~\ref{f:O3_new}) emission from H~II regions in the LMC. 

The [\ion{O}{iii}] emission in our NTT image    
extends outside the X-ray PWN, and the faint [\ion{O}{iii}] glow fills 
the space out to filament F1. This can certainly not be explained 
by worse spatial resolution of the ground-based 
observations, and further shows that we are losing some important 
information in the narrow band HST filters. 
Extended [\ion{O}{iii}] emission was discussed in \citet{Morse06}, and
was traced out to a radius of $\sim 8\arcsec$. As filament F1 lies at a projected distance
of $\approx 8\farcs5$ from the pulsar, the [\ion{O}{iii}] glow is therefore more extended
than argued for by \citet[][see also below]{Morse06}. As shown in Fig.~\ref{f:O3_diff}, there is in particular
a broad region of [\ion{O}{iii}] glow to the southwest which extends well outside 
the ring with a radius of 10\arcsec. We return to this below whether or not this glow is intrinsic to 0540, or emanates from
other nebulae. 

The [\ion{O}{iii}] glow in 0540 contrasts the situation for the Crab nebula, where
the filamentary structure terminates with a shell-like structure. Observations of the filamentary 
ejecta in the Crab \citep[e.g.,][]{Blair97,Hes98} suggest that the filaments are the result of 
Rayleigh-Taylor instabilities at the interface between the synchrotron nebula and the swept-up 
ejecta. The emission comes from the cooling region behind the shock driven into
the extended remnant by the pressure of the PWN. The same most likely applies to 0540 
\citep[cf.][see also Section 3.4]{Williams08}, but in the Crab there is no external glow \citep[e.g.][]{Tziamtzis09}, 
presumably simply because of lack of fast SN ejecta \citep{Yang15}. Further comparing with
the Crab, the protruded [\ion{O}{iii}] glow in the south-western direction in 0540, bears
similarities with the Crab chimney \citep[e.g.,][]{rfy08}, especially when viewed through the 
HST/WFPC2/F502N filter (see Fig.~\ref{f:O3_new}).

\subsection{Two-dimensional spectroscopy}\label{Spec}

As seen from Fig.~\ref{f:OIIIim}, slits 1 and 2 cross the SNRC in two 
almost orthogonal directions. In both directions the angular sizes 
of the SNRC are about ten times larger than the $\sim$1\arcsec\ seeing value 
in our NTT and VLT spectral observations. This enables us to spectrally resolve   
different spatial parts of the SNRC projected on the slits, i.e., to create space-velocity
maps of the encapsulated SN ejecta.
In these space-velocity maps the continuum emission was subtracted using the {\sf IRAF} 
{\sf background} task along the spatial axis to reveal the kinematic structure 
from the emission lines. We caution that the uneven LMC
background introduces some uncertainty in the derived velocity structure,
especially for H$\alpha$, at velocities embracing the LMC redshift, i.e., 
at $\sim 270\pm150 \kms$. In Fig.~\ref{f:O3_new}, where we include results for [\ion{O}{iii}] from
\citet{Sand13}, we have used LMC redshift as the reference velocity, and will continue to do so as default 
throughout the paper, unless otherwise remarked. 

The slit (``slit 2") in Fig.~\ref{f:velom1} (top panel) is oriented almost along the continuum-emitting
elongated structure of the central PWN, discussed in detail by \citet{nlun11}. 
In Fig.~\ref{f:velom1} we have used a zoomed-in and rotated version of our NTT/EMMI [\ion{O}{iii}] difference image
in the right panel of Fig.~\ref{f:O3_new} for reference, as this image 
is free from continuum sources, to show how slit 2 probes 0540.

The second panel from the top in Fig.~\ref{f:velom1} shows the 
space-velocity distribution of the [\ion{O}{iii}]~$\lambda5007$ emitting material
encapsulated by slit 2. The space-velocity 
structure of  [\ion{O}{iii}]~$\lambda$5007 is unaffected by contamination 
from [\ion{O}{iii}]~$\lambda$4959 which is shifted in velocity to [\ion{O}{iii}]~$\lambda$5007 
by $2\,870 \kms$.

The [\ion{O}{iii}] emission is dominated by a component
with an average redshift velocity of $\sim400~\kms$ (relative to LMC redshift). 
There is a slight asymmetry with the northeast part being redshifted 
with a few hundred $\kms$ more than the southwestern part. From the 3D structure of [\ion{O}{iii}]
outlined by \citet{Sand13}, the dominating centra of [\ion{O}{iii}] emission in our
space-velocity map are parts of two separate dominating ring-like structures in the ejecta.
They are clearly displayed in Fig.~\ref{f:O3_new}.

To highlight the fainter [\ion{O}{iii}] emission, we have constructed a similar plot
to that in Fig.~\ref{f:velom1}, but for wider velocity and spatial ranges. This is
shown in Fig.~\ref{f:OIIIim_slit2} where we have again used our NTT/EMMI [\ion{O}{iii}] difference
image, but this time a rotated version of that in Fig.~\ref{f:O3_diff} to highlight fainter emission.
Fig.~\ref{f:OIIIim_slit2} also includes 
 [\ion{O}{iii}]~$\lambda$4959, which is $\approx 3$ times fainter than 
 [\ion{O}{iii}]~$\lambda$5007, as expected. Both components of the doublet are shown to
more easily evaluate high-velocity features, as well as the structure of the
faint glow outside the SNRC. We have carefully subtracted stars, the PWN and LMC \ion{H}{ii} regions
to  trace the weakest features. Despite this, artifacts due to over-subtraction of the
LMC background emission are seen,
%cosmic ray particle effect,  
but are not crucial to conclusions about the space-velocity structure of
the [\ion{O}{iii}] glow. 

As seen, the faint glow has a completely different structure
to that of the core, as its maximum redshift, close to $+2\,100 \kms$,
occurs in the southwestern part, whereas the maximum blueshift, $\sim -1\,300 \kms$, 
is in northeastern part. The structures apparent for the two [\ion{O}{iii}] line components 
do not overlap, and have counterparts in both components that increase the 
reliability of the velocity structure. It is evident that the SNRC emission and
the outer [\ion{O}{iii}] glow form two distinctly different ejecta components, which is fully consistent
with the findings of \citet{Sand13} (see also Fig.~\ref{f:O3_new}), although the 3D cube
of \citet{Sand13} did not include the highest velocities on the receding side. 

We also note that the $+2\,100 \kms$ component has a continuation from $\sim 3-4\arcsec$
southwest of the pulsar, and further along the slit in the same direction, all the way out to the edge 
of the frame at $9\arcsec$ from the pulsar, where the glow has a velocity just redward
of LMC rest velocity, but perhaps also connects to emission on the blue side. It appears as if the glow 
is part of an incomplete shell structure with stronger emission on the receding side to the southwest and on the 
approaching side to the northeast. This is also consistent with the VIMOS/IFU image in Fig.~\ref{f:O3_new}. 
The glow continues further to the southwest, but the signal-to-noise is too low in the NTT/EMMI 
spectrum to trace the velocity outside the frame of Fig.~\ref{f:OIIIim_slit2}. Signal-to-noise
was also too low in the study of \citet{Mat80} to probe possible broad-line emission
in this direction (at PA~$=60\degr$).

Turning to [\ion{S}{ii}]~$\lambda\lambda$6716,6731 in Fig.~\ref{f:velom1} we note 
that it has a smaller extent to the southwest than [\ion{O}{iii}], but similar extent to
the northeast. Although not shown here, we see a similarly small extent 
in [\ion{Ar}{iii}]~$\lambda$7136. The two line components of [\ion{S}{ii}] blend 
together which makes the real space-velocity structure of each component more 
difficult to disentangle than for [\ion{O}{iii}]. In \citet{Sand13} we devised a way to separate the
components, and at the same time create an electron density map from the relative intensities of 
the two [\ion{S}{ii}] line components, and we highlight this here in Sect. 3.2.2. The 
[\ion{S}{ii}] lines follow the trend for [\ion{O}{iii}] in Fig.~\ref{f:velom1}, 
i.e., there is a general redshift towards the northeast compared to the southwest. The 
spectral structure range between a blueshift of $\sim 800 \kms$ and a redshift of $\sim 1\,100 \kms$.
%The intensity contours in the top panel of Fig.~\ref{f:velom1} is for [\ion{S}{ii}] as
%derived by \citet{Sand13}, and facilitates the interpretation. For example, the local peak
%of [\ion{S}{ii}] emission $1\farcs0 - 1\farcs5$ northeast of the pulsar has no counterpart
%in [\ion{O}{iii}].

The H$\alpha$ space-velocity structure (second panel from the bottom of 
Fig.~\ref{f:velom1}) is less organized, mainly due to subtraction of the 
uneven LMC background in H$\alpha$, but there is also a hint of subtraction 
residuals due to [\ion{N}{ii}]~$\lambda\lambda$6548,6583. The bright emission 
between $700-1\,200 \kms$ actually falls on top of the
[\ion{N}{ii}]~$\lambda$6583 line from the LMC.  
H$\alpha$ or [\ion{N}{ii}] was
a serious matter of discussion in \citet{Kirshner89}. To underline this, 
there is no strong [\ion{O}{iii}] or [\ion{S}{ii}] emission just north of the pulsar along the slit
reaching out to $\sim +1\,900 \kms$), whereas the H$\alpha$ plot shows emission there. 
We agree with \citet{Morse06} that [\ion{N}{ii}]~$\lambda$6583 from 0540
contributes at those wavelengths. That we expect H$\alpha$ from 0540 at all 
mainly rests on the results from slit 1 which clearly displays several Balmer lines 
(see below). This was reported for the first time in \citet{Ser05}, and this has led to 
the interpretation that 0540 stems from a Type II SN explosion with a zero-age main-sequence
mass of $\sim20 \Msun$ \citep{Chevalier06,Williams08,nlun11}. 
%We can, unfortunately, not constrain [\ion{N}{ii}] in a meaningful way
%from the lack of [\ion{N}{ii}]~$\lambda$5755 (Sect. 3.4)

%----------------------  Table 2 --------------------
%\input{table_vel_1.tex}
\begin{table*}[htb]
\caption{Dereddened line fluxes relative to [\ion{O}{iii}]~$\lambda$5007 and line velocities in \snr.}%
\label{t:FluxVel}
\begin{tabular}{lcrlccrlcrc}
\hline\hline
  & \multicolumn{3}{c}{VLT (PA=88\degr)} &  & \multicolumn{3}{c}{NTT (PA=22\degr)} & & \multicolumn{2}{c}{Others$^d$} \\ \cline{2-4}\cline{6-8}\cline{10-11}
Line & Measured & flux$^a$ & velocity$^b$ & & Measured & flux$^c$ & velocity$^b$ & & flux & velocity$^b$ \\
 & (\AA) & & (km s$^{-1}$) & & (\AA) & & (km s$^{-1}$) & & & (km s$^{-1}$) \\
\hline
{[\ion{O}{ii}]~$\lambda\lambda$3726,3729}$^e$    & 3735.0 & {59.6} & {600 $\pm$ 90} & &        & & & & 52/61& {732 $\pm$ 80} \\
{[\ion{Fe}{vii}]~$\lambda$3759} & 3764.4 & {1.5} & {460 $\pm$ 30} & &        & & & &  &  \\
{[\ion{Fe}{v}]~$\lambda\lambda$ 3783$-$3797}  & 3797.8 & {1.7} & {690 $\pm$ 60}  & &        & & & &  &  \\
{[\ion{Ne}{iii}]~$\lambda$3869$+$\ion{H}{i},\ion{He}{i}~$\lambda$3889}$^f$ & 3878.2 & {3.3} & {710 $\pm$ 80} & &        & & & & $<$2/7.3) \\
%{[\ion{Fe}{ii}]~$\lambda$3935}        & 3944.7 & {0.6:}& {700 $\pm$ 110} & &        & & & & & \\
{[\ion{Ne}{iii}]~$\lambda$3967$+$\ion{H}{i}~$\lambda$3970}$^f$ & 3976.1 & {1.6} & {630 $\pm$ 120} & &        & & & & -/1.7 & \\
{\ion{He}{i}~$\lambda$4026}        & 4034.0 & {0.4:} & {580 $\pm$ 90} & &        & & & &  &  \\
{[\ion{S}{ii}]~$\lambda\lambda$4069,4076$^g$}    & 4078.1  & {3.5} & {700 $\pm$ 90} & &        & & & & 4/4.8  & \\
{\ion{H}{i}~$\lambda$ 4102} & 4106.0 & {0.9:} & {310 $\pm$ 50} & &        & & & &  & \\
{\ion{He}{i}~$\lambda$ 4143} & 4154.5 & {0.7:} & {780 $\pm$ 200} & &        & & & &  & \\
{[\ion{Fe}{v}]~$\lambda$4227}        & 4235.7 & {1.6} & {450 $\pm$ 170} & &        & & & & 6.7/- & \\
{[\ion{Fe}{ii}]~$\lambda$4287}       & 4295.3 & {0.9:} & {550 $\pm$ 180} & &        & & & & & \\
{[\ion{O}{iii}]~$\lambda$4363$+$H$\gamma$}       & 4368.7 & {5.7} & {380 $\pm$ 190} & &        & & & & 9.3/4.4  & \\
{[\ion{Fe}{ii}]~$\lambda$4414} &4418.2 & {0.3:} & {250 $\pm$ 150}  & &        & & & & & \\
{[\ion{Fe}{ii}]~$\lambda\lambda$4452,4458,4475 + \ion{He}{i}~$\lambda$4471} & 4473.6 & {1.4} & {640 $\pm$ 250} & &        & & & & & \\
{\ion{Mg}{i}]~$\lambda$4571} & 4576.8 & {0.4:} & {380 $\pm$ 90} & &        & & & & & \\
{[\ion{Fe}{iii}]~$\lambda$4658}      & 4665.5 & {1.9} & {480 $\pm$ 100} & &{$\sim$4670}&{::$^i$} & & & 2: & \\
{H$\beta$}                   & 4869.2 & {3.3} & {490 $\pm$ 70 } & &{$\sim$4870}&{::$^i$} & & & {$<$2}/1.9& \\
{[\ion{Fe}{iii}]~$\lambda$4881} & 4890.7 & {0.3:} & {590 $\pm$ 80} & &        & & & & & \\
{[\ion{O}{iii}]~$\lambda$4959}       & 4970.4 & {33.2} & {700 $\pm$ 120} & & 4970.7 & { 37} & {700 $\pm$ 50} & &33/33 & {515 $\pm$ 100}\\
{[\ion{Fe}{iii}]~$\lambda$4986}      & 4987.6 & {1.4:} & {100 $\pm$ 300} & & & & & &  & \\
{[\ion{O}{iii}]~$\lambda$5007}       & 5017.9 & {100}   & {660 $\pm$ 110} & & 5016.3 & {100} & {560 $\pm$ 50} & &100/100 & {461 $\pm$ 120}\\
{[\ion{Fe}{ii}]~$\lambda$5044}        & 5052.6 & {0.3:}& {540 $\pm$ 110} & &        & & & & & \\
{[\ion{Fe}{ii}]~$\lambda$5159$+$[\ion{Fe}{vii}]~$\lambda$5159}       & 5171.0 & {1.6} & {700 $\pm$ 200} & &{$\sim$5160}&{2:}&{100 $\pm$ 200} & & & \\
{[\ion{Fe}{iii}]~$\lambda$5270}      & 5280.8 & {1.2} & {590 $\pm$ 120} & &{$\sim$5280}&{2:}& {500 $\pm$ 200} & & & \\
{[\ion{Fe}{vii}]~$\lambda$5721}      & 5730.6 & {0.8:} & {520 $\pm$ 90} & &{}&{} & & &  & \\
{\ion{He}{i}~$\lambda$5876}          & 5881.7 & {0.8:} & {310 $\pm$ 100} & &{}&{} & & & -/1.3 & \\
{[\ion{O}{i}]~$\lambda$6300}         &        &  	&		  & & 6313.2 & {  7} & {630 $\pm$ 20} & & 6.7/4.4 & \\
{[\ion{O}{i}]~$\lambda$6363}         &        &  	&		  & & 6374.3 & {  4:}& {530 $\pm$ 70} & & -/1.2  & \\
{H$\alpha$$^h$}                   &        &  	&		  & & 6577.8 & { 17} & {680 $\pm$ 130} & & 36/19.3 & \\
{[\ion{S}{ii}]~$\lambda\lambda$6717,6731}$^g$&        &  	          & & & 6738.1 & { 99} & {320 $\pm$ 50} & & 67/45.1 & \\
{[\ion{Ar}{iii}]~$\lambda$7136}      &        &  	&		  & & 7151.0 & {  9} & {630 $\pm$ 40} & & 10.7/- & {630 $\pm$ 100}\\
{[\ion{O}{ii}]~$\lambda$7325}        &        &  	&		  & & 7344.7 & { 11} & {810 $\pm$ 120} & & 8/- & \\
{[\ion{Ni}{ii}]~$\lambda$7378}       &        &  	&		  & & 7390.5 & {  9} & {520 $\pm$ 50} & & 6.7/- & \\
\hline
\end{tabular} \\
\begin{tabular}{ll}
$^a$~The flux of [\ion{O}{iii}]~$\lambda$5007 is 4.1~$\times~10^{-14}$ ergs cm$^{-2}$ s$^{-1}$. & $^b$~Not corrected for the LMC redshift ($\sim 270 \kms$). \\
$^c$~The flux of [\ion{O}{iii}]~$\lambda$5007 is 5.0~$\times~10^{-14}$ ergs cm$^{-2}$ s$^{-1}$. & $^d$~\citet[][PA=77\degr]{Kirshner89}~/~\citet[][PA=124\degr]{Morse06}.\\
$^e$~[\ion{O}{ii}]~$\lambda$3727.5 was used for velocity estimate. & $^f$~Velocity estimated for [\ion{Ne}{iii}].\\
$^g$~Velocity for [\ion{S}{ii}]~$\lambda$4069 and [\ion{S}{ii}]~$\lambda$6731, respectively. & $^h$~Blended with [\ion{N}{ii}]~$\lambda\lambda$6548,6583. \\ 
$^i$~Line detected, flux uncertain. & \\
\end{tabular} \\
\end{table*}

%-----------------------Table3-----------------------------
%\input{table_vel_2.tex}
\begin{table*}[hbt]
\caption{Dereddened line fluxes relative to H$\beta$ and line velocity in 
H II regions along the VLT ``slit 1''.}%
\label{t:FluxVel_2}
%\begin{center}
\begin{tabular}{@{\hspace{1mm}}l@{\hspace{1mm}}c@{\hspace{1mm}}rc@{\hspace{1mm}}cc@{\hspace{1mm}}rc@{\hspace{1mm}}cc@{\hspace{1mm}}rc@{\hspace{1mm}}}
\hline\hline
  & \multicolumn{3}{c}{(F1) 8\arcsec~from pulsar} &  & \multicolumn{3}{c}{(F2) 25\arcsec~from pulsar} & & \multicolumn{3}{c}{(F3) 62\arcsec~from pulsar} \\ \cline{2-4}\cline{6-8}\cline{10-12}
Line & Measured & flux$^a$ & velocity$^b$ & & Measured & flux$^c$ & velocity$^b$ & & Measured & flux$^d$ & velocity$^b$ \\
 & (\AA) & & (km s$^{-1}$) & & (\AA) & & (km s$^{-1}$) & & (\AA) & & (km s$^{-1}$) \\
\hline
{[O II]~$\lambda\lambda$3726, 3729}$^e$ & 3730.8 & {478.3} & {270 $\pm$ 70} & & 3729.7 & 756.3 & {180 $\pm$ 10} & & 3730.6 & 529.8 & {250 $\pm$ 60} \\
{H I~$\lambda$3771}               &	   &	     &		      & &        &       &                & & 3773.5 &   5.2 & {230 $\pm$ 90} \\
{H I~$\lambda$3798}               &	   &	     &		      & & 3799.7 &  20.1 & {140 $\pm$ 50} & & 3800.7 &   5.6 & {220 $\pm$ 90} \\
{H I~$\lambda$3835}               &	   &	     &		      & & 3838.7 &  23.2 & {260 $\pm$ 50} & & 3838.7 &   8.0 & {260 $\pm$ 80} \\
{[Ne~III]~$\lambda$3869}          & 3872.9 & {37.1}  & {320 $\pm$ 60} & & 3871.8 & 146.3 & {230 $\pm$ 50} & & 3872.1 &  21.1 & {260 $\pm$ 10} \\
{H I~$+$He~I$\lambda$3889}        & 3894.6 & { 28.0} & {430 $\pm$ 150}& & 3892.2 &  26.9 & {260 $\pm$ 20} & & 3892.3 &  20.6 & {270 $\pm$ 10} \\
{Ca II~$\lambda$3934}           & 3939.2 & { 26.3} & {420 $\pm$ 110}& & 3941.3 &  12.5 & {580 $\pm$ 180}& & 3937.6 & 4.3:&{120 $\pm$200} \\
{[Ne III]~$+$~Ca II~$+$ H I}$^f$& 3972.6 & { 35.9} & {300 $\pm$ 70} & & 3971.5 &  68.9 & {210 $\pm$ 40} & & 3972.9 &  25.1 & {320 $\pm$ 50} \\
{He I~$\lambda$4026}              &        &         &                & &        &       &                & & 4031.2 &  4.2:& {370 $\pm$ 90} \\
{[S II]~$\lambda$4069}$^g$        & 4072.6 & { 26.8} & {	    } & & 4072.9 &  11.9 & { } & & 4072.6 &   4.0 & {   	    } \\
{[S II]~$\lambda$4076}$^g$        & 4082.9 & { 13.9} & {	    } & & 4080.9 &   5.5 & { } & & 4080.9 &   2.3 & {   	    } \\
{H$\delta$}                       & 4107.3 & { 24.3} & {410 $\pm$ 50} & & 4105.2 &  34.6 & {260 $\pm$ 30} & & 4105.7 &  26.8 & {300 $\pm$ 30} \\
{[Fe V]~$\lambda$4227}            & 4231.7 & { 10.0} & {320 $\pm$ 70} & & 4231.0 &  17.6 & {270 $\pm$ 40} & &	     &       & {            } \\
{[Fe II]~$\lambda$4244}           & 4248.3 & {  8.0} & {250 $\pm$ 100}& &        &       &                & &        &       & {            } \\
{[Fe II]~$\lambda$4287}           & 4293.2 & {  7.3} & {400 $\pm$ 90} & &        &       &                & & 4291.5 &  2.3:& {290 $\pm$ 90} \\
{H$\gamma$}                       & 4346.2 & { 44.4} & {400 $\pm$ 50} & & 4344.2 &  55.7 & {260 $\pm$ 20} & & 4344.8 &  47.1 & {300 $\pm$ 40} \\
{[O III]~$\lambda$4363}           & 4367.1 & { 25.5} & {270 $\pm$ 50} & & 4367.0 &  79.3 & {270 $\pm$ 10} & & 4367.2 &   6.3& {280 $\pm$ 70} \\
{He~I~$\lambda$4471}              & 4478.7 & { 11.9:}& {480 $\pm$ 100}& &        &       &                & & 4476.2 &   4.9 & {320 $\pm$ 60} \\
{[Mg I]~$\lambda$4563}            &        &         &                & &        &       &                & & 4567.5 &   2.4:& {320 $\pm$ 90} \\
{Mg I]~$\lambda$4571}             & 4574.4 & {  6.4} & {220 $\pm$ 100}& &        &       &                & & 4576.0 &  2.2:& {320 $\pm$ 110} \\
{[Fe III]~$\lambda$4658}          & 4663.2 & { 19.0} & {330 $\pm$ 80} & & 4661.8 &  16.3 & {250 $\pm$ 30} & & 4662.7 &  11.5 & {300 $\pm$ 40} \\
{He II~$\lambda$4686}             & 4691.3 & { 14.3} & {360 $\pm$ 90} & & 4690.1 &  36.3 & {280 $\pm$ 50} & &        &       & {            } \\
{[Fe III]~$\lambda$4702}          &        &         &                & &        &       &                & & 4706.5 &   3.4 & {320 $\pm$ 60} \\
{[Fe III]~$+$~[Ar IV]~$+$~[Ne IV]}$^j$& 4713.8 & { 14.9} & {300 $\pm$ 70} & & 4717.3 & 15.0 & {380 $\pm$ 90} & &     &       & {            } \\
{He I~$\lambda$4713}            &        &         &                & &        &       &                & & 4718.6 &   1.5:& {340 $\pm$ 90} \\
{[Ne IV]~$\lambda$4726}           &        &         &                & & 4730.2 &   8.3 & {290 $\pm$ 30} & &        &       & {            } \\
{[Fe III]~$\lambda$4734}          & 4738.2 & { 16.5} & {270 $\pm$ 80} & &        &       &                & & 4738.9 &   1.0:& {320 $\pm$ 100} \\
{[Ar IV]~$\lambda$4740}           &        &         &                & & 4744.3 &   8.8 & {260 $\pm$ 30} & &        &       & {            } \\
{[Fe III]~$\lambda$4755}          & 4760.6 & {  6.0} & {360 $\pm$ 110}& &        &       &                & & 4759.2 &   2.0:& {290 $\pm$ 50} \\
{[Fe III]~$\lambda$4769}          &        &         &                & &        &       &                & & 4774.8 &   1.8:& {340 $\pm$ 60} \\
{[Fe III]~$\lambda$4778}          & 4781.7 & {  7.0} & {240 $\pm$ 100}& &        &       &                & & 4782.4 &   0.9:& {290 $\pm$ 80} \\
{[Fe II]~$\lambda$4814}           & 4820.8 & {  7.1} & {400 $\pm$ 90} & &        &       &                & &        &       & {            } \\
{ H$\beta$}                       & 4867.9 & {100  } & {410 $\pm$ 60} & & 4865.4 &  100  & {260 $\pm$ 10} & & 4866.2 & 100   & {310 $\pm$ 20} \\
{[Fe III]~$\lambda$4881}          & 4886.6 & {  6.9} & {350 $\pm$ 80} & & 4885.0 &  7.1  & {250 $\pm$ 50} & & 4885.9 &   3.4:& {300 $\pm$ 70} \\
{He I~$\lambda$4922}              &        &         &                & &        &       &                & & 4927.4 &   3.0:& {340 $\pm$ 70} \\
{[O III]~$\lambda$4959}           & 4965.0 & {121.1} & {370 $\pm$ 20} & & 4963.4 & 455.0 & {270 $\pm$ 10} & & 4963.9 &  75.6 & {300 $\pm$ 30} \\
{[Fe III]~$\lambda$4986}          & 4991.9 & {  8.2} & {350 $\pm$ 50} & & 4990.3 & 15.8  & {260 $\pm$ 60} & & 4990.9 &  14.5 & {300 $\pm$ 50} \\
{[O III]~$\lambda$5007}           & 5013.2 & {358.8} & {390 $\pm$ 20} & & 5011.5 &1370.2 & {290 $\pm$ 10} & & 5011.9$^h$& 222.2 & {310 $\pm$ 10} \\
{[Fe II]~$\lambda$5112}           & 5120.2 & {  9.9} & {500 $\pm$ 90} & &        &       &                & &        &       & {            } \\
{[Fe II]~$\lambda$5159}           & 5165.4 & { 16.9} & {390 $\pm$ 100}& &	 &	 & {            } & & 5162.9 &  2.5:& {240 $\pm$ 100} \\
{[N I]~$\lambda\lambda$5198-5200} & 5211.0 & {  8.7:}& {680 $\pm$ 240}& &        &       &                & & 5201.9 &   2.5:&{160 $\pm$ 140} \\
{[Fe III]~$\lambda$5270}          & 5274.6 & { 19.2} & {240 $\pm$ 80} & & 5276.0 &   7.6 & {320 $\pm$ 70} & & 5275.9 &   6.4 & {310 $\pm$ 80} \\
{[Fe XIV]~$\lambda$5303$+$[Ca V]~$\lambda$5309}& 5311.3 & {  9.8} & {300 $\pm$ 100} & &        &       &   & &        &       & {            } \\
{[Fe II]~$\lambda$5334}           & 5340.2 & {  5.9} & {380 $\pm$ 90} & &        &       &                & &        &       & {            } \\
{[Ar X]~$\lambda$5534}            & 5537.9 & {  8.0} & {210 $\pm$ 60} & &        &       &                & &        &       & {            } \\
{[Fe VII]~$+$[Fe II]~$\lambda$5721}& 5724.6 & { 15.2}& {270 $\pm$ 80} & &        &       &                & &        &       & {            } \\
{He I~$\lambda$5876}              & 5883.4 & { 13.5} & {400 $\pm$ 70} & & 5881.2 &  11.9 & {290 $\pm$ 60} & & 5881.8 &  12.2 & {310 $\pm$ 70} \\
\hline
\end{tabular}\\
\begin{tabular}{ll}
$^a$~The flux of H$\beta$ is 3.4~$\times~10^{-16}$ ergs cm$^{-2}$ s$^{-1}$ & $^b$~Not corrected for the LMC redshift ($\sim 270 \kms$)\\
$^c$~The flux of H$\beta$ is 8.1~$\times~10^{-16}$ ergs cm$^{-2}$ s$^{-1}$ & $^d$~The flux of H$\beta$ is 2.1~$\times~10^{-15}$ ergs cm$^{-2}$ s$^{-1}$\\
$^e$~[O II]~$\lambda$3727.5 was used for velocity estimate & $^f$~For velocity estimate used $\lambda$3968.67\\
$^g$~Deblending for [S II] was carried out assuming & $^h$~He I~$\lambda$5015 detected in the red wing of [O III]~$\lambda$5007, \\
Gaussian line profiles & but flux is uncertain.\\
\end{tabular} \\
%\end{center}
\end{table*}

%-----------------------Table4-----------------------------
%\input{table_vel_3.tex}
\begin{table*}[htb]
\caption{Dereddened line fluxes relative to H$\beta$ and line velocities in 
H~II regions along the VLT ``slit 3''.}%
\label{t:FluxVel_3}
\begin{tabular}{lcrcccrc}
\hline\hline
  & \multicolumn{3}{c}{(F4)} & & \multicolumn{3}{c}{(F5)}\\ \cline{2-4} \cline{6-8}
Line & Measured & flux$^a$ & velocity$^b$ & & Measured & flux$^c$ & velocity$^b$\\
 & (\AA) & & (km s$^{-1}$) & & (\AA) & & (km s$^{-1}$)\\
\hline
{[O II]~$\lambda\lambda$3726, 3729}$^d$ & 3729.2 & { 299.6} & {140$\pm$ 50} & & 3729.5 & 377.1 & {160$\pm$ 60}\\
{H~I~$\lambda$3798}           & 3802.1 & 12.6  & {330$\pm$ 80} & & 3802.0 & 7.4  & {330$\pm$ 50}\\
{H~I~$\lambda$3835}           & 3837.7 &  7.1  & {180$\pm$ 90} & &        &      & {           }\\
{[Ne III]~$\lambda$3869}      & 3871.6 & 48.1  & {220$\pm$ 20} & & 3871.5 & 80.3 & {210$\pm$ 20}\\
{H~I~$+$~He~I~$\lambda$3889}    & 3891.7 & 18.8  & {220$\pm$ 30} & & 3892.7 & 24.9 & {290$\pm$ 30}\\
{[Ne III]~$+$~H~I~$+$~Ca~II}$^e$ & 3971.9 & 31.5  & {240$\pm$ 50} & & 3971.4 & 38.3 & {210$\pm$ 40}\\
{[S II]~$\lambda\lambda$4069, 4076}    & 4072.5 & 10.6 &  & & 4071.5 & 8.3 & \\
{H$\delta$}                   & 4104.8 & 29.8  & {230$\pm$ 40} & & 4104.9 & 24.3 & {230$\pm$ 40}\\
{[Fe V]~$\lambda$4227}        & 4230.5 & 3.3:  & {230$\pm$ 110}& & 4230.0 & 6.7  & {200$\pm$ 90}\\
{H$\gamma$}                   & 4343.9 & 46.3  & {250$\pm$ 20} & & 4343.9 & 44.4 & {240$\pm$ 20}\\
{[O III]~$\lambda$4363}       & 4366.5 & 17.0  & {230$\pm$ 30} & & 4366.6 & 41.2 & {240$\pm$ 30}\\
{He~I~$\lambda$4471}          & 4475.2 &  4.0  & {250$\pm$ 90} & & 4472.6 &  5.4 & {80$\pm$ 140}\\
{[Fe III]~$\lambda$4658}      & 4661.5 &  8.0  & {230$\pm$ 80} & & 4661.7 & 10.5 & {230$\pm$ 50}\\
{He II~$\lambda$4686}         & 4690.1 &  8.3  & {290$\pm$ 80} & & 4689.8 & 13.6 & {270$\pm$ 60}\\
{[Ar IV]~$\lambda$4711~$+$~[Ne IV]~$\lambda$4714}$^f$ &  &  &  & & 4716.5 &  5.7 & {240$\pm$ 70}\\
{[Ar IV]~$\lambda$4740}                               &  &  &  & & 4743.3 &  3.3 & {200$\pm$ 100}\\
{[Fe III]~$\lambda$4756}                              &  &  &  & & 4758.1 &  3.2 & {210$\pm$ 110}\\
{ H$\beta$}                   & 4865.1 & 100   & {240$\pm$ 10} & & 4865.1 & 100  & {240$\pm$ 10}\\
{[O III]~$\lambda$4959}       & 4963.3 & 186.0 & {260$\pm$ 10} & & 4963.1 & 282.2& {260$\pm$ 20}\\
{[Fe III]~$\lambda$4986}      & 4989.9 &  7.1  & {230$\pm$ 80} & & 4990.8 & 10.3 & {290$\pm$ 60}\\
{[O III]~$\lambda$5007}       & 5011.2 & 553.9 & {270$\pm$ 10} & & 5011.1 & 851.8& {260$\pm$ 10}\\
{[Fe II]~$\lambda$5159}       & 5163.4 &  2.5: & {270$\pm$ 100}& & 5163.7 &  4.2 & {290$\pm$ 90}\\
{[Fe III]~$\lambda$5270}      & 5276.3 &  3.3: & {330$\pm$ 120}& & 5275.6 &  4.8 & {290$\pm$ 80}\\
{[Fe XIV]~$\lambda$5303~$+$~[Ca V]~$\lambda$5309}& 5312.8 & 11.5 & {380$\pm$ 110} & & 5308.9 & 7.3 & {170$\pm$ 90}\\
{He I~$\lambda$5876}          & 5880.3 & 12.0  & {240$\pm$ 50} & & 5880.8 & 12.4 & {270$\pm$ 40}\\
\hline
\end{tabular} \\
\begin{tabular}{ll}
$^a$~The flux of H$\beta$~is 1.8~$\times~10^{-15}$ ergs cm$^{-2}$ s$^{-1}$ & $^b$~Not corrected for LMC redshift ($\sim 270 \kms$)\\
$^c$~The flux of H$\beta$~is 1.0~$\times~10^{-15}$ ergs cm$^{-2}$ s$^{-1}$ & $^d$~[O II]~$\lambda$3727.5 was used for velocity estimate\\
$^e$~Velocity estimated using 3968.67 & $^f$~Velocity estimated using mean value \\ 
\end{tabular} \\
\end{table*}

Moving to the space-velocity diagrams for slit 1, these are shown in the images in 
Fig.~\ref{f:velom2}. We have again used a zoomed-in and rotated (this time by 2\degr) version 
of our NTT/EMMI [\ion{O}{iii}] difference image in the right panel of Fig.~\ref{f:O3_new} for reference.
The [\ion{O}{iii}]~$\lambda$5007 VLT/FORS image along slit 1 (Fig.~\ref{f:velom2}) has a 
``pepper-slice"-like shape that also reflects the expansion of the SNRC. Like for slit 2, 
the emission is heavily redshifted, but less skewed in the space-velocity diagram. The 
line can be traced between about $-700 \kms$ and $+1\,400 \kms$.
It is easy to imagine a relatively symmetric (but redshifted) shell 
with filamentary structures reaching inward toward the center around the 
average velocity of the shell, i.e., $\sim 400 \kms$. However, judging from 
the 3D structure derived by \citet{Sand13} and shown in Fig.~\ref{f:O3_new}, 
the eastern and western parts of the ``pepper"
come from the same two ring-like features for the SNRC discussed for slit 2 above.

H$\beta$ (panel two from the bottom of Fig.~\ref{f:velom2}) is partly corrupted by 
the LMC background subtraction, but can be seen to have a surprisingly similar 
structure to that of [\ion{O}{ii}]~$\lambda\lambda$3727,3729 
(bottom panel of Fig.~\ref{f:velom2} in which the velocity scale is centered on $3727.5 \AA$). 
Both H$\beta$ and the [\ion{O}{ii}] lines lack
emission at the highest velocities ($\gsim 1\,200 \kms$) west of the pulsar. 
The different structures in [\ion{O}{ii}] and [\ion{O}{iii}] reveal different ionization 
conditions, especially in the western part along the slit. As a matter of fact, 
the [\ion{S}{ii}] 3D structure of \citet{Sand13} resembles those of H$\beta$ and [\ion{O}{ii}]
in Fig.~\ref{f:velom2}, which may argue for that the differences between the [\ion{O}{iii}]
and [\ion{S}{ii}] are more likely to be due to an effect of different levels of ionization rather
than abundance effects. Moreover, the similarity between H$\beta$ and [\ion{O}{ii}] 
further argues for contamination of [\ion{N}{ii}] in the H$\alpha$ image in Fig.~\ref{f:velom1}.

Like the NTT data in Fig.~\ref{f:OIIIim_slit2}, the VLT data in Fig.~\ref{f:OIIIim_slit1} show 
a glow of  [\ion{O}{iii}]~$\lambda$5007 emission outside the inner part of the SNRC. The 
glow extends out to a radius of about $8\arcsec-10\arcsec$, and seems to connect 
to filament F1 in the west, which in turn appears to emit at a velocity close to LMC rest velocity. The blue 
circle in the upper panel marks a 10\arcsec\ radius from the pulsar.
The space-velocity plot in Fig.~\ref{f:OIIIim_slit1} reveals four 
interesting features: fast blueshifted (at least to $-1\,600\kms$) faint emission around the 
pulsar position, fast redshifted (at least to $+1\,800\kms$) faint emission to the west of 
the pulsar, no redshifted emission at the pulsar position that was not already revealed 
by the bright emission in Fig.~\ref{f:velom2}, and an intricate low-velocity structure
which seems to form a loop to the east and a stream to the west possibly
connecting to regions around filament F1. F1 itself stands out clearly, which could indicate that 
it is just an H~II-region projected onto 0540. However, as we will see in Sect. 3.3.2, this is not
the correct interpretation. In Section 3.3.2. we will also discuss [\ion{Fe}{iii}]~$\lambda$4986, which 
would introduce emission in Fig.~\ref{f:OIIIim_slit1} at $\sim -1\,250 \kms$. Contamination by 
\ion{He}{i}~$\lambda$5016 would introduce emission at $\sim +530 \kms$, and could be responsible
for the ``horn" sticking out on the eastern side in the space-velocity plot, but a similar feature is seen for
[\ion{O}{iii}]~$\lambda$4959, so background subtractions do not leave imprints from \ion{He}{i}~$\lambda$5016 
in Fig.~\ref{f:OIIIim_slit2}, contrary to the  VLT/VIMOS/IFU images of \citet[][cf. Fig.~\ref{f:O3_new}]{Sand13}.
We highlight that the VLT/VIMOS/IFU images neither cover F1 nor the low-velocity feature to the east. 

The impression from slit 1 is the same as from slit 2, i.e, there appears to be glow from fast 
[\ion{O}{iii}]-emitting ejecta with a tilt towards red on the western side and towards the blue 
on the eastern side. "Glow" (or at least fast [\ion{O}{iii}]-emitting ejecta) can be seen in 
the VIMOS/IFU image in Fig.~\ref{f:O3_new} to be more pronounced for ejecta moving towards 
us than away from us, and this agrees with the fast blue-shifted ejecta in Fig.~\ref{f:OIIIim_slit1} with
a projected center slightly to the east of the pulsar position.

The blue circle drawn in, e.g., Fig.~\ref{f:OIIIim_slit1} has a radius of 10\arcsec\ 
and corresponds to a maximum velocity of freely coasting [\ion{O}{iii}]-emitting ejecta with velocity
\begin{equation}
    v_{\rm max}([\ion{O}{iii}]) = 2150~\left(\frac{D_{\rm LMC}} {50}\right)~\left(\frac{t_{\rm yr}} {1100}\right)^{-1} \kms,
\label{eq:velocity1}
\end{equation}
where $D_{\rm LMC}$ is the distance to 0540 in kpc and $t_{\rm yr}$ is the age of 
0540 in years. As the space-velocity spectra do not really reveal [\ion{O}{iii}] at velocities $\gsim 2\,000 \kms$ (taking 
the spectral resolution into account), one may come to the conclusion that there is no support for space motions of 
 [\ion{O}{iii}]-emitting gas much in excess of $2\,000 \kms$ (relative to LMC). However, for both slits 1 and 2, the maximum velocity 
 on the red side occurs $\sim 4\arcsec$ west and southwest of the pulsar, respectively.  The true space motion of ballistic ejecta, 
 $v_{\rm true}$, if originating from a central position in the SNRC, is therefore 
\begin{equation}
    v_{\rm true}=\sqrt{ v_{\rm obs}^2 + \left( \frac{r_{\rm obs}} t\right)^2},
\label{eq:velocity2}
\end{equation}
where $r_{\rm obs}$ is the projected distance from the pulsar at which the maximum observed redshift 
velocity, $v_{\rm obs}$, of [\ion{O}{iii}], emission occurs, and $t$ is the age since explosion. With $D = 50$~kpc
and $r_{\rm obs} \approx 3.0\EE{18}$~cm (corresponding to  $4\arcsec$), one obtains 
$v_{\rm true}([\ion{O}{iii}]) \sim 2\,090~(2\,060) \kms$ for $t_{\rm yr} = 1\,100~(1\,200)$ and $v_{\rm obs} = 1\,900 \kms$ for slit 1, and
$v_{\rm true}([\ion{O}{iii}]) \sim 2\,270~(2\,240) \kms$ for $t_{\rm yr} = 1\,100~(1\,200)$ and $v_{\rm obs} = 2\,100 \kms$ for slit 2, 
respectively. This is several hundred $\kms$ faster than $v_{\rm max}([\ion{O}{iii}])$ for the SNRC on the red side, and also faster 
than the glow on the blue side, although confusion with [\ion{O}{iii}]~$\lambda$4959 causes uncertainty there.

To guide the eye, we have inserted two ellipses in Fig.~\ref{f:OIIIim_slit1}. Both are for constant $v_{\rm true}$ in all
azimuthal directions carved out by the slit, and they are both tuned to give $v_{\rm obs} \pm1\,900 \kms$ at $4\arcsec$ west 
of the pulsar (as for slit 1). The blue and red curves are for $t_{\rm yr} = 1\,100$ and $t_{\rm yr} = 1\,200$, respectively, and 
$D_{\rm LMC}=50$.
Both curves encapsulate F1, but reach zero velocity at $\pm 9\farcs7$ and  $\pm 10\farcs4$, respectively. If $v_{\rm true}$ is indeed
the same for the full region encapsulated by the slit west of the SNRC, Figs.~\ref{f:O3_diff} and~\ref{f:OIIIim_slit1} point
to an age of $1\,100 \lsim t_{\rm yr} \lsim 1\,200$. At the position of F1 (i.e., at $\sim 8\farcs5$), the red side of  
[\ion{O}{iii}]~$\lambda$5007 should reach $800 \kms$ for $t_{\rm yr} = 1\,100$, and $1\,200 \kms$ for $t_{\rm yr} = 1\,200$.
For a non-accelerating scenario determining $v_{\rm true}$ (as in Eq.~\ref{eq:velocity2}), the pulsar seems to be significantly 
younger than the spin-down age, i.e., $\sim 1600$~years. We return to the pulsar age in Section 3.4.3.

Looking at other parts of the space-velocity structure of the [\ion{O}{iii}]~$\lambda$5007 glow for slit 1, we note that $v_{\rm true}$ 
appears to be significantly lower than for the receding region west of the SNRC, perhaps except for the very eastern part, and 
towards the center of the SNRC on the blue side. The absence of [\ion{O}{iii}]~$\lambda$5007 at high velocities on the
red side towards the SNRC is remarkable, This could be due real differences in the azimuthal distribution of [\ion{O}{iii}]-emitting 
gas, but it could in principle also be due dust absorption screening of the red side behind the SNRC. We find this explanation less likely
though, since the dust would have to be located between the SNRC and fast (unseen due to dust) ejecta on the blue side,
while there is no dust counterpart on the approaching side. Moreover, one does not see this effect in the Crab, despite it containing
large amounts of dust \citep{gom12}. In the Crab the dust is concentrated to filaments with modest filling factor which in the 
optical are notably strong in [\ion{O}{iii}].

The picture that emerges for the [\ion{O}{iii}] glow seen through slits 1 and 2 is that it does not come a spherically symmetric shell 
surrounding the SNRC, but from a much less complete structure. On the blue side it appears to be concentrated to the ``wall"-like
structure centered just west of the pulsar position and reaching out to $\sim -1\,600 \kms$ seen and marked as ``Blue Wall" in 
Fig.~\ref{f:O3_new} and shown by \citet{Sand13}. On the red side, the fastest glow is in the southwestern and western 
parts $\sim 4 \arcsec$ from the pulsar reaching
space velocities well in excess of $2\,000 \kms$. In the eastern part, velocities covered by slits 1 and 2 are less extreme, especially on 
the receding side. However, our slits do not cover the southeastern part. Here we make use of the long-slit observations by
\citet{Morse06} (described more in Section 3.3.1) as they cover the southeastern and northwestern parts of the [\ion{O}{iii}] glow 
region. Figure~5a of \citet{Morse06} shows a clear asymmetry for the [\ion{O}{iii}] glow not covered by our slits 1 and 2, or the
VIMOS/IFU data of \citet{Sand13}. The figure of \citet{Morse06} shows strong [\ion{O}{iii}] glow on the approaching side to the 
northwest, as well to the southeast for the receding side. Apart from the ``wall"-like structure on the blue side (which is also evident
as the blue-ward emission in the VIMOS/IFU spectrum in. Fig~\ref{f:Slit_vs_Full}), it appears as if the 
incomplete shell of [\ion{O}{iii}] glow is devoid of emission on the receding northern side, and on the approaching southern side. 
%There is also no obvious link between the SNRC and the structure giving rise to the [\ion{O}{iii}] glow.

%-----------------------figure11------------------------------
\begin{figure*}[hbt]
\begin{center}
\includegraphics[width=175mm,angle=0, clip]{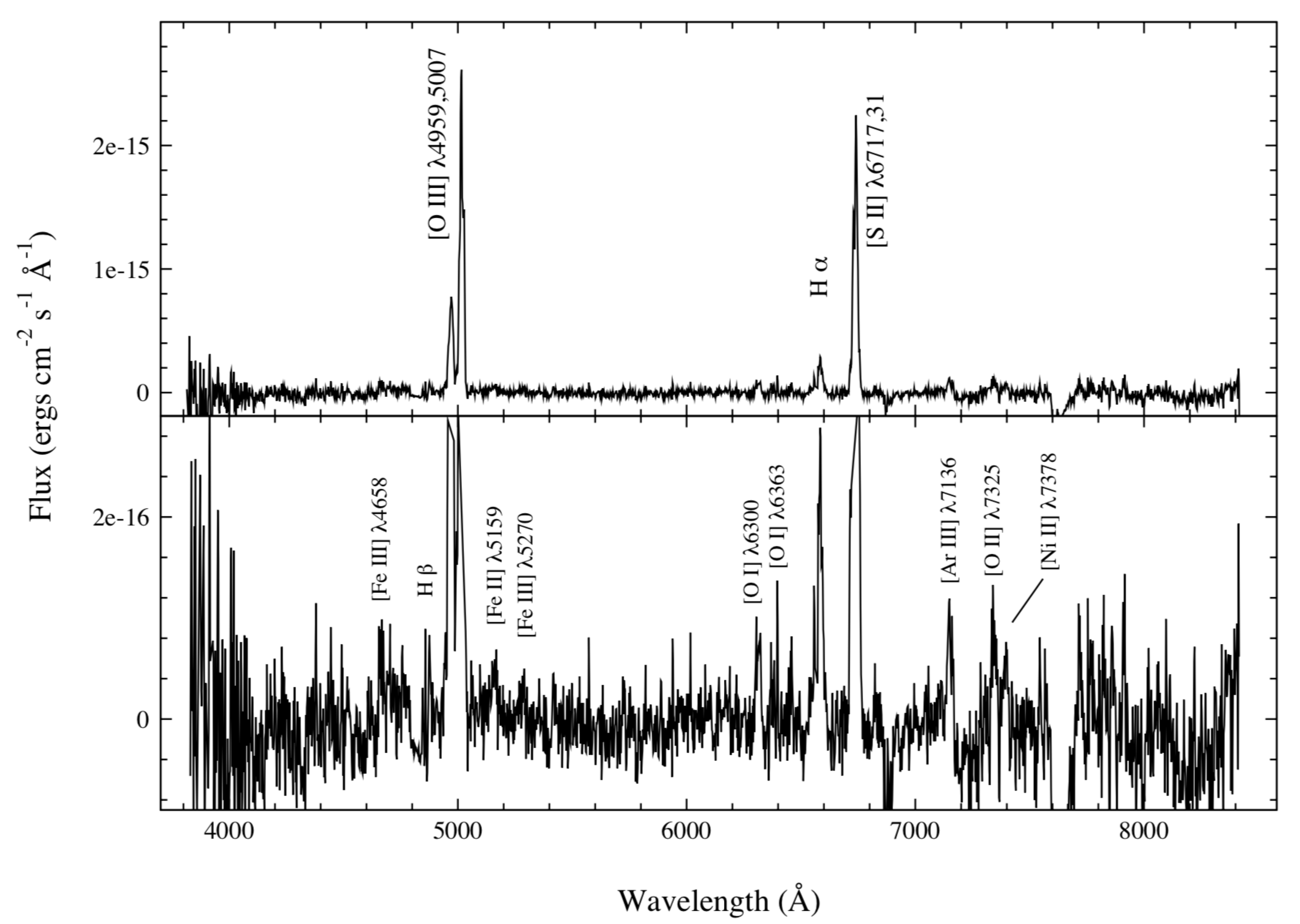}
\end{center}  
\caption{Spectrum of the central part of the~\snr\ obtained with ESO$/$NTT 
(PA=22\degr). The lower panel has an expanded flux scale to highlight weaker lines. 
The spectrum has been dereddened using $E(B-V)=0.19$ and R=3.1. ``H$\alpha$" is a blend of
H$\alpha$ and [\ion{N}{ii}]~$\lambda\lambda$6548,6583.
} \label{f:Snrs_spa}
\end{figure*}

%-----------------------figure12------------------------------
\begin{figure*}[hbt]
\begin{center}
\includegraphics[width=170mm,angle=0, clip]{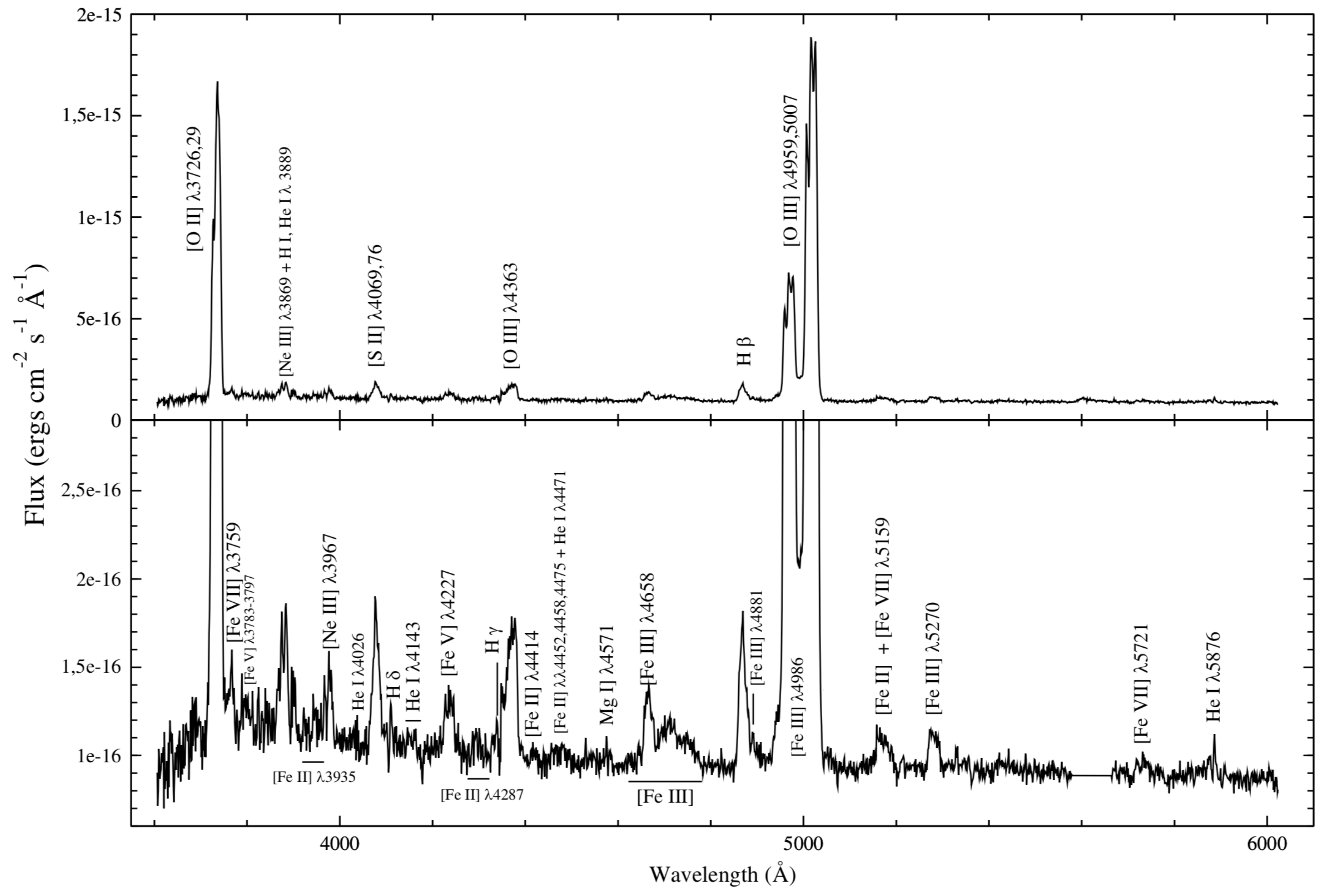}
\end{center}  
\caption{Spectrum obtained with ESO$/$VLT (PA=88\degr), and again 
(like in the Fig.~\ref{f:Snrs_spa}) the lower panel is plotted
to bring out weaker lines. The spectrum has been dereddened using
$E(B-V)=0.19$ and R=3.1. Note the significantly higher signal-to-noise in the
VLT spectrum compared to Fig.~\ref{f:Snrs_spa}, which made it possible to detect new lines 
in~\snr\, as reported in \citet{Ser05} and here. Of particular importance are the [\ion{Ne}{iii}] 
lines and H$\beta$ (see text). 
} \label{f:Snrs_spb}
\end{figure*}

\subsubsection{Density from [\ion{S}{ii}]~$\lambda\lambda$6716,6731}

The two components of the [\ion{S}{ii}] doublet in Fig.~\ref{f:velom1} have a difference in wavelength
corresponding to a velocity shift of $\approx 640 \kms$. In 0540 they blend together
because of the velocity broadening of the emitting gas. There are,
however, a few positions along the slit for which the blend is small.
One such position is marked with a vertical solid line in Fig.~\ref{f:velom1}, along 
detector row 29 in our notation. We discussed the deblending of 
[\ion{S}{ii}]~$\lambda\lambda$6716,6731 along this row in \citet{Ser05}, assuming 
that the line intensity ratio $R_{\rm  [\ion{S}{ii}]} = \frac{I(\lambda6716)}{I(\lambda6731)}$ 
is the same for all parts of the supernova remnant encapsulated by the slit row.
%We show in Fig.~\ref{f:SIIprofile29} the intensity profile (dashed line) along this 
%detector row, i.e, in velocity space. Zero velocity is for the rest wavelength of 
%[\ion{S}{ii}]~$\lambda$6731 (in this plot uncorrected for LMC redshift). 
%In \citet{Ser05}. 
%!TEX encoding = UTF-8 Unicodebriefly, but we do it here in greater detail.  
%We have assumed that the line intensity ratio 
%$R_{\rm  [\ion{S}{ii}]} = \frac{I(\lambda6716)}{I(\lambda6731)}$ 
%is the same for all parts of the supernova remnant encapsulated by detector 
%row 29. 
As $R_{\rm  [\ion{S}{ii}]}$ is density sensitive \citep[e.g.,][]{ost06}, 
we effectively assumed that the density is constant in this part 
of the remnant. 
%For such a situation, we only need a good template line profile (which is 
%thus the same for both components), and then adjust the relative strengths 
%of the two line components to fit the total blended profile. A reasonable 
%assumption (from an inspection of Fig.~\ref{f:velom1}) is that only the $\lambda$6731
%component contributes at $\gsim +1\,000 \kms$, and the  $\lambda$6716 component
%at $\lsim -500 \kms$. Furthermore, the separation between the peaks of the
%total observed line profile is very similar to the wavelength shift between 
%the two components, i.e., $\approx 640 \kms$, so one can be rather confident in that
%\citet{Ser05} found that $R_{\rm  [\ion{S}{ii}]}  < 1$, which 
%especially
%since a similar separation between the main peaks is also seen for other 
%detector rows. The fact that  $R_{\rm [\ion{S}{ii}]} < 1$ 
%immediately points to an electron density, $n_{\rm e} \gsim 5\EE2 \cm3$ \citep{ost06}.
The debelendig in \citet{Ser05} resulted in $R_{\rm [\ion{S}{ii}]} = 0.7$,
and the multilevel model for [\ion{S}{ii}] described 
in \citet{Mar00} was used
%and also used by \citet{Sand13}. Adopting $R_{\rm [\ion{S}{ii}]} = 0.7\pm0.1$ 
to obtain $n_{\rm e} = (1.4-4.3)\EE3 \cm3$ for the temperature $T=10^4$ K, 
and $n_{\rm e} = (1.8-5.3)\EE3 \cm3$ for $T=2\EE4$ K. 

As we discuss in Sect. 3.4, the temperature in [\ion{S}{ii}]-emitting gas is probably $T \sim 15\,000$~K.
In addition, a more careful background subtraction along row 29 than the preliminary one 
in \citet{Ser05} gives $R_{\rm [\ion{S}{ii}]} = 0.85\pm0.10$,
translating into $n_{\rm e} = (0.9-2.0)\EE3 \cm3$. This is fully consistent with densities found in the 
Crab Nebula for which [\ion{O}{ii}] and[\ion{S}{ii}] line ratios indicate electron
densities in the range $4\EE2 - 4\EE3 \cm3$ for the various
filaments observed  \citep{DF85}. 
For the region mapped out by our detector row 29, \citet{Sand13} estimate 
$n_{\rm e} \lsim 1\EE3 \cm3$. As emphasized by these authors there are, however, 
pockets of ejecta with $n_{\rm e} \sim (1-2)\EE3 \cm3$ in the SNRC when spectra 
from individual fibers are studied; their density plot was smoothed over a $2\times2$ pixel kernel.
Inspection of the individual fibre spectra of \citet{Sand13} also reveals background LMC
residuals, which introduces a bias to underestimate electron densities from [\ion{S}{ii}].
The reason is the small field of view of VIMOS which makes background subtraction cumbersome.
Our estimate of $n_{\rm e} \sim (1-2)\EE3 \cm3$ for
detector row 29 are therefore not in conflict with \citet{Sand13}. For a more complete discussion 
about the variation of electron density from [\ion{S}{ii}] within the SNRC we refer to 
\citet{Sand13}, with the caveat in mind that the electron densities in that paper coud be
somewhat underestimated in general. In addition, we emphasize that the 
densities derived for detector row 29 and in \citet{Sand13} are average densities
along the line-of-sight for the SNRC, unlike the situation in the Crab for which more
detailed 3D studies can be done more easily due its proximity \citep[e.g.,][]{Charle10,Mart21}.
In the analysis in Sect. 3.4 for the SNRC we will use  $n_{\rm e} =10^3 \cm3$ unless otherwise specified.

\subsection{One-dimensional spectroscopy}

\subsubsection{The central part of 0540}

Previous optical spectroscopic studies of the SNRC were carried out by \citet{Mat80}, 
\citet{Kirshner89} and \citet{Morse06}, and we reported preliminary results of our NTT/EMMI 
and VLT/FORS observations in \citet{Ser05}. We will mainly compare with \citet{Kirshner89} 
and \citet{Morse06}. 
\citet{Kirshner89} used a slightly larger slit ($1\farcs5$) than us, and positioned
their slit to cross the SNRC at PA~$=77\degr$. \citet{Morse06} used an even
wider slit ($2\arcsec$) at PA$=124\degr$, and we discussed some of their results already in
Sect. 3.2. 

We extracted 1D spectra of the SNRC from our spectral images using the {\sf IRAF}
procedure {\sf apall} and spatial extents of 10\arcsec\ and 8\arcsec\ centered
on the pulsar for slits ``1" and ``2", respectively. The extracted windows
correspond to the observed extents of the SNRC along the respective slit
directions. The extracted spectra are shown in Fig.~\ref{f:Snrs_spa}
and \ref{f:Snrs_spb}. 

Line fluxes were derived by integrating over each line profile. (No
Gausian fitting can be done since the line profiles are strongly
non-Gaussian.) This gives accurate fluxes for strong lines 
like [\ion{O}{ii}] and [\ion{O}{iii}] as found from a test where we varied the
background level up and down by $1\sigma$ from the mean value; the
statistical error of the strongest lines is less than 5\%. For faint lines
like the [\ion{Fe}{ii}] lines and \ion{Mg}{i}] the flux uncertainty can be up to 40\%.
In Table~\ref{t:FluxVel} we have marked the fluxes of such lines by
a colon.

A list of identified lines and their measured fluxes, central wavelengths and
velocities is presented in Table~\ref{t:FluxVel}. The results of \citet{Kirshner89} and \citet{Morse06} are
also included for comparison. All our fluxes have been dereddened 
using $E(B-V)=0.19$ and R=3.1, as was also done by Kirshner et al., while \citet{Morse06}
used $E(B-V)=0.20$. A detailed discussion on the extinction towards 0540
can be found in \citet{Ser04}. The fluxes were then normalized to 
[\ion{O}{iii}]$\wl 5007$. We note that  \citet{Kirshner89} and \citet{Morse06} normalized to the sum 
of [\ion{O}{iii}]$\wl4959$ and [\ion{O}{iii}]$\wl5007$. In Table~\ref{t:FluxVel}
we therefore simply assumed a $1:3$ line ratio of those two components, and renormalized all 
fluxes in \citet{Kirshner89} and \citet{Morse06} to [\ion{O}{iii}]$\wl 5007$. 
Guided by the VIMOS/IFU observations of \citet{Sand13}, who 
concentrated on the two doublets [\ion{O}{iii}]$\wll 4959,5007$ and [\ion{S}{ii}]$\wll 6716,6731$,
%(cf. the top panels of Figs.~\ref{f:velom1} and \ref{f:velom2})
one expects line fluxes for the various
slit orientations shown in Table~\ref{t:FluxVel} to be different just because
of different parts of the SNRC being caught by the slits. 

Our 1D spectra of the SNRC are consistent with those of the other studies
but also reveal many differences. The higher sensitivity of the VLT
observations allowed us in \citet{Ser05} to report many new lines not detected 
in previous studies, and we report more here. The most important findings in \citet{Ser05}
were [\ion{Ne}{iii}]$\wll$3869,3967 and Balmer lines of hydrogen all the way down to 
\ion{H}{i}~$\lambda$3889 (H$\zeta$). While the neon lines constrain the supernova, the Balmer 
lines show that the previously detected emission around H$\alpha$ is at least partly due 
to H$\alpha$, and not only to [\ion{N}{ii}]~$\lambda$6583, or any other line as discussed by 
\citet{Car98}. We do not try to separate H$\alpha$ from 
%We the most likely blend, i.e., 
[\ion{N}{ii}]$\wll$6548,6583. However, from the H$\alpha$ panel of Fig.~\ref{f:velom1} it is clear 
that [\ion{N}{ii}] is present with a velocity structure that looks similar to that of [\ion{S}{ii}], and with a 
velocity for [\ion{N}{ii}]$\wl$6583 that reaches at least $+1\,000 \kms$ in the LMC rest frame.
\citet{Morse06} estimated an intensity ratio of $I_{{\rm H}\alpha} / I_{[\ion{N}{ii}]} \sim 1.1$
for their slit position.

The [\ion{Ne}{iii}]$\wll$3869,3967 lines, and [\ion{O}{iii}]~$\lambda$4363 are all
contaminated by \ion{H}{i} lines, which leads to an overestimate of the fluxes of
these lines not accounted for in Table~\ref{t:FluxVel}, and for [\ion{O}{iii}]~$\lambda$4363 fluxes in 
previous investigations. Another complication is that all \ion{H}{i} lines suffer from 
influence of the uneven LMC background which must be compensated for. We 
discuss this further Sect.~\ref{lin_mesh}.

Using unblended lines we estimate that the mean velocity of the central 
region of the 0540 system is +500$\pm$55 $\kms$ for slit ``1" 
and +380$\pm$60 $\kms$ for slit ``2", respectively, when the LMC redshift 
of $270 \kms$ is accounted for. Within errors both values are consistent with
each other, and a mean velocity of the two gives +440$\pm$80 $\kms$.  
This is $\sim 100 \kms$ higher than the value of  \citet{Kirshner89}, but
the two investigations are consistent with each other, despite different slit orientations.

The redshift of the SNRC of 0540 is indeed large. An explanation could be that the 
0540 system is the result of an asymmetric explosion, as discussed by \citet{Ser04}.
The SNRC of 0540 is not unique in that sense. 
\citet{BalHek78} found that the oxygen-rich SNR NGC 4449 also moves at 
a velocity of $\sim 500 \kms$ relative to its surrounding H~II regions. Like 
0540, this remnant is also oxygen-rich. Over a larger scale, 0540 may not be 
all that asymmetric as revealed by the [\ion{O}{iii}] glow from ejecta moving at high
speed ($\gsim 1\,500 \kms$) both towards and away from us discussed in Sect. 3.2.
\citep[see also][]{Morse06,Sand13}.

%-----------------------figure13------------------------------
\begin{figure*}[htb]
\begin{center}
\includegraphics[width=170mm,angle=0, clip]{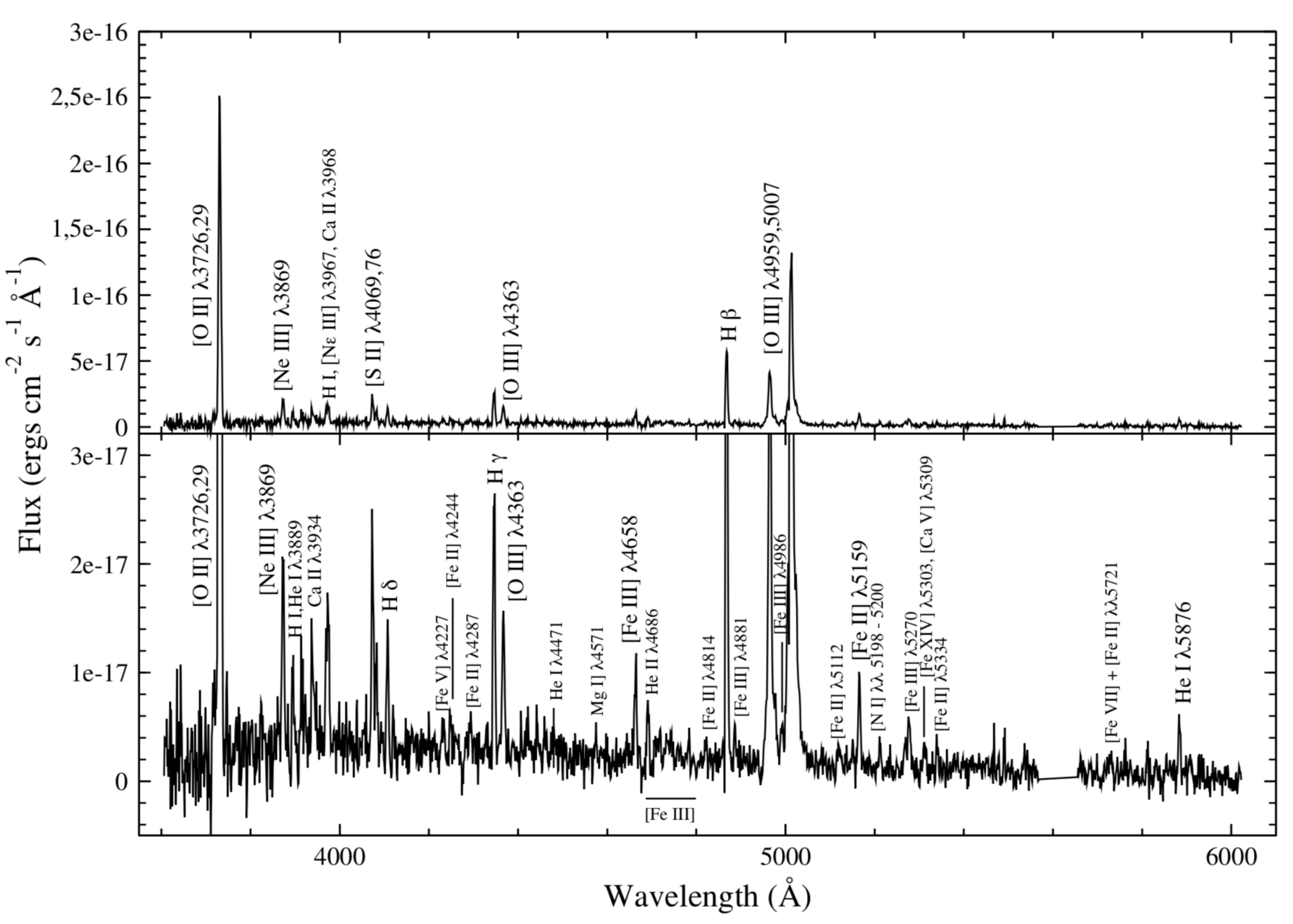}
\end{center}  
\caption{Spectrum of filament F1 in the vicinity of~\snr. 
The filament is situated only~8\arcsec\ west of the pulsar 
(see Figs.\ref{f:OIIIim} and~\ref{f:OIIIim2}). 
As in Figs.~\ref{f:Snrs_spa} and \ref{f:Snrs_spb}, we have also 
plotted the spectra with expanded flux scales to highlight weaker lines.
The spectrum was dereddened using $E(B-V)=0.19$ and R=3.1. Note the broad 
base of the [\ion{O}{iii}]~$\lambda\lambda$4959,5007 lines skewed to the red 
as compared to the narrow components, reaching at least +1\,700$\kms$. 
%A broad base is also hinted in H$\beta$. 
No such features are seen in spectra of the emission from filaments F2, F3, F4
or F5 (see Fig.~\ref{f:HIIreg_spb}) as marked in Fig.~\ref{f:OIIIim}. 
Line fluxes of all filaments are given in Tables~\ref{t:FluxVel_2} and \ref{t:FluxVel_3}.
} \label{f:HIIreg_spa}
\end{figure*}

%-----------------------figure14------------------------------
\begin{figure*}[htb]
\begin{center}
\includegraphics[width=170mm,angle=0, clip]{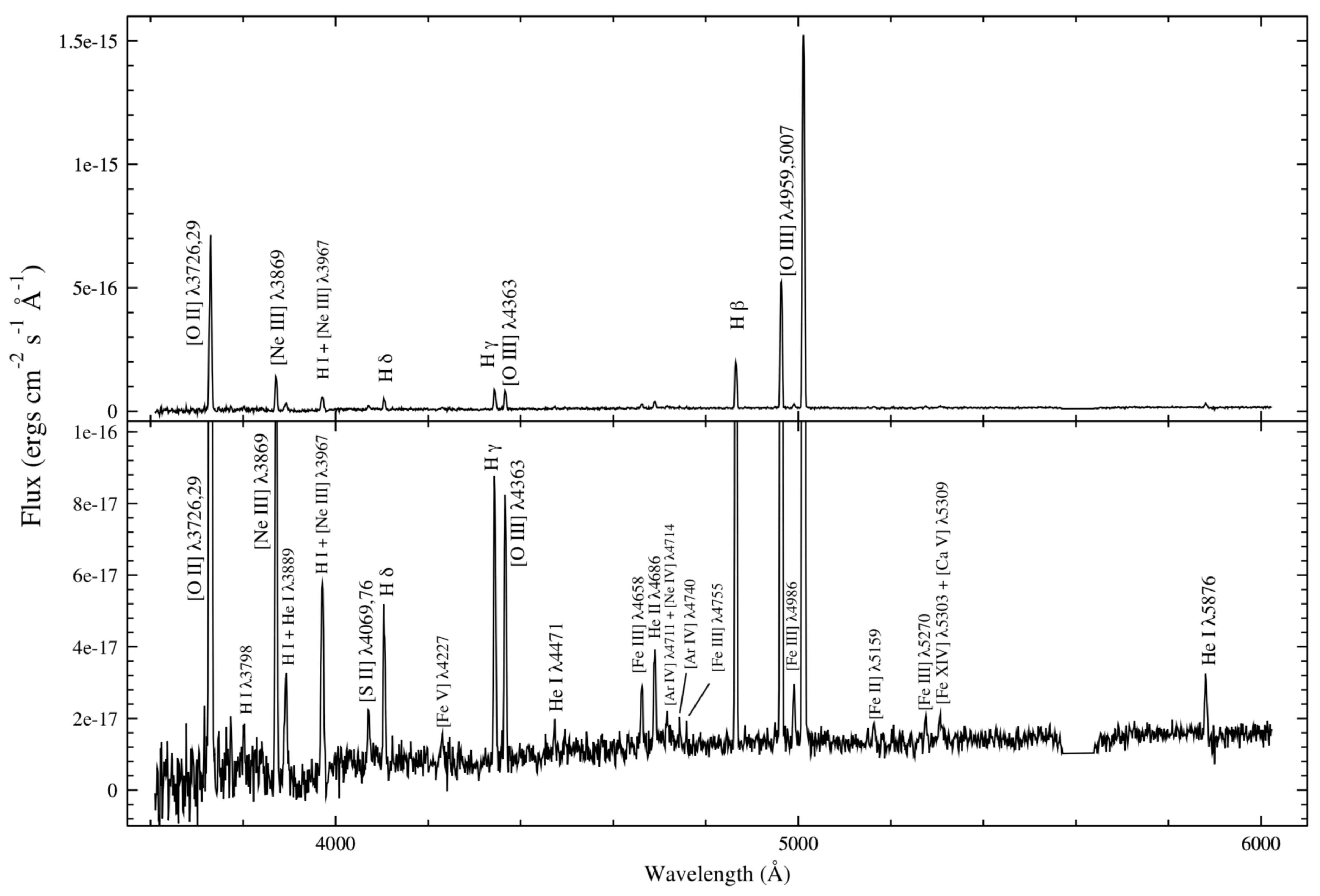}
\end{center}  
\caption{%Spectra of H~II regions in the vicinity of~\snr. 
Spectrum of an H~II region along the VLT slit ``3'' at a projected distance 
of~25\arcsec\ to the west of the pulsar (called filament F5 in 
Figs.~\ref{f:OIIIim} and~\ref{f:OIIIim2}). As in Fig.~\ref{f:Snrs_spa}, we have also 
plotted spectra with expanded flux scales to highlight weaker lines.
The spectrum have been dereddened using $E(B-V)=0.19$ and R=3.1. 
} \label{f:HIIreg_spb}
\end{figure*}

\subsubsection{Filaments}

\snr\ is near the LMC H~II region DEM 269 and the OB association LH 104
\citep[see][and references therein, as well as below.]{Kirshner89} This is consistent with 
that we detect narrow emission lines that vary spatially in strength along our slits. 
The lines are spectrally unresolved (except for one case, see below).

To study whether or not some of these emitting regions are affected
by the SNR activity, we looked at filaments detected along the
slits. We mainly considered those with an [\ion{O}{iii}] temperature in clear excess
of $10^4$~K, as is usually observed in shocks or photoionization heated
regions of known SNRs. We assumed the same extinction for the filaments
as for the remnant, and used the standard ratio 
$R_{\rm [\ion{O}{iii}]} = \frac{I(\lambda4959)+I(\lambda5007)}{I(\lambda4363)}$
for the temperature \citep[e.g.,][]{ost06}. We also looked for lines of highly ionized 
ions as tracers of influence by the SNR.
We detected a few filaments of interest along the VLT slits ``1" and ``3", and they are 
marked as F1, F2, F4 and F5 in Fig.~\ref{f:OIIIim}. We also mark filament F3,
and use this as reference for a presumed LMC H~II region.

The typical size of the filaments
along the slit is  $\sim 3 \arcsec$. As in the case of the SNRC we have 
constructed 1D spectra for each filament. The dereddened spectra of F1, which
is closest to the SNRC in projection, and F5, which appears close to the 
outer shock front detected in X-rays, are shown in
Figs.~\ref{f:HIIreg_spa} and \ref{f:HIIreg_spb}, respectively.
Spectra of the other filaments are similar to the ones presented. Lists
of identified spectral lines for F1-F5 with measured fluxes, central wavelengths and
velocities, are presented in Tables~\ref{t:FluxVel_2} and~\ref{t:FluxVel_3}.
We return to a more detailed discussion of the filaments in Sect. 3.4.2.

%%%%%%%%%%%%%%%%%%%%%%%%%%%%%%%%%%%%%%%%%%%%%%%%%%%%%%%%%%%%%%%%%%%%%%%%%%
%______________________________________________________________
\subsection{Line profiles and intensities}\label{lin_mesh}
\subsubsection{The central part}\label{lin_mesh_snrc}
%______________________________________________________________
%%%%%%%%%%%%%%%%%%%%%%%%%%%%%%%%%%%%%%%%%%%%%%%%%%%%%%%%%%%%%%%%%%%%%%%%%%
Supernova remnants are good laboratories to test the theory of stellar
evolution, explosive nucleosynthesis, radiation processes and shock physics. 
To test the first two items, an estimate of elemental abundances of the ejecta
is essential. As the information from SNRs mainly rests on collisionally excited
lines, we therefore also require knowledge about the density and temperature
of the emitting gas.

The observed strength of the spectral lines allow us to deduce this information. 
The prime thermometer is $R_{\rm [\ion{O}{iii}]}$.
%the $\frac{[\ion{O}{iii}]\wll 4959,5007}{[\ion{O}{iii}]\wl 4363}$ ratio.
Because we see [\ion{O}{iii}]$\wll 4959,5007$ and [\ion{O}{iii}]$\wl$4363 in all VLT spectra of
0540 and the filaments, we are able to estimate the temperature of the [\ion{O}{iii}] emitting plasma in both the
SN ejecta and the filaments.
%-------------------------------------------------------------
%-----------------------figure15------------------------------
\unitlength=1mm
\begin{figure}[htb]
\begin{center}
\includegraphics[width=180mm, angle=90, clip]{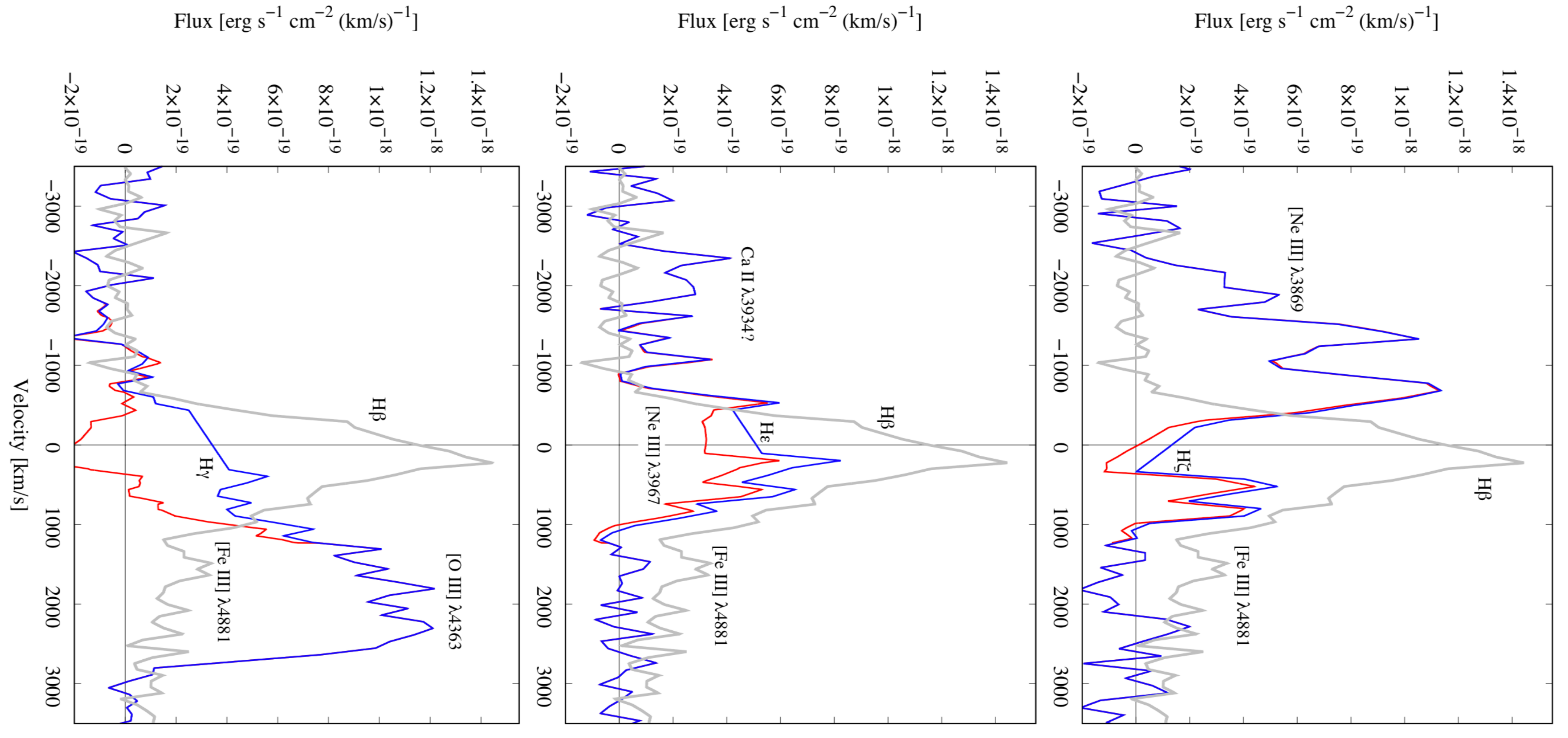}
\end{center}
\caption{Line profiles of the Balmer lines from 0540 in velocity space relative to LMC 
for the VLT spectrum in Fig.~\ref{f:Snrs_spb}. {\it Top panel:} H$\zeta$ blended with  
[\ion{Ne}{iii}]$\wl$3869, 
{\it Middle panel:} H$\epsilon$ blended with [\ion{Ne}{iii}]$\wl$3967, and 
{\it bottom panel:} H$\gamma$ blended with [\ion{O}{iii}]$\wl$4363. In all panels we also show 
H$\beta$ (grey) and  the continuum level (black). 
H$\beta$ was used as a template to subtract the expected emission from H$\gamma$,
H$\epsilon$ and H$\zeta$ using Case B line ratios \citep{Brock71}. The blue and red solid lines show
[\ion{O}{iii}]$\wl$4363 and [\ion{Ne}{iii}]$\wll$3869,3967 before and after deblending, respectively. 
} \label{f:velo_temp}
\end{figure}
%------------------------------------------------------------
%-----------------------figure16------------------------------
\unitlength=1mm
\begin{figure}[htb]
\begin{center}
\includegraphics[width=90mm, angle=0, clip]{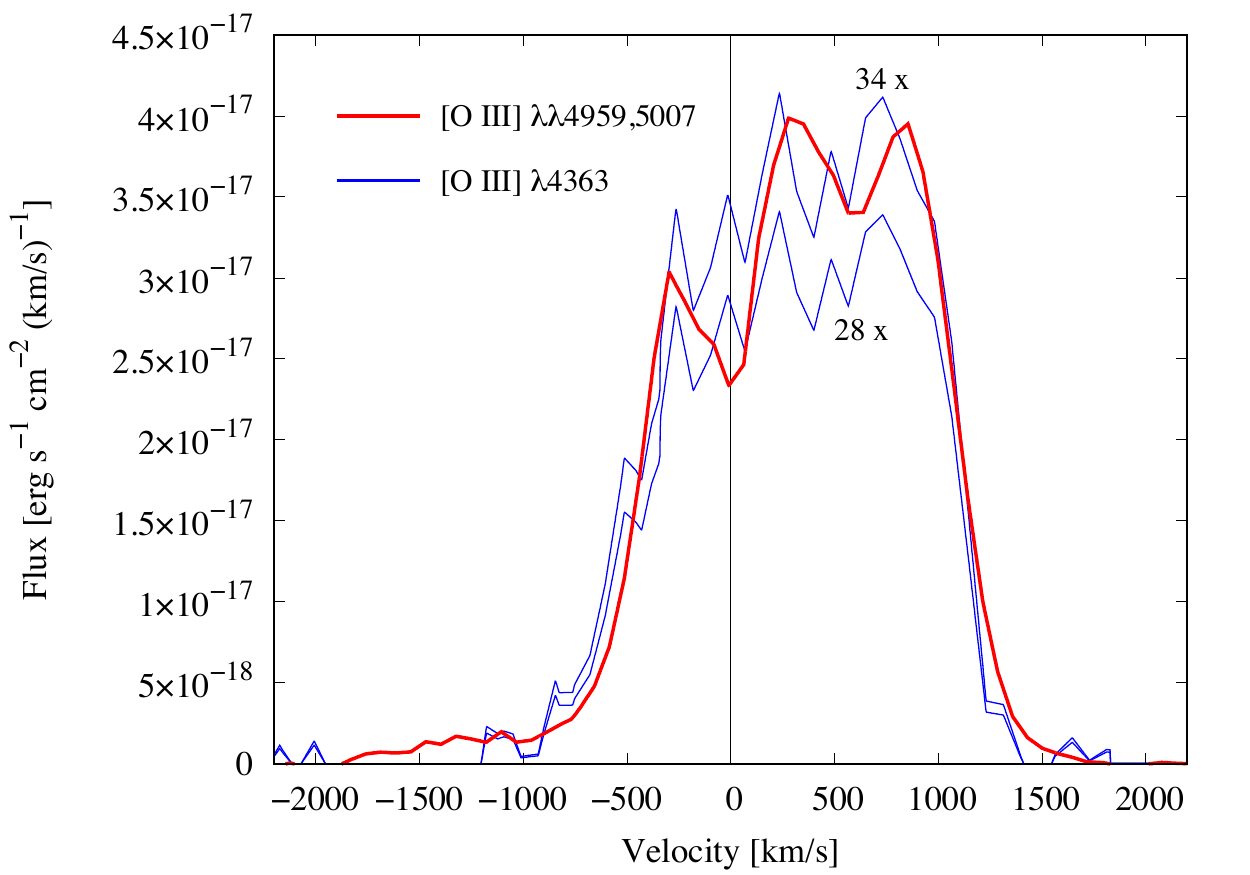}
\end{center}
\caption{Line profiles of [\ion{O}{iii}]~$\lambda\lambda$4959,5007 and [\ion{O}{iii}]$\wl$4363 of the emission from 0540 as observed 
through the VLT slit. 
The velocity is relative to LMC. The [\ion{O}{iii}]~$\lambda\lambda$4959,5007 profile
was created from a cleaning procedure described in the text, and is for the sum of the two line components. 
The [\ion{O}{iii}]$\wl$4363 is the blue-red profile in the bottom panel of Fig.~\ref{f:velo_temp}. The flux of [\ion{O}{iii}]$\wl$4363
has been multiplied by 28 and 34 to match the various parts of the line profiles. There is a tendency of relatively weaker 
[\ion{O}{iii}]$\wl$4363 on the blue side, which could mean hotter  [\ion{O}{iii}]-emitting plasma on the receding side of 0540 (see
text). 
%The [\ion{O}{iii}]~$\lambda\lambda$4959,5007 emission stretches at least out to $\pm 1\,800 \kms$. 
} \label{f:velo_O3_prof}
\end{figure}
%------------------------------------------------------------
%-----------------------figure17------------------------------
\unitlength=1mm
\begin{figure}[htb]
\begin{center}
\includegraphics[width=90mm, angle=0, clip]{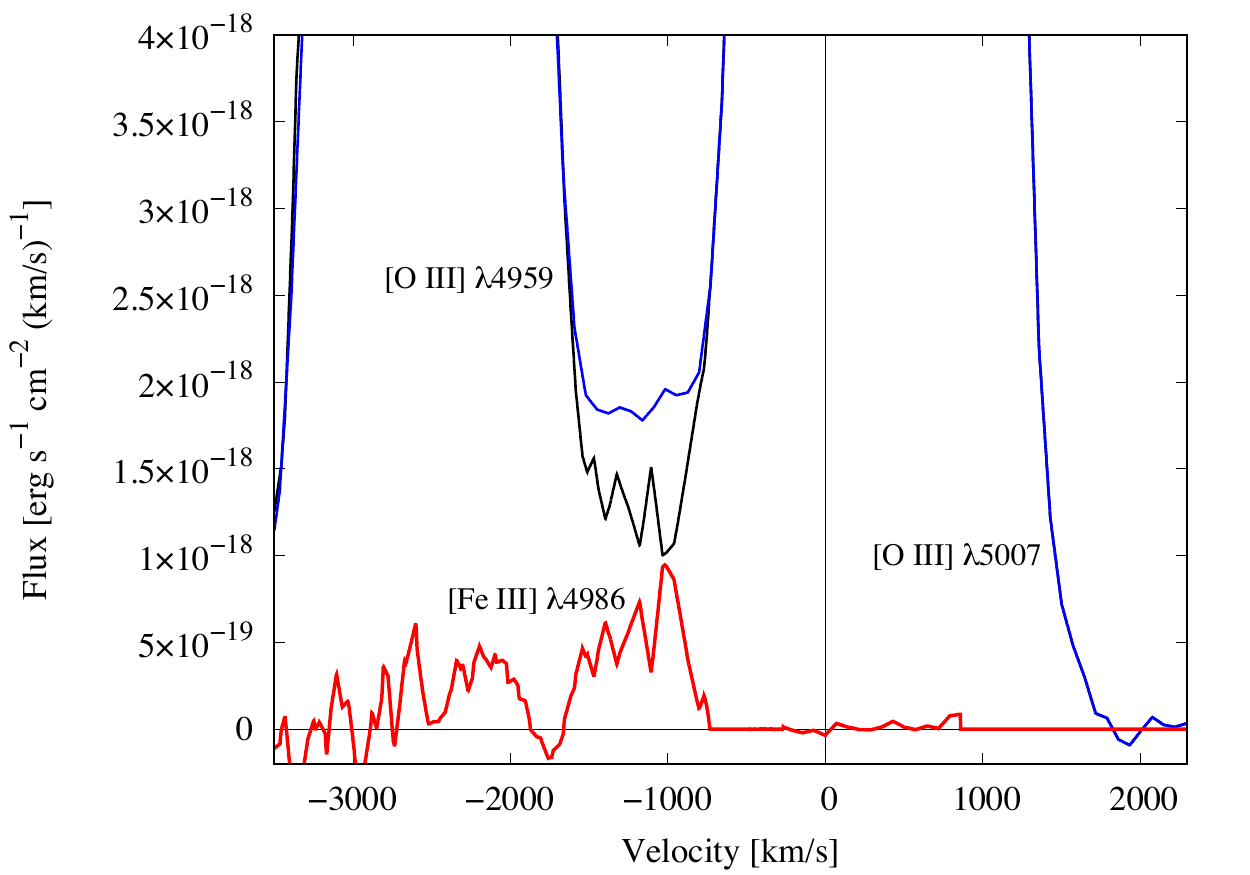}
\end{center}
\caption{Sum of the ``cleaned" [\ion{O}{iii}]$\lambda\lambda$4959,5007 profiles (cf. Fig.~\ref{f:velo_O3_prof}) of the emission from 0540 
(black) compared to the observed emission (blue). The velocity scale is for the $\wl$5007 component relative to LMC. The difference 
(red) is attributed to [\ion{Fe}{iii}]$\wl$4986. The maximum 
velocity on the red side of [\ion{O}{iii}]~$\wl$5007 is $\approx +1\,800 \kms$, which may be less than on the blue side 
($\approx -1\,900 \kms$, see Fig.~\ref{f:velo_O3_prof}).
} \label{f:Fe_find}
\end{figure}
%------------------------------------------------------------

%-----------------------figure18------------------------------
\unitlength=1mm
\begin{figure}[htb]
\begin{center}
\includegraphics[width=90mm, angle=0, clip]{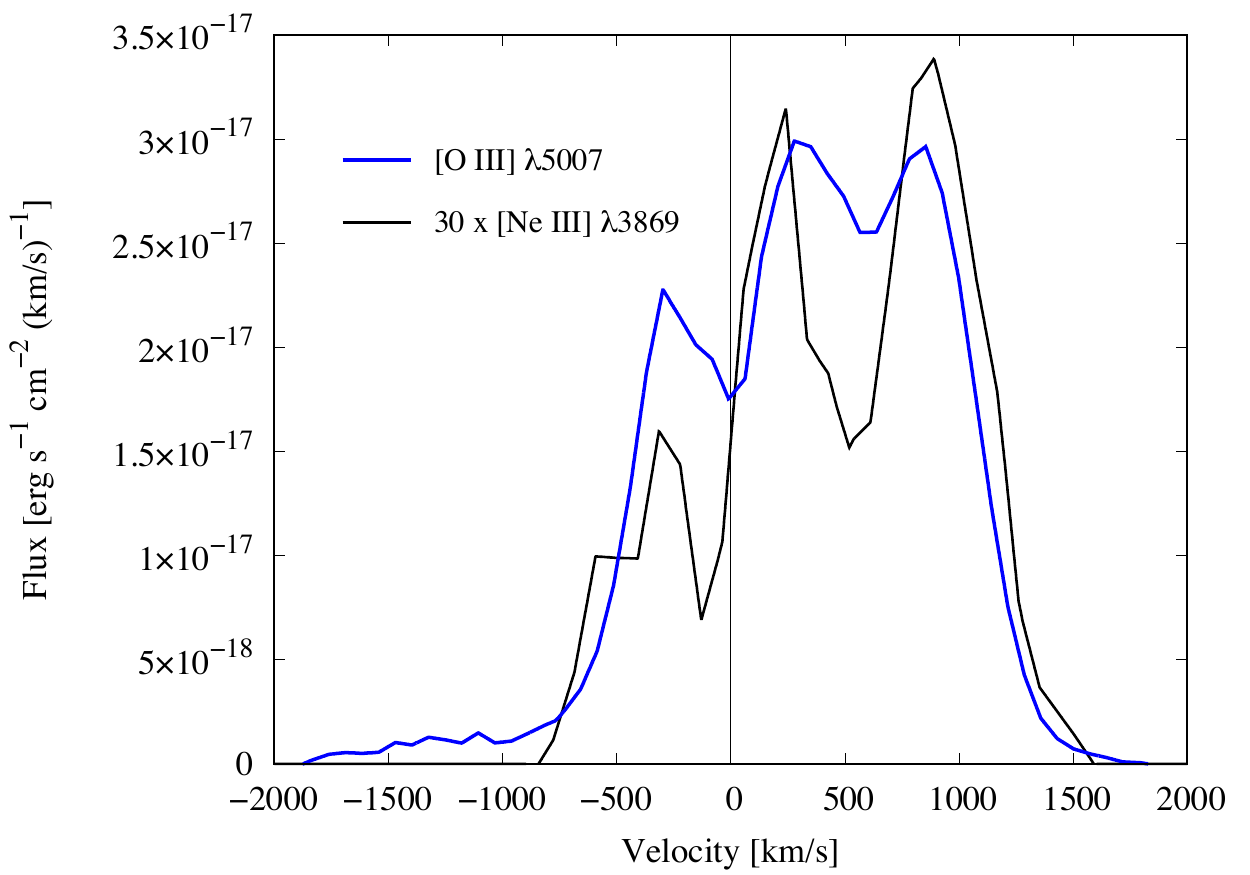}
\end{center}
\caption{Same profile for [\ion{O}{iii}] as in Fig.~\ref{f:velo_O3_prof}, but only for [\ion{O}{iii}]~$\lambda$5007 (blue) together 
with the ``cleaned" [\ion{Ne}{iii}]~$\lambda$3869 profile in Fig.~\ref{f:velo_temp}. The  [\ion{Ne}{iIi}] profile was multiplied by a 
factor of 30 to match [\ion{O}{iii}]~$\lambda$5007. Note the similar widths and structures of the two profiles. Noise is too high
for the [\ion{Ne}{iii}] line to trace it out the highest velocities on the blue side.
} \label{f:O_Ne}
\end{figure}
%------------------------------------------------------------

%-----------------------figure19------------------------------
\unitlength=1mm
\begin{figure}[htb]
\begin{center}
\includegraphics[width=90mm, angle=0, clip]{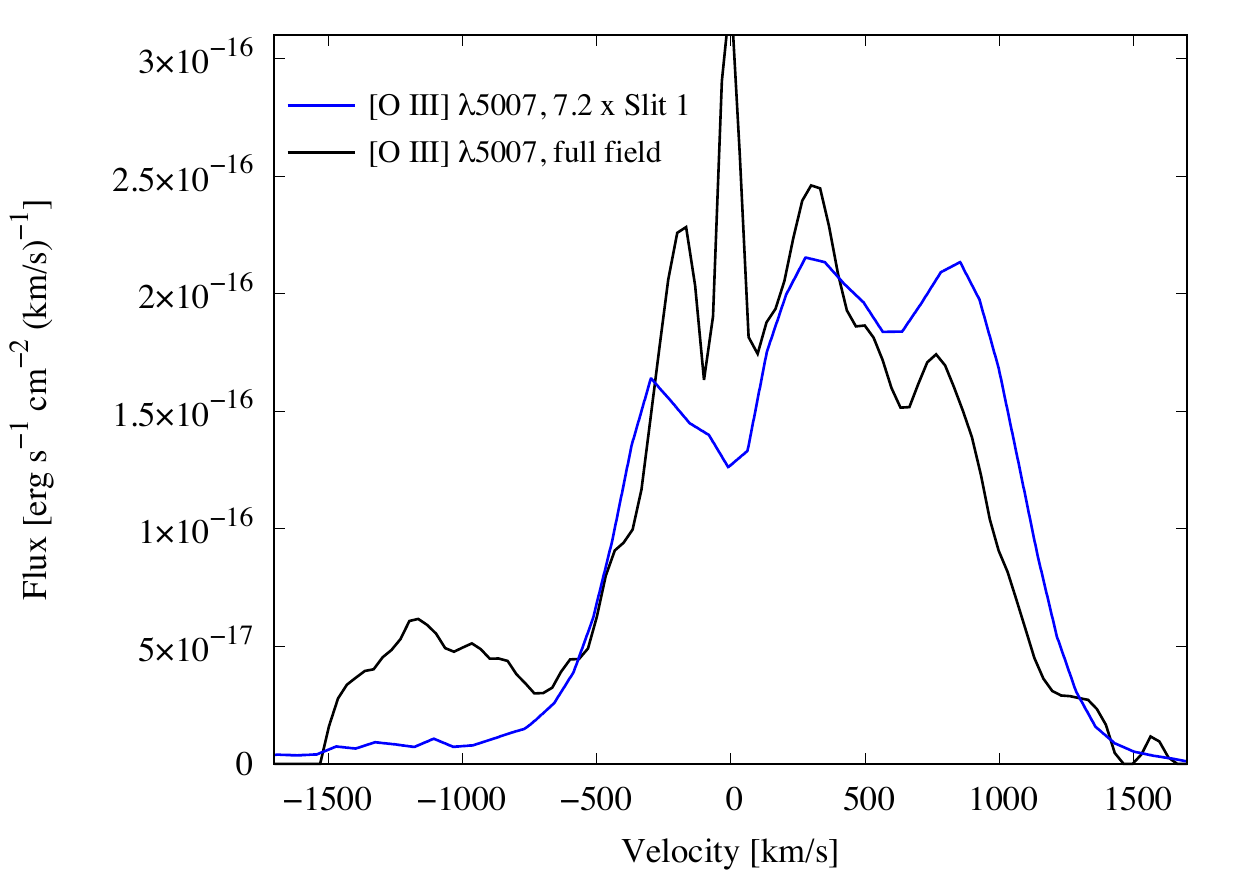}
\end{center}
\caption{Same profile for [\ion{O}{iii}]$\lambda\lambda$4959,5007 as in Fig.~\ref{f:velo_O3_prof} (blue) together with the
[\ion{O}{iii}]$\lambda\lambda$4959,5007 profile for the full $13\arcsec \times 13\arcsec$ field as observed by \citet{Sand13}.
The slit 1 profile was multiplied by a factor of 7.2 to match the integrated emission at $\geq -750 \kms$. Note the strong
[\ion{O}{iii}] emission for $\leq -750 \kms$ for the larger field-of-view compared to the slit 1 observations.
This emission corresponds to about 10\% of the total [\ion{O}{iii}] emission. A minor fraction ($\sim 10\%$) of the emission at 
$\leq -750 \kms$ could be due to [\ion{Fe}{iii}]~$\lambda$4986, as displayed in Fig.~\ref{f:Fe_find}, but most of it comes from
the approaching ``wall" described in Sect. 3.2.
} \label{f:Slit_vs_Full}
\end{figure}
%------------------------------------------------------------

%-----------------------figure20------------------------------
\unitlength=1mm
\begin{figure}[htb]
\begin{center}
\includegraphics[width=90mm, angle=0, clip]{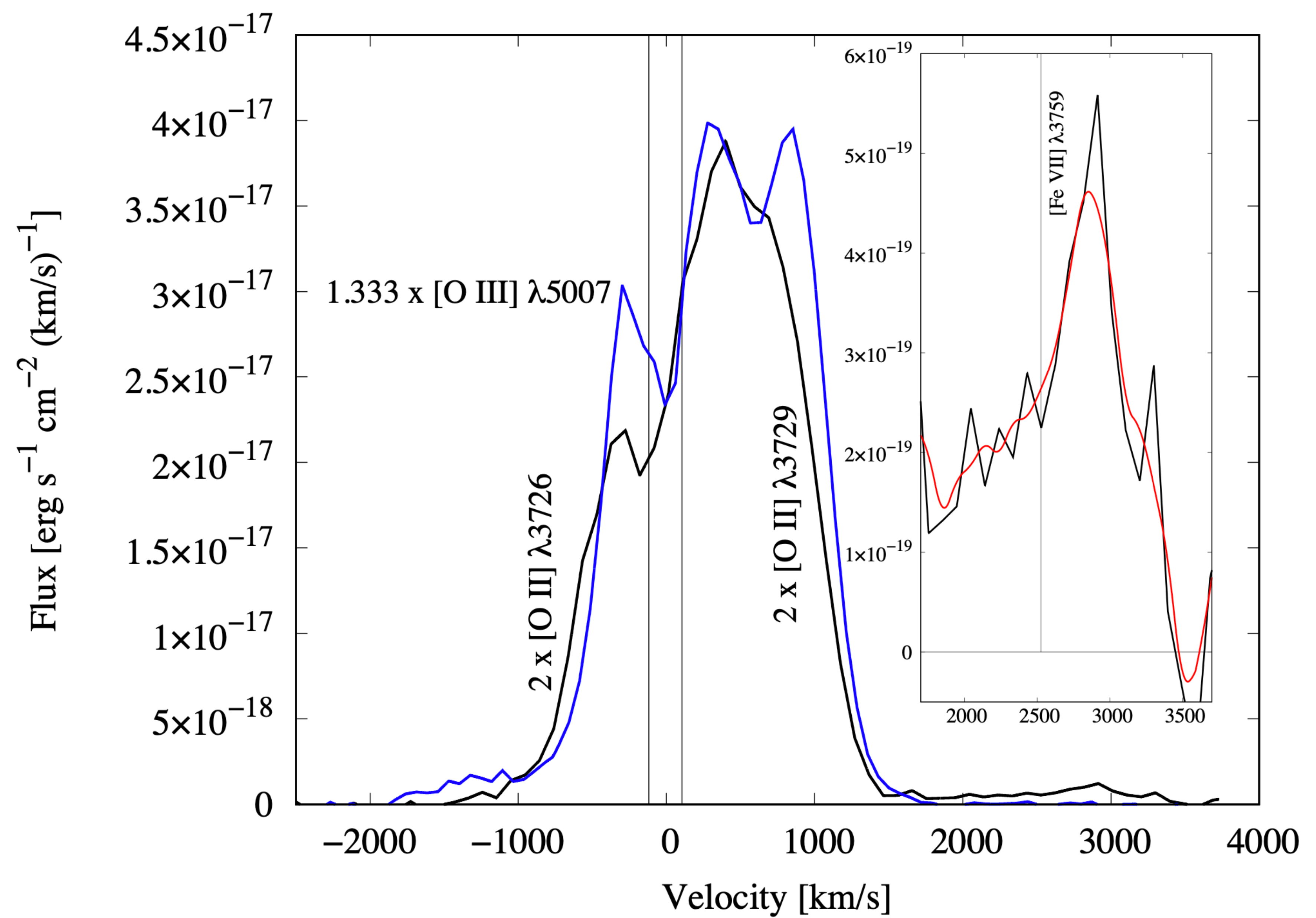}
\end{center}
\caption{Same profile for [\ion{O}{iii}]$\lambda\lambda$4959,5007 as in Fig.~\ref{f:velo_O3_prof} (blue) together with the
combined [\ion{O}{ii}]~$\lambda\lambda$3726,3729 profile. The velocity for the [\ion{O}{ii}] lines is centered on 3727.5~\AA. 
The  [\ion{O}{ii}] profile was multiplied by a factor of 2 to match $1.33\times$~[\ion{O}{iii}]~$\lambda$5007. 
Note that [\ion{O}{ii}]~$\lambda$3726 only reaches $\approx -1\,400 \kms$ on the blue side (when the $118 \kms$ difference
between [\ion{O}{ii}]~$\lambda$3726 and 3727.5~\AA\ is accounted for). On the red side of [\ion{O}{ii}]~$\lambda$3729 the emission 
is blended  with [\ion{Fe}{vii}]~$\lambda$3759 as shown by the inset. The red curve is Savitzky-Golay fitting to the noisy [\ion{Fe}{vii}] 
line profile.
} \label{f:O2_O3}
\end{figure}
%------------------------------------------------------------

%-----------------------figure21------------------------------
\unitlength=1mm
\begin{figure}[htb]
\begin{center}
\includegraphics[width=90mm, angle=0, clip]{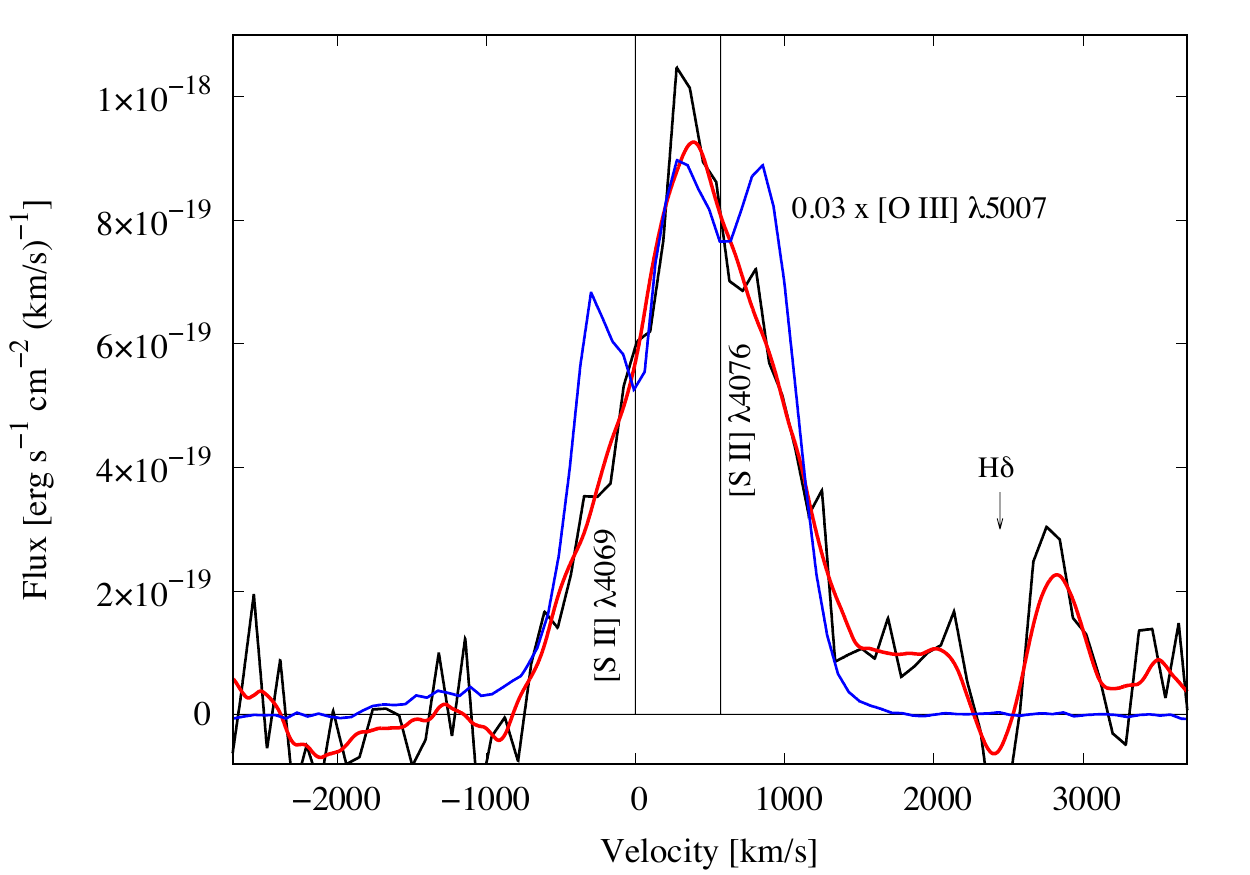}
\end{center}
\caption{Same in Fig.~\ref{f:O_Ne}, but for [\ion{S}{ii}]~$\lambda$$\lambda$4069,4076 (black, and smoothed in red) instead of 
[\ion{Ne}{iii}]~$\lambda$3869. The [\ion{O}{iIii}] profile was multiplied by a factor of 0.03 to match [\ion{S}{ii}]. The sulphur doublet was
velocity-centered on the $\lambda$4069 component. LMC rest frame velocities for both components are marked by vertical thin lines.
The red side of the [\ion{S}{ii}] doublet is affected by H$\delta$. (The dip in H$\delta$ is due to over-subtraction of the LMC background.)
} \label{f:O_S}
\end{figure}
%------------------------------------------------------------

%-----------------------figure22------------------------------
\unitlength=1mm
\begin{figure}[htb]
\begin{center}
\includegraphics[width=90mm, angle=0, clip]{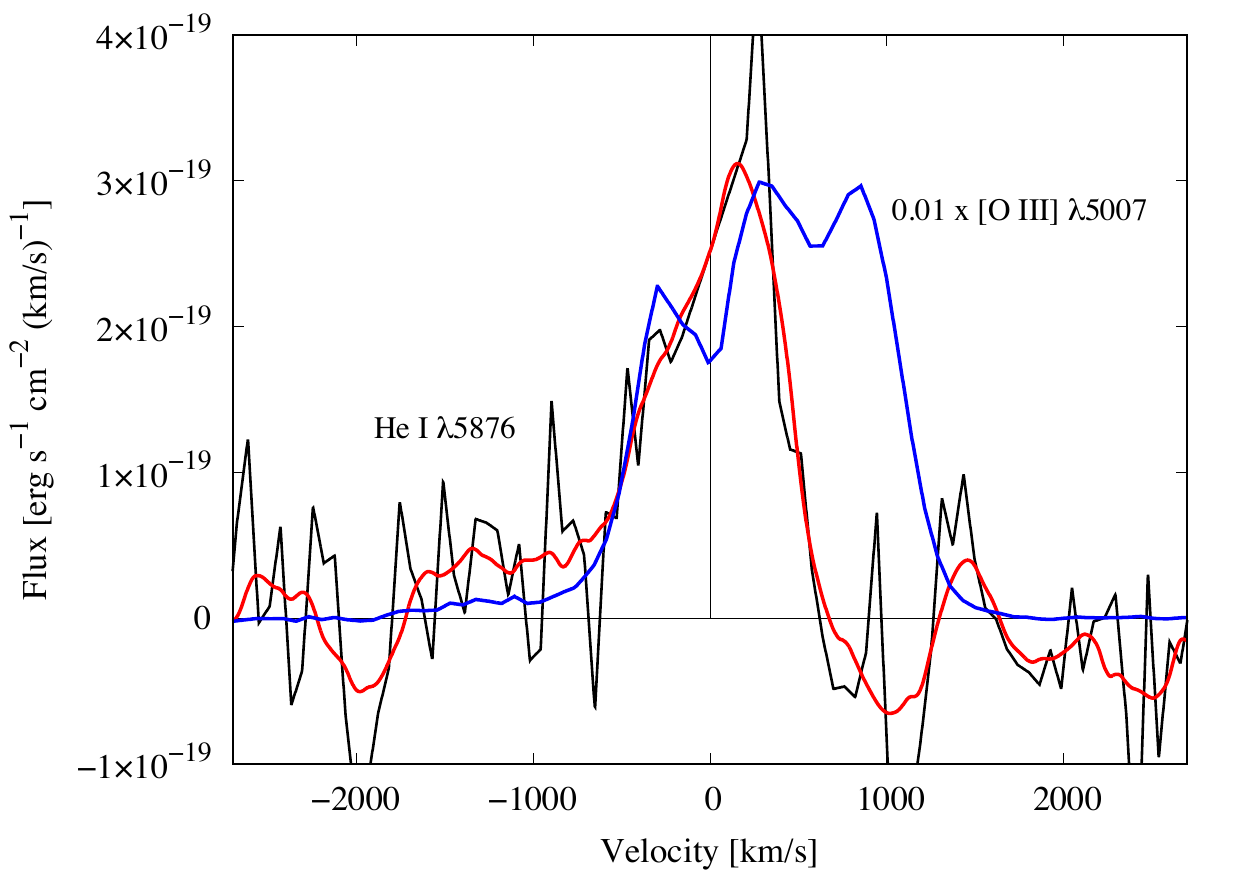}
\end{center}
\caption{Same in Fig.~\ref{f:O_Ne}, but for \ion{He}{i}~$\lambda$5876 (black, and smoothed in red) instead of 
[\ion{Ne}{iii}]~$\lambda$3869. 
The [\ion{O}{iIii}] profile was multiplied by a factor of 0.01 to match the \ion{He}{i} line. LMC rest frame velocitiy is marked by a 
vertical thin line. The red side of the \ion{He}{i} line appears depreciated compared to [\ion{O}{iIii}], similar to 
 H$\beta$ in Fig.~\ref{f:velo_temp}.
} \label{f:He_5876}
\end{figure}
%------------------------------------------------------------

Unfortunately, for the SNRC with its broade lines, [\ion{O}{iii}]$\wl$4363 is blended with H$\gamma$.
This may lead to over-estimating the [\ion{O}{iii}] temperature.
To get the real [\ion{O}{iii}] temperature we subtracted H$\gamma$ from the line profile 
of the observed [\ion{O}{iii}]$\wl$4363$+$H$\gamma$ blend. As a template 
profile for all \ion{H}{i} lines we used H$\beta$. As shown in Fig.~\ref{f:velo_temp}, 
the H$\beta$ template is affected by [\ion{Fe}{iii}]$\wl$4481, but this has very small 
impact on the process of removing the bluer Balmer lines.
A further complication is that all hydrogen lines suffer from background contamination
of emission around LMC rest velocity. We approximated this for all \ion{H}{i} lines by
a linear approximation of the flux across the corrupted frequency range.

The expected flux of H$\gamma$ relative to H$\beta$ was estimated from 
Case B theory \citep{Brock71}. The bottom panel of Fig.~\ref{f:velo_temp} shows how we 
were able to remove H$\gamma$ from the [\ion{O}{iii}]$\wl$4363$+$H$\gamma$ blend.
As a spin-off, the line subtraction confirms the values we used for the dereddening, 
i.e., $E(B-V)=0.19$ and $R=3.1$; when the Balmer line was removed, continuum level
was attained. 

After the H$\gamma$ subtraction, the flux of [\ion{O}{iii}]$\wl$4363 relative 
to [\ion{O}{iii}]$\wl$5007 (set to 100) is $\sim 4.3$ (instead of 5.7, if H$\gamma$ is unaccounted
for). This gives $R_{\rm [\ion{O}{iii}]} \sim 31$. We can, however, refine this analysis
by comparing the deblended profile of [\ion{O}{iii}]$\wl$4363 to that of a
template profile for [\ion{O}{iii}]$\wll$4959,5007. The template for the latter
was made from a similar procedure to what was done in \citet{Sand13}, i.e., the blue part
of [\ion{O}{iii}]$\wl$4959 and the red part of [\ion{O}{iii}]$\wl$5007 were used
as initial guesses for the template. With the known intensity ratio of 3 for 
$I(\lambda5007)/I(\lambda4959)$, we constructed the combined
[\ion{O}{iii}]$\wll$4959,5007 template shown in red in Fig.~\ref{f:velo_O3_prof}.
The template is for the sum of the two line components.
In the same figure we have also included the ``clean" [\ion{O}{iii}]$\wl$4363
profile from Fig.~\ref{f:velo_temp}. Once multiplied by a factor of 28, and once
by 34. In general, the [\ion{O}{iii}]$\wl$4363 profile is similar to that of the 
[\ion{O}{iii}]$\wll$4959,5007 template, although noise limits the usefulness
of the [\ion{O}{iii}]$\wl$4363 profile in its wings.

We thus find $R_{\rm [\ion{O}{iii}]} = 31\pm3$, with a tendency for a lower ratio in the blue 
part of the line profile compared to the red. The corresponding [\ion{O}{iii}]
temperature range for the SNRC is $23\,500 \pm 1\,800$~K (cf. Table~\ref{t:Temper4}).
This is less than $\sim 34\,000$~K, but consistent with $\approx 25\,500$~K estimated by 
\citet{Kirshner89} and \citet{Morse06}, resepctively. In neither of those two studies
H$\gamma$ was accounted for. The slightly larger ratio we find for $R_{\rm [\ion{O}{iii}]}$
for the receding side of the remnant (compared to LMC rest velocity) could
be due to a lower temperature in this part of the SNRC, or intrinsic dust reddening. 
The latter would not be surprising considering  that dust in the Crab nebula is concentrated
to [\ion{O}{iii}]-emitting filaments, and the pathlength through dust to the receding
side of the SNRC is likely to be longer than to ejecta moving toward us. To
compensate for the factor $\sim 1.2$ in larger  $R_{\rm [\ion{O}{iii}]}$-value
for the red side, $E(B-V)$ would have to be $\sim 0.5$ instead of 0.19. This
would depress the red side of the line profiles severely, which is not what we
see. Dust is therefore not a likely reason for the large $R_{\rm [\ion{O}{iii}]}$-value 
on the red side of the [\ion{O}{iii}] lines.

Fig.~\ref{f:velo_O3_prof} shows that the blue sides of [\ion{O}{iii}]$\wl$4959
and [\ion{O}{iii}]$\wl$5007 nearly reach $-1\,900 \kms$, which was difficult
to disentangle from Fig.~\ref{f:OIIIim_slit1} due to blending. For the red side, the 
maximum velocity is $+1\,700 \kms$, which is consistent with the space-
velocity results. Fig.~\ref{f:Fe_find} shows that the construction of the
[\ion{O}{iii}]$\wll$4959,5007 template reveals a residual that we 
attribute to [\ion{Fe}{iii}]$\wl$4986, a line that is also seen in the spectra
of the filaments and other SNRs \citep{FesHur96}.

The [\ion{Ne}{iii}]$\wll$3869,3967 doublet is also
affected by blending with Balmer lines. The same subtraction routine was 
applied for both components of [\ion{Ne}{iii}], and in Fig.~\ref{f:velo_temp} 
we show how the Balmer lines H$\zeta$ and H$\epsilon$ were removed
from their blends with [\ion{Ne}{iii}]$\wl$3869 and [\ion{Ne}{iii}]$\wl$3968,
respectively. As seen from the figure, [\ion{Ne}{iii}]$\wl$3869 is insignificantly 
contaminated by Balmer line emission, whereas [\ion{Ne}{iii}]$\wl$3967
is more affected. 

%-----------------------figure23------------------------------
\unitlength=1mm
\begin{figure*}[htb]
\begin{center}
\includegraphics[width=163mm, angle=0, clip]{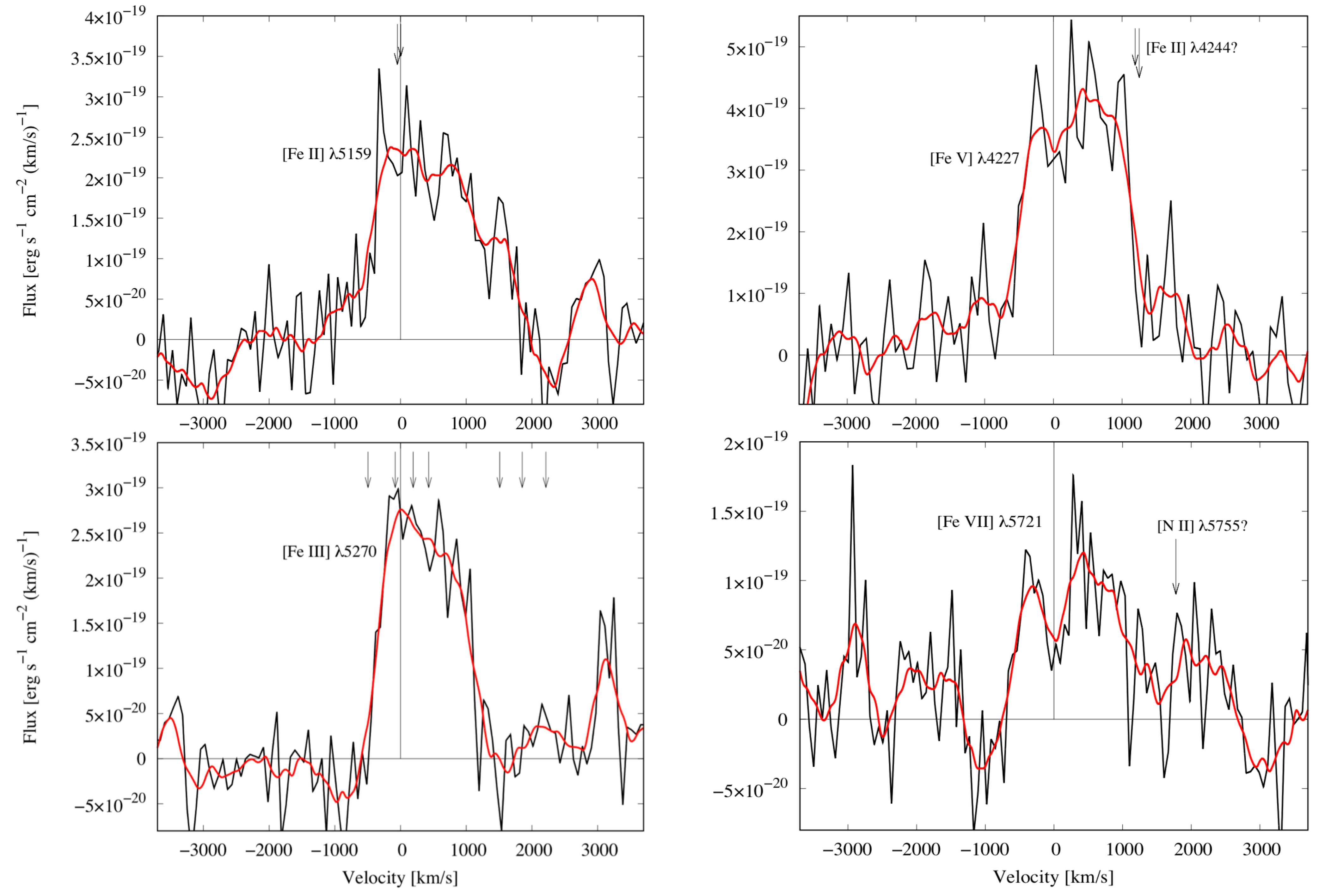}
\end{center}
\caption{Same in Fig.~\ref{f:He_5876}, but for several forbidden iron lines (black, and smoothed in red) instead of 
\ion{He}{i}~$\lambda$5876. The velocities are centered on LMC rest frame velocities of [\ion{Fe}{ii}]~$\lambda$5159 (upper left panel),  
[\ion{Fe}{iii}]~$\lambda$5270 (lower left), [\ion{Fe}{v}]~$\lambda$4227 (upper right) and [\ion{Fe}{vii}]~$\lambda$721 (lower right). 
In each panel grey arrows mark velocities for other nebular lines seen in SNRs \citep{FesHur96}. In particular, 
[\ion{Fe}{ii}]~$\lambda$5159 could be blended with (and even dominated by) [\ion{Fe}{vii}]~$\lambda$5159. Note the broad wings of
[\ion{Fe}{v}]~$\lambda$4227, possibly reaching $\pm 2\,000 \kms$.
} \label{f:Fe_lines}
\end{figure*}
%------------------------------------------------------------

A direct comparison between [\ion{Ne}{iii}]$\wl$3869 and [\ion{O}{iii}]$\wl$5007
is shown in Fig.~\ref{f:O_Ne}. The profile of the neon line is fairly noisy, but it is
consistent with the general features and maximum velocities of [\ion{O}{iii}]$\wl$5007.
The ratio of the two lines is $I_{[\ion{O}{iii}]\wl5007} / I_{[\ion{Ne}{iii}]\wl3869} \sim 30$.
With a temperature in the  $23\,500 \pm 1\,800$~K, and assuming similar ionization
structures of O$^{2+}$ and Ne$^{2+}$, our multilevel atom models \citep[e.g.,][]{Mar00,Matt10},
updated with the recent atomic data in the CHIANTI database \citep{Del21},
result in a mass ratio $M_{\rm Ne}/M_{\rm O} \sim 0.07$, which should probably be 
considered a lower limit since the ionization zone of O$^{2+}$ is likely to be wider than that
of Ne$^{2+}$. In this context, we highlight the observations of \citet{Morse06}. They detected 
[\ion{Ne}{iii}]$\wl$3869, but neither got a $3\sigma$ detection of [\ion{Ne}{iii}]$\wl$3967, nor 
corrected for Balmer line contamination. Their flux values translate into 
$M_{\rm Ne}/M_{\rm O} \sim 0.13$ for the temperature we find from [\ion{O}{iii}].  

\citet{Williams08} modeled the emission from 0540, and in those models
they assumed the abundance ratio Ne/O~$=0.2$ (i.e., a mass 
ratio of 0.25), and obtained $I_{[\ion{Ne}{iii}]\wl3869} / I_{[\ion{O}{iii}]\wl5007} = 0.093$.
This is almost three times the value we obtain from our observations; if
the $M_{\rm Ne}/M_{\rm O}$ ratio in their model is lowered to 0.085 (i.e., close to our value of 0.07), 
their predicted line ratio should agree with our observations. The question is how likely
a value of $M_{\rm Ne}/M_{\rm O} \sim 0.1$ could be. It is, for example, about half the mass 
ratio for the inner ejecta found in a recent model for the nucleosynthesis yields for the central part of 
an exploding $11.8 \Msun$ star \citep{Siever20}, but the models of \citet{CL13}, which take rotation of the
progenitor into account, yield $M_{\rm Ne}/M_{\rm O} = 0.063~(0.053)~(0.21) \Msun$ for 
progenitor masses of $13~(15)~(20) \Msun$. For non-rotating progenitors \citet{CL13} find 
$M_{\rm Ne}/M_{\rm O} = 0.22~(0.46)~(0.32) \Msun$. Judging from this, rotation could
be important as $M_{\rm Ne}/M_{\rm O}$ in those models agree with the values derived
from observations.

In the models of \citet{Williams08}, infrared lines observed with {\sl Spitzer Space Telescope} were included.
To compare the IR lines to optical lines, these authors multiplied the fluxes in \citet{Morse06}
by a factor of 4 to account for the 2\arcsec\ slit in the optical as the infrared observations
observed the full region. We use the same procedure, but to put the slit compensation on firmer footing,
we make use of the VLT/VIMOS observations of \citet{Sand13}. In Fig.~\ref{f:Slit_vs_Full} we show
the line profile of [\ion{O}{iii}]$\wl$5007 for the full $13\arcsec \times 13\arcsec$ field observed with VIMOS,
and compare this to $7.2\times$ the template [\ion{O}{iii}]$\wl$5007 profile from our  slit 1 observations. With the
7.2 factor, the integrated emission of the two profiles is the same for velocities $\gsim -750 \kms$. About 10\% of the 
VIMOS [\ion{O}{iii}]$\wl$5007 flux falls at velocities below $-750 \kms$, and is mainly attributed to [\ion{O}{iii}] glow not 
captured efficiently by the slits used by  \citet{Morse06} and us. We therefore use the factor of 7.2 for most lines  For
lines with suspected glow we use 8.0 since the {\sl Spitzer} field-of-view is likely to capture much of the glow. (The corresponding 
numbers for our NTT/EMMI slit, i.e., slit 2, is 6.3 and 6.6, respectively.)  As \citet{Morse06} obtain $\approx 1.6$ times larger 
[\ion{O}{iii}]$\wl$5007 flux through their $2\arcsec \times 11\arcsec$ slit than we do through our $1\arcsec \times 10\arcsec$ slit, 
the conversion factor \citet{Williams08} should have used  for most lines is 4.5, instead of their 4. For [\ion{O}{iii}]$\wl$5007,
and other possible lines with glow, a conversion factor of 5 would have been more appropriate.

\citet{Williams08} measure $(7.29\pm0.56)\EE{-14}$~erg~cm$^{-2}$s$^{-1}$ for [\ion{Ne}{iii}]~15.6$\mu$. 
If we use our VLT observations, multiplied by 8, the [\ion{Ne}{iii}]$\wl$3869 flux is 
$1.1\EE{-14}$~erg~cm$^{-2}$s$^{-1}$, i.e., $I_{[\ion{Ne}{iii}]~15.6\mu} / I_{[\ion{Ne}{iii}]\wl3869} \sim 6.7$,
which is twice as high as in the model of \citet{Williams08}. From our multilevel modeling it
is obvious that the [\ion{Ne}{iii}]~15.6$\mu$ flux cannot come from the same hot gas as the 
 [\ion{O}{iii}] lines we observe, since the ratio for this at 23\,000~K is 
 $I_{[\ion{Ne}{iii}]~15.6\mu}/I_{[\ion{Ne}{iii}]\wl3869} \sim 0.14$. An intensity ratio of 6.7 is only obtained for
 temperatures $\lsim 7\,000$~K. Although there are several uncertainties, the main explanation must 
 be that [\ion{Ne}{iii}]~15.6$\mu$ predominantly comes from regions with much lower temperatures than 
 that emitting the optical [\ion{O}{iii}] and [\ion{Ne}{iii}] lines. The most likely region is gas heated by photionization
 rather than by shocks, as shock-heated material would be hotter. Extinction due to internal dust may be a 
 complementary explanation as this would not affect the IR line.
 
Fig.~\ref{f:Slit_vs_Full} emphasizes the redshift of all [\ion{O}{iii}] emission from the full SNRC, and shows that the line profile
through slit 1 surprisingly well represents the emission at most velocities. The most obvious difference
is the strong emission at velocities $\lsim -750 \kms$ (which is also evident from Fig.~\ref{f:O3_new}, and which is mainly due to the
approaching ``wall" described in Sect. 3.2) as well as slightly more emission 
at $\gsim +750 \kms$. The peak around LMC rest velocity is at least to some extent artificial due to difficulty with
removing LMC background emission due to the small field-of-view of VIMOS. The VIMOS data are not sensitive enough
to trace velocities $\gsim \mid{\pm 1\,500}\mid \kms$.

Turning to ions of lower degree of ionization observed through slit 1, we show [\ion{O}{ii}]$\wll$3726,3729 in Fig.~\ref{f:O2_O3},
[\ion{S}{ii}]$\wll$4069,4076 in Fig.~\ref{f:O_S}, and \ion{He}{i}$\wl$5876 in Fig.~\ref{f:He_5876}. For [\ion{O}{ii}]
and [\ion{S}{ii}] we have centered the velocity on 3727.5 \AA\ and the strongest component of [\ion{S}{ii}] (i.e., $\wl$4069),
respectively. As already discussed in Sect. 3.2, [\ion{O}{ii}] has a profile similar to that of H$\beta$, and [\ion{S}{ii}]$\wll$4069,4076 
is similarly narrow, i.e., primarily extends from $\sim -700 \kms$ to $\sim +1\,000 \kms$, although there is a blue wing
of [\ion{O}{ii}] that reaches $\sim -1\,400 \kms$. This is not seen for [\ion{S}{ii}]$\wll$4069,4076 in Fig.~\ref{f:O_S}, but
the low signal-to-noise does not allow us to draw firm conclusions on that part of the line profile. Interestingly enough, 
the full view SNRC observations of [\ion{S}{ii}]$\wll$6716,6731 by \citet[][their Figure 2]{Sand13} does show a blue wing between 
$-1\,200 \kms$ and $-700 \kms$. Our choice to center the [\ion{O}{ii}] doublet on 3727.5 \AA\ is because this does not introduce
any bias for either of the two doublet components. What is intrinsically assumed by centering on 3727.5 \AA\ is that we assume
that $I_{[\ion{O}{ii}]\wl3729} / I_{[\ion{O}{ii}]\wl3726} \approx 1.11$, which means $n_{\rm e} \sim 350~(400) \cm3$ for 
$T \sim 20\,000~(30\,000)$~K. According to our multilevel model, the intensity ratio decreases from $\approx 1.35$ to $\approx 0.38$ 
as electron density increases from $10^2 \cm3$ to $10^4 \cm3$. 

Our observed $I_{[\ion{S}{ii}]\wll4069,4076} / I_{[\ion{O}{ii}]\wll3726,3729}$
ratio is $\sim 0.06$. For the temperature interval $10\,000 - 15\,000$~K and density interval $500 - 1\,000 \cm3$,  our multilevel 
model atoms translate the observed ratio into an abundance ratio of S/O$=0.05\pm0.01$, and thus a mass ratio
$M_{\rm S} / M_{\rm O} = 0.10\pm0.02$. The models of \citet{CL13} predict a ratio of $0.05-0.06$ for rotating progenitors
for progenitor masses between $13-20 \Msun$, and between $0.05-0.13$ for non-rotating (with 0.05 for $20 \Msun$). The more
recent model of \citet{Siever20} predicts $M_{\rm S} / M_{\rm O} = 0.085$ from their $11.8 \Msun$ model for the inner ejecta and
$0.12 \Msun$ for all ejecta. \citet{Williams08} assumed $M_{\rm S} / M_{\rm O} = 0.05$ in their models, and underproduced 
[\ion{S}{iii}]~18$\mu$ and [\ion{S}{iv}]~10$\mu$ by factors of $\sim 1.4$ and $\sim 2.8$, respectively, when we adjust their
results with more accurate slit compensations. It therefore appears as if their model would fit the {\sl Spitzer} data
better with $M_{\rm S} / M_{\rm O} \sim 0.1$, which is consistent with both \citet{CL13} and \citet{Siever20}, albeit
perhaps not for non-rotating low-mass progenitors in the case of \citet{CL13}. However, the model of \citet{Williams08}
gives an $I_{[\ion{S}{ii}]\wll4069,4076} / I_{[\ion{O}{ii}]\wll3726,3729}$ ratio which is $\sim 0.09$, and that is on the high side 
compared to our observations already with $M_{\rm S} / M_{\rm O} = 0.05$. 

$M_{\rm S} / M_{\rm O}$ can also be estimated from our slit 2 data for which we have
$I_{[\ion{S}{ii}]\wll6716,6731} / I_{[\ion{O}{ii}]\wll7319-7331} \sim 9.0$. This is similar to $\approx 8.4$ by \citet{Kirshner89}, and
10 in the model of \citet{Williams08} assuming S/O~$=0.1$). Our multilevel atom modeling shows that a temperature 
$\gsim 20\,000$~K is needed to excite O$^{+}$ to emit [\ion{O}{ii}]$\wll$7319-7331 at the observed relative intensity. In particular, 
for the [\ion{O}{iii}] temperature $T \sim 23\,000$~K, $I_{[\ion{S}{ii}]\wll6716,6731} / I_{[\ion{O}{ii}]\wll7319-7331} \sim 7$ for 
S/O~$=0.1$ and electron densities estimated by \citet{Sand13}. The ratio rises steeply below 20\,000~K. 

Another way to estimate a temperature in the [\ion{O}{ii}]-emitting regions is from the ratio 
$I_{[\ion{O}{ii}]\wll3726,3729} / I_{[\ion{O}{ii}]\wll7319-7331}$, but for this we have to combine slits 1 and 2. Using the slit correction
factors 7.2 and 6.3, respectively, we obtain $\sim 6.2$, which is very close to $\sim 6.5$ by \citet{Kirshner89}, but very different
from the modeled flux by \citet{Williams08}, which is $\approx 14.7$. Our multilevel model suggests very high temperatures (well
above $30\,000$~K) to get a ratio below 8.0, which appears unlikely. Since our flux agrees with that of  \citet{Kirshner89}, a
possible explanation could be that [\ion{O}{ii}]$\wll$7319-7331 is contaminated by [\ion{Ca}{ii}] in both our slit 2 and the
data in \citet{Williams08}. Such contamination (at 20\% level) was already discussed by \citet{Kirshner89}.

%-----------------------figure24------------------------------
\unitlength=1mm
\begin{figure*}[htb]
\begin{center}
\includegraphics[width=187mm, angle=0, clip]{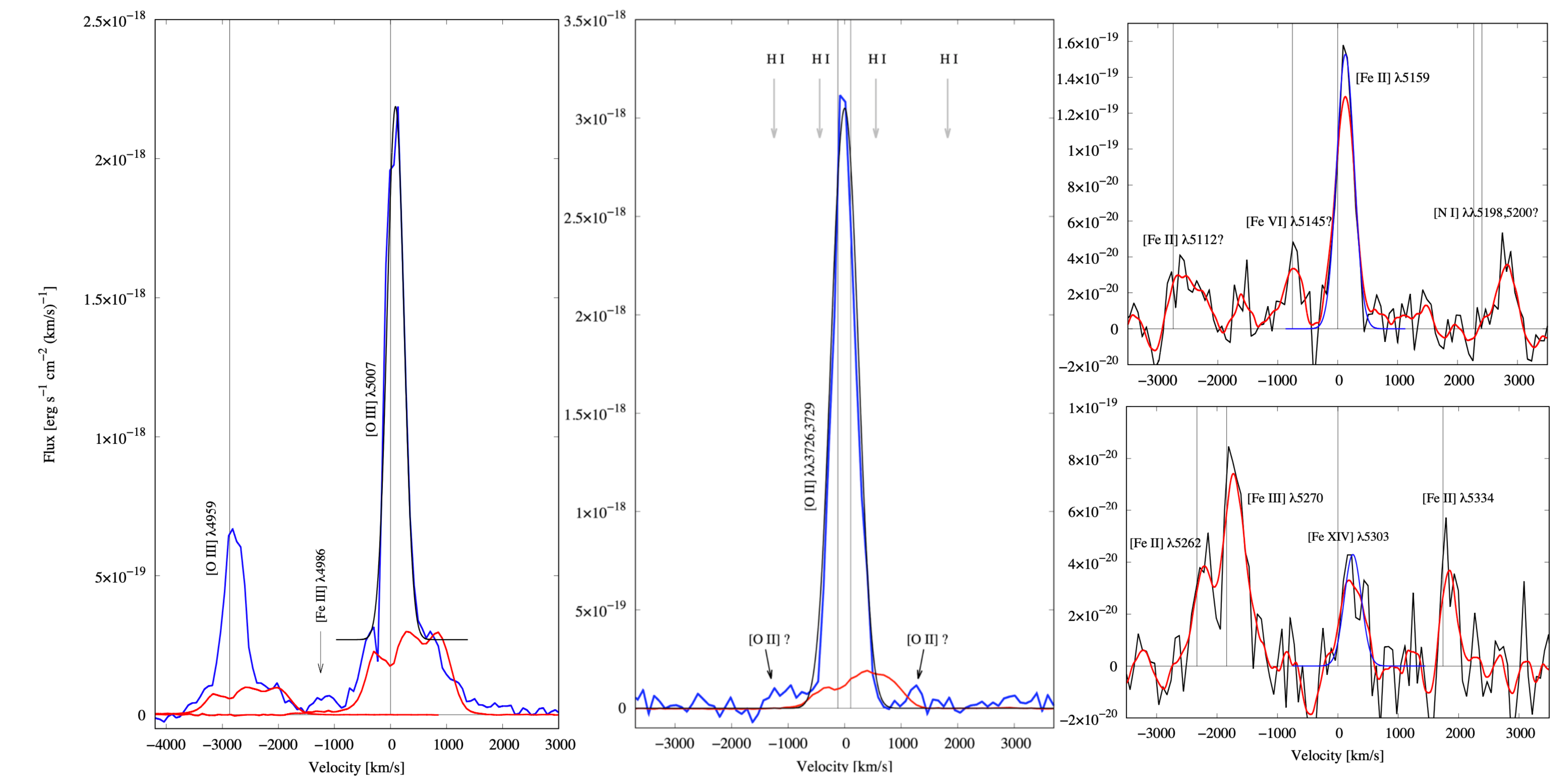}
\end{center}
\caption{Spectra of forbidden lines from F1 (cf. Figs.~\ref{f:O3_diff} and \ref{f:HIIreg_spa}). All spectra are in the LMC rest 
frame and the velocities are for [\ion{O}{iii}]~$\lambda$5007 (left), 3727.5 \AA\ (middle), [\ion{Fe}{ii}]~$\lambda$5159 
(upper right) and [\ion{Fe}{xiv}]~$\lambda$5303 (lower right). For the oxygen lines, the line profiles of the emission from the SNSC in
Figs.~\ref{f:velo_O3_prof} and \ref{f:O2_O3}, multiplied by a factor of 0.01, have been included for comparison (in red).
The narrow spectral lines are unresolved, as highlighted for [\ion{O}{iii}]~$\lambda$5007 and the [\ion{O}{ii}] doublet,
where the solid black lines show Gaussians with the expected FWHM of the instrument at  5007 \AA\ and 3727.5  \AA, 
i.e., $\sim 355 \kms$ and $\sim 476 \kms$, respectively. Note the broad bases seen for [\ion{O}{iii}] and possible broad base wings
for [\ion{O}{ii}]~$\lambda\lambda$3726,3729. Lines of the \ion{H}{i} Balmer series are marked in the middle panel. For the iron lines,
Gaussians have also been drawn (in blue) according to the expected resolution. Several other lines have also been marked/indicated. 
Red marks Savitzky-Golay fitting to the spectrum.
} \label{f:F1_specs}
\end{figure*}
%------------------------------------------------------------
 
 A combination of slits 1 and 2 can also be used to estimate an [\ion{S}{ii}]- temperature  from 
 $I_{[\ion{S}{ii}]\wll6716,6731} / I_{[\ion{S}{ii}]\wll4069,4076}$, which with our slit compensations is $\sim 30$. This is 
 much higher than $\sim 17$ by \citet{Kirshner89}, and inconsistent with $\sim 9.4$ by \citet{Morse06} as well as the
 model by \citet{Williams08} which predicts $\sim 7.4$. Our multilevel models show some density dependence, but 
 intensity ratios such as those by \citet{Morse06} and \citet{Williams08} require temperatures at least as high as our
 derived [\ion{O}{iii}] temperature from slit 1, whereas our ratio and that of \citet{Kirshner89} indicate temperatures of
 well below $10\,000$~K and around $15\,000$~K, respectively. Comparing observations through different slits, and at
 different epochs, admittedly increases the risk of systematic errors, so to make a consistency check we have used 
 the full SNRC [\ion{S}{ii}] spectra of \citet{Sand13} to make artificial observations of [\ion{S}{ii}] through slits 1 and 2. From
 this exercise we find that the slit compensation for [\ion{S}{ii}]$\wll$6716,6731 is 6.5 for slit 1 and 7.7 for slit 2 (instead of the 
 values 7.2 and 6.3, as indicated by [\ion{O}{iii}]). This is not large enough to change our 
 $I_{[\ion{S}{ii}]\wll6716,6731} / I_{[\ion{S}{ii}]\wll4069,4076}$ ratio by much (and actually makes it even larger). 
 
 The source of inconsistency for $I_{[\ion{S}{ii}]\wll6716,6731} / I_{[\ion{S}{ii}]\wll4069,4076}$ could either lie in the large relative 
 [\ion{S}{ii}]$\wll$6716,6731 flux we get for slit 2 compared  to others (cf. Table~\ref{t:FluxVel}), and/or the lower flux we measure for 
 [\ion{S}{ii}]$\wll$4069,4076. The latter is not unreasonable since [\ion{S}{ii}]$\wll$4069,4076 is affected by H$\delta$. 
 From \citet{Brock71} we estimate that H$\delta$ would amount to $\sim 25\%$ of H$\beta$, which would decrease the 
 $I_{[\ion{S}{ii}]\wll6716,6731} / I_{[\ion{S}{ii}]\wll4069,4076}$ ratio accordingly, if not corrected for. This effect, of course, does not
 affect the modeled results of \citet{Williams08}. Regarding [\ion{S}{ii}]$\wll$6716,6731 we can do a direct test against the VIMOS 
 observations of \citet{Sand13} by placing artificial slits on the VIMOS images to simulate slits 1 and 2. From figure 3 
 of \citet{Sand13} it can be seen that $I_{[\ion{S}{ii}]\wll6716,6731} / I_{[\ion{O}{iii}]\wl5007}$ is never higher than $\sim 0.8$, and 
 for slits 1 and 2 we get $\sim 0.47$ and $\sim 0.35$, respectively. Even if we allow for slight displacement of slit 2 due to 
 atmospheric dispersion (which was not accounted for in the NTT/EMMI observations), the ratio will not increase by more than 
 $\sim 20\%$. Careful 
 check of possible erroneous background subtraction of [\ion{S}{ii}]$\wll$6716,6731, together with the fact that the flux of other 
 lines through slit 2 agrees with the results of  \citet{Kirshner89} and \citet{Morse06}, leaves us no choice other than to suggest that 
 the [\ion{S}{ii}$]\wll$6716,6731 flux could vary with time. If we use a ratio of 0.35 from a simulated slit 2, we obtain 
 $I_{[\ion{S}{ii}]\wll6716,6731} / I_{[\ion{S}{ii}]\wll4069,4076} \sim 12$, which is more aligned with what other groups get, and
would suggest an [\ion{S}{ii}] temperature of $\sim 20\,000$~K. The question remains regarding our high [\ion{S}{ii}$]\wll$6716,6731 
flux. We note that there are NTTT/EMMI spectra covering [\ion{S}{ii}$]\wll$6716,6731 from 1995 January, i.e., only one year before 
our NTT spectrum, which were shown in \citet{Car98}. Unfortunately, these authors did not provide any fluxes. As there is activity going 
on in the south-western part of the PWN, at least between 1999-2007, the NTT spectra of  \citet{Car98} could prove decisive 
regarding temporal variations in [\ion{S}{ii}$]\wll$6716,6731. As discussed in Sect. 3.4.2, variations of elements other than 
oxygen on relative short time scale are conceivable.  

Figure~\ref {f:He_5876} shows that \ion{He}{i}~$\lambda$5876 is very concentrated to within $\pm 500 \kms$, in a fairly similar
fashion to H$\beta$ and [\ion{S}{ii}]. This means that hydrogen was deeply mixed into the core region 
during the explosion. Such mixing is expected from 3D explosion models \citep[e.g.,][]{Hammer2010}, and the amount of hydrogen 
mixed into the SNRC is sensitive to hydrodynamic instabilities at the composition interfaces \citep[e.g.,][]{Wong15}.
A crude He/H ratio for the SNRC can be estimated from recombination theory, assuming similar ionization zones for H and He.
The observed flux ratio of \ion{He}{i}~$\lambda$5876 to H$\beta$ for our slit 1 is $\sim 0.24$, with at least 40\% uncertainty.
For a temperature in the range $10\,000-20\,000$~K this translates into a mass ratio of He$^+$/H$^+$ which is $\sim 0.8$. As He$^0$ 
roughly has the same ionization potential as S$^+$, and [\ion{S}{ii}] emission dominates over [\ion{S}{iii}] in 0540 \citep{Kirshner89},
a larger fraction of helium is likely to be neutral than that of hydrogen, and $M_{\rm He} / M_{\rm H}$ is likely to be larger than 0.8 
for the SNRC of 0540. The fraction of fully ionized helium could in principle be tested with \ion{He}{ii}~$\lambda$4686, but it sits in a spectral 
range with strong [\ion{Fe}{iii}] lines, and is not detected. SN~1987A may serve as interesting comparison for $M_{\rm He} / M_{\rm H}$,
and we note that \citet{Jerk11} estimate that both H and He have $\sim 1.2 \Msun$ each in the central $2000 \kms$ region of SN~1987A, 
i.e., the mass ratio is $M_{\rm He} / M_{\rm H} \approx 1$, which is consistent with what we estimate for the SNRC of 0540.

Through slit 2 we also detect [\ion{Ar}{iii}]~$\lambda$7136 and [\ion{Ni}{ii}]~$\lambda$7378 with intensities relative to 
[\ion{O}{iii}]$\wl$5007 similar to what was reported by \citet{Kirshner89} (cf. Table~\ref{t:FluxVel}). On the other hand, the
modeled [\ion{Ar}{iii}]~$\lambda$7136 flux by \citet{Williams08} was more than twice as high as the observed relative
fluxes of the line. This could suggest that Ar/O is less than the value of 0.1 assumed by \citet{Williams08}. Tuning the model 
results of \citet{Williams08} with the observed  [\ion{Ar}{iii}]~$\lambda$7136 fluxes then suggest $M_{\rm Ar} / M_{\rm O} \sim 0.02$.
The nucleosynthesis yields in the model of \citet{Siever20} give the same value, i.e., $M_{\rm Ar} / M_{\rm O} \sim 0.02$ for the 
total yield of Ar and O. The ratios in \citet{CL13} are $\sim 0.01$ for $13-20 \Msun$ rotating, and $20 \Msun$ for non-rotating 
progenitors, whereas the ratio is higher, $\sim 0.02$, for $13 \Msun$ non-rotating progenitors. Our multilevel model atoms suggest 
$M_{\rm Ar} / M_{\rm O} = 0.015 (0.030)$ for equal ionization zones of Ar$^{2+}$ and O$^{2+}$ in the SNRC and for temperatures of
20\,000 (30\,000)~K. The observed level of [\ion{Ar}{iii}]~$\lambda$7136 is therefore consistent with that expected from
elemental yields in progenitor models and with [\ion{O}{iii}] temperatures. 

As shown in Fig.~\ref{f:Fe_find} ([\ion{Fe}{iii}]$\wl$4986), Fig.~\ref{f:O2_O3} ([\ion{Fe}{v}]$\wl$3726) and 
Fig.~\ref{f:Fe_lines} ([\ion{Fe}{ii}]$\wl$5159, [\ion{Fe}{iii}]$\wl$5270, [\ion{Fe}{v}]$\wl$4227 and [\ion{Fe}{vii}]$\wl$5721) we detect 
several forbidden iron lines through slit 1, with decent signal-to-noise. 
We also detect [\ion{Fe}{ii}]~$\lambda\lambda$4287,4414,5044, [\ion{Fe}{iii}]~$\lambda$4658,4881,4986,
[\ion{Fe}{v}]~$\lambda\lambda$3783-3797 and [\ion{Fe}{vii}]~$\lambda$5721. The strongest line is 
[\ion{Fe}{iii}]$\wl$4658, but it blends with several other [\ion{Fe}{iii}] lines, as do also several [\ion{Fe}{ii}] lines around $\sim 4474~\AA$.
Through slit 2 we detect [\ion{Fe}{ii}]$\wl$5159 and [\ion{Fe}{iii}]$\wll$4658,5270, but with poor signal-to-noise. 

We have chosen to concentrate on presumably unblended lines observed through slit1.
%It is beyond the scope in this paper to make an abundance analysis of iron. 
In general, the line profiles are similar to those of the [\ion{O}{ii}] lines. In particular this is the case 
for [\ion{Fe}{iii}]~$\lambda$5270 and the core of [\ion{Fe}{v}]~$\lambda$4227, whereas [\ion{Fe}{vii}]~$\lambda$5721 is noisy, 
and perhaps also affected by [\ion{N}{ii}]~$\lambda$5755 in the far red wing. In Fig.~\ref{f:Fe_lines} we have marked the 
velocities corresponding to other lines detected in SNRs \citep{FesHur96} which could perhaps contribute to the lines
we have highlighted. For example, [\ion{Fe}{ii}]~$\lambda$4247 could contribute to the red wing of [\ion{Fe}{v}]~$\lambda$4227, and
our identification of [\ion{Fe}{ii}]~$\lambda$5159 could instead be dominated by [\ion{Fe}{vii}]~$\lambda$5159, in
particular since we also detect [\ion{Fe}{vii}]~$\lambda$5721. To test this we have used our nine-level model atom for Fe${^6+}$,
including data from \citet{Berr00} and \citet[][and references therein]{Del21}. For $n_{\rm e} = 10^3 \cm3$ and $T = 30\,000$~K,
 the expected strongest [\ion{Fe}{vii}] lines are (in order and normalized to [\ion{Fe}{vii}]~$\lambda$5721): 
 [\ion{Fe}{vii}]~$\lambda$6089, [\ion{Fe}{vii}]~$\lambda$3759, [\ion{Fe}{vii}]~$\lambda$3587, [\ion{Fe}{vii}]~$\lambda$4990 
 and [\ion{Fe}{vii}]~$\lambda$5159 with relative strengths 1.9, 1.4, 1.0, 0.7 and 0.5, respectively. Unfortunately, 
[\ion{Fe}{vii}]~$\lambda$3587 and [\ion{Fe}{vii}]~$\lambda$6089 are outside the spectral range of our slit 1 observations, 
although the latter line can, after smoothing, be hinted in the NTT/EMMI spectrum. [\ion{Fe}{vii}]~$\lambda$3759 is clearly detected 
through slit 1, and [\ion{Fe}{vii}]~$\lambda$4990 may add to [\ion{Fe}{iii}]$\wl$4986 in Fig.~\ref{f:Fe_find}. Judging from the weak 
measured flux of [\ion{Fe}{vii}]~$\lambda$5721, and that [\ion{Fe}{vii}]~$\lambda$5159 is not expected to be stronger, we conclude 
that the spectral line at $\sim 5171$~\AA\ (and $\sim 5160$~\AA\ through slit 2) is dominated by [\ion{Fe}{ii}]~$\lambda$5159. 
The line has a possible broad blue wing, which is also hinted for [\ion{Fe}{v}]~$\lambda$4227 (cf. Fig.~\ref{f:Fe_lines}). 
However, since no such wing is expected for [\ion{Fe}{ii}] lines, it is probably not real. Deeper observations are needed to check 
the wing of the [\ion{Fe}{v}] line.

\subsubsection{Masses and abundances - a summary}\label{mass_abund}
In the previous section we estimated the relative abundances in 0540. We now summarize this and estimate
the total ejecta mass of the elements we have discussed. The mass of O$^{2+}$ ions in the SNRC emitting 
[\ion{O}{iii}]~$\lambda$5007 is
\begin{equation}
\begin{split}
M({\rm O}^{2+}) \approx 
3.2\EE{-5}~
\left( 
\frac{f_{[\ion{O}{iii}]\lambda5007}}{10^{-13}~{\rm erg~cm}^{-2}~{\rm s}^{-1}} 
\right) 
\left(
\frac{D}{50~{\rm kpc}} 
\right)^{2} \\
\left( 
\frac{n_{\rm e}}{10^{3}~\cm3} 
\right)^{-1} 
\left( 
\frac{j_{[\ion{O}{iii}]\lambda5007}}{10^{-21}~{\rm erg~cm}^{3}~{\rm s}^{-1}~{\rm sr}^{-1}} 
\right)^{-1}
~\Msun.
\end{split}
\label{eq:O3flux}
\end{equation}
Here $f_{[\ion{O}{iii}]\lambda5007}$ is the dereddened flux for one of the slit positions with a slit correction included to account for
the full SNRC. For the SNRC we get 
$f_{[\ion{O}{iii}]\lambda5007} \sim 3.0\times10^{-13}~{\rm erg~cm}^{-2}~{\rm s}^{-1}$. $j_{[\ion{O}{iii}]\lambda5007}$ 
is the emissivity of the $\lambda$5007 transition, which for $T = 23\,500$~K is 
$\sim 1.1\EE{-21}~{\rm erg~cm}^{3}~{\rm s}^{-1}~{\rm sr}^{-1}$. For $n_{\rm e} = 10^{3}~\cm3$ one then gets 
$M({\rm O}^{2+}) \sim 9\EE{-5} \Msun$. 

A similar estimate for $M({\rm O}^{+})$ from [\ion{O}{ii}]~$\lambda\lambda$3726,3729 gives
$M({\rm O}^{+}) \sim 5\EE{-5} \Msun$ when we use $T = 20\,000$~K (and from this 
$j_{[\ion{O}{iii}]\lambda\lambda3726,3729} \sim 1.1\EE{-21}~{\rm erg~cm}^{3}~{\rm s}^{-1}~{\rm sr}^{-1}$).
A combined mass of $M({\rm O}^{+}) + M({\rm O}^{2+}) \sim 1.4\EE{-4} \Msun$ may seem like a 
surprisingly low mass considering that about two solar masses of oxygen should be present in 0540 (if similar to, e.g., SN~1987A).
However, this estimate is only for the amount of oxygen in the SNRC that has been recently 
shocked. Those shocks are thin due to intense radiative cooling. The cooling time, calculated from a simple six-level model atom of 
pure O$^{2+}$ at $T = 30\,000$~K with $n_{\rm e} = 10^{3}~\cm3$, is $\sim 0.02$~years, and at  $T = 50\,000$~K it is 
$\sim 0.024$~years. The similar numbers for pure O$^{+}$ at $T = 20\,000$~K and $T = 30\,000$~K  are $\sim 0.023$~years and 
$\sim 0.02$~years. The real cooling time is longer if the shocked gas starts out hotter than $T = 50\,000$~K, although
this is not so important since the cooling function for pure oxygen peaks around $\sim 2\EE5$~K \citep[cf.][]{BS90}, and falls
monotonically towards lower temperatures. More important is that the cooling is counteracted by photoionization of ambients 
photons from the PWN. With this in mind, a lower limit on the cooling time is $\sim 0.05$ years. Thus, $\sim 1.4\EE{-4} \Msun$ 
has to be replenished every $\gsim 0.05$ years, which means $\sim 0.3 \Msun$ of shocked oxygen in $\gsim 100$ years.

\citet{Williams08} did shock modeling for a similar situation where they assumed a low-velocity shock of $20 \kms$ aided 
by photoionization of PWN photons to account for the heavy-element ejecta lines. This is much lower than the shock velocity in 
\citet{BS90} for Cas~A, although that study only included preionization by photons produced by the shock. The oxygen
 column density of the  emitting region in \citet{Williams08} is $\sim 10^{14}$~cm$^{-2}$, and the mass flux of ejecta flowing into the 
low-velocity shocks is $\sim 0.01 \Msun$~yr$^{-1}$, which appears consistent with our estimate of 
$\lsim 0.003~X_{\rm other} \Msun$~yr$^{-1}$, 
where $X_{\rm other} > 1$ is the correction factor to include all elements, and not only oxygen.

The mass fractions we estimated in Sect. 3.4.1 for the SNRC of 0540, namely ${\rm O:Ne:S:Ar} = 1:0.07:0.10:0.02$ and 
${\rm H:He} = 1:\gsim0.8$, are thus  sampled over a very small fraction of the inner ejecta at any given time. How representative these numbers 
are for all inner ejecta is not very clear. However, one could imagine that there could be fluctuations on moderate time scales if the 
low-velocity shocks run into regions which are not fully microscopically mixed. This could be the reason for the apparent elevated sulphur 
abundance indicated by our NTT/EMMI observations. 

 %-----------------------figure25-----------------------------
\unitlength=1mm
\begin{figure}[htb]
\begin{center}
\includegraphics[width=100mm, angle=270, clip]{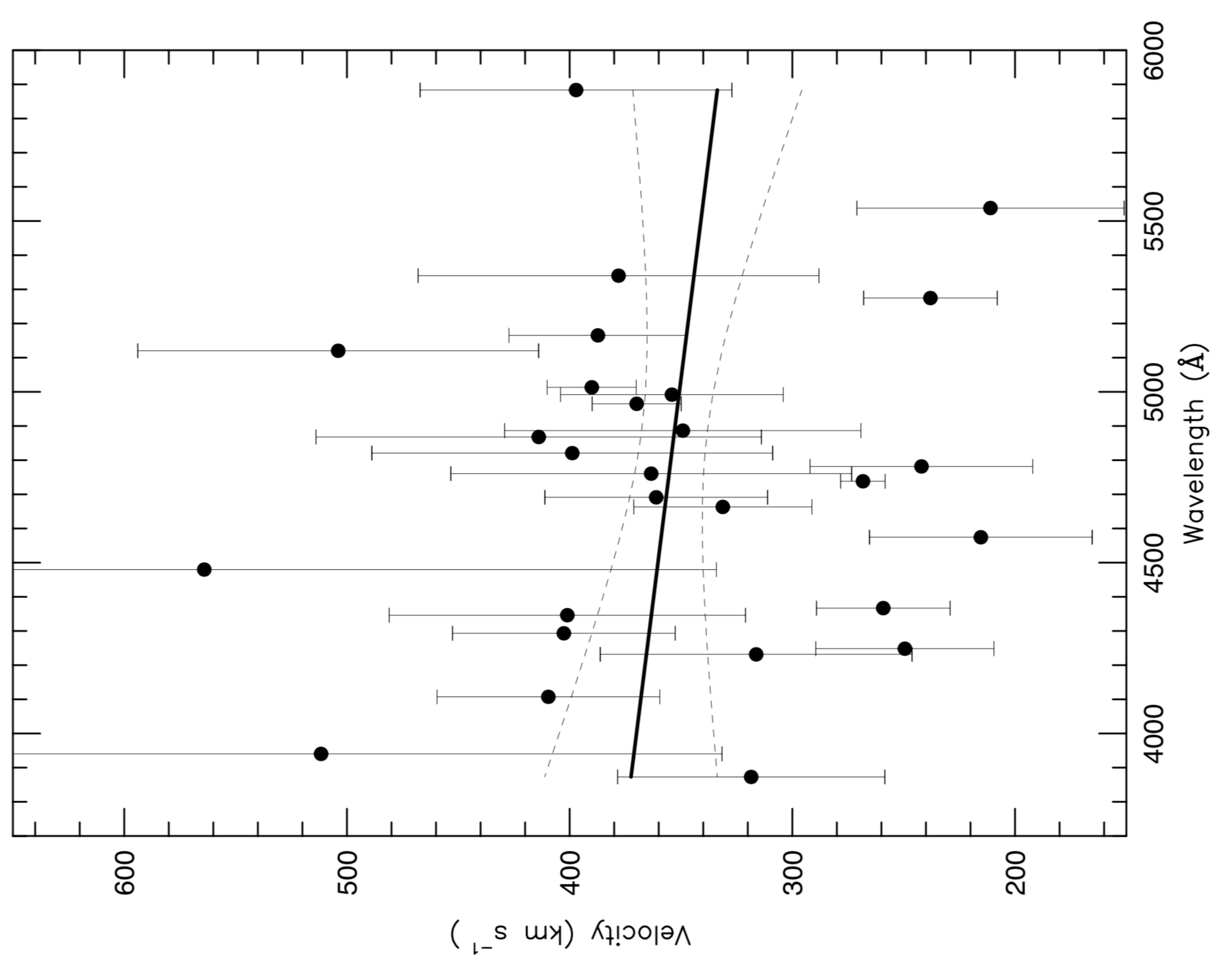}
\end{center}
\caption{Velocities estimated for emission lines from Filament F1. No doublets/multiplets are included except
 [\ion{O}{ii}]$\wll$3726,3729. As can be seen, the wavelength calibration is stable across the spectrum. Interestingly,
 there is a group of lines with LMC redshift, although the majority has a redshift close to $\sim +120 \kms$ compared to
 LMC redshift. See Table~\ref{t:FluxVel_2} and text for further details.
} \label{f:F1_velo}
\end{figure}
%------------------------------------------------------------

\subsubsection{The halo and beyond}\label{lin_mesh_outside}
Added to the mass of the SNRC is the mass of ejecta in the unshocked freely moving supernova ejecta. 
According to \citet{Williams08}, the inner part of this is photoionized to O$^{2+}$, so we use Equation~\ref{eq:O3flux} with
$n_{\rm e} = 1 \cm3$ and $f_{[\ion{O}{iii}]\lambda5007} \sim 1.5\times10^{-13}~{\rm erg~cm}^{-2}~{\rm s}^{-1}$,
which corresponds to the integrated broad base of [\ion{O}{iii}] emission in Fig.~\ref{f:Slit_vs_Full}, and assuming that the 
VIMOS/IFU field-of-view misses roughly one third of the halo emission. For the photoionized halo we further assume a
temperature of $\sim 12\,000$~K, so that  $j_{[\ion{O}{iii}]\lambda5007} \sim 4\EE{-22}~{\rm erg~cm}^{3}~{\rm s}^{-1}~{\rm sr}^{-1}$. 
With these assumptions, the mass of O$^{2+}$ in the halo, $M({\rm O}^{2+})_{\rm halo} \sim 0.12 \Msun$. The mass is most likely 
higher since $n_{\rm e} = 1 \cm3$ only stems from the 4 a.m.u. at the inner boundary of the unshocked SN ejecta estimated
by \citet{Williams08}. The average electron density could be $n_{\rm e} < 1 \cm3$ in the unshocked [\ion{O}{iii}]\ emitting halo.
According to \citet{Williams08}, enough ionizing photons are produced to ionize $\sim 1 \Msun$ halo ejecta.
In a mixed abundance model by \citet{BL00} for SN~1987A, which contains $\sim 2.0 \Msun$ of oxygen with a drop
in mass fraction just short of $2\,000 \kms$, the mass fraction of oxygen falls from $\sim 0.3$ from $\sim 1\,300 \kms$ to well
below 0.1 at $2\,000 \kms$. Choosing 0.1 for the 0540 halo makes our estimate for $M({\rm O}^{2+})_{\rm halo}$ consistent 
with the ionized mass estimated by \citet{Williams08}, and if the total oxygen mass is similar to that in SN~1987A, roughly
half of the oxygen is now confined to the filaments in the SNRC. 

In SN~1987A, the total mass inside $2\,000 \kms$ is dominated by He, O and H \citep[cf.][]{BL00}. We may assume the same
situation for 0540. For a density of 4 a.m.u. at the inner boundary of the halo, corresponding to $V_{\rm halo,inner} \sim 1\,200 \kms$, 
and an outer velocity for the ionized halo of $V_{\rm halo,outer} \gsim 1\,900 \kms$ (as is indicated by our results here), the total mass 
in this shell is $\sim 6.8 \Msun$, assuming a flat density profile of the unshocked ejecta \citep[cf.][]{Williams08}. This 
compares well with the mixed model for SN~1987A by \citet{BL00} in which the total mass inside the radius corresponding to this 
velocity is about $5-6 \Msun$. It therefore seems that 0540 and SN~1987A could be fairly similar, and perhaps stem from progenitors 
with roughly similar zero-age-main sequence masses. The main difference between the dynamics in these events is of course the 
influence of a pulsar, which, if existent in SN~1987A \citep[e.g.,][]{Cigan19,Greco21,Alp21}, plays a much smaller role than in 0540.
On top of that, one has the 3D effects stemming from the explosion hydrodynamics, and possibly pre-explosion rotation.
In our mass estimates for the halo we have compared with models assuming 1D, but from Sect. 3.1.1. it is evident that the
halo is far from spherically symmetric. This obviously introduces uncertainties in our mass estimates for the halo.

Moving further away from the SNRC, we introduced in Sect. 3.3.2 five interesting filaments which we have named F1-F5. 
For all filaments, the spectral lines are unresolved, except for F1, which displays
broad bases of the [\ion{O}{iii}]$\wll$4959,5007 line profiles, skewed to the red as 
compared to the positions of the narrow and strong (unresolved) peaks of these lines (cf. Fig.~\ref{f:F1_specs}). 
The minimum and maximum velocities of the base (relative to LMC) are at least $-900 \kms$ and $+1\,300 \kms$,
with a possible extension in the red to even larger velocities. No such features are seen in spectra 
of F2-F5. The presence of the broad base is consistent with the fact that F1 is 
projected onto the region showing the faint [\ion{O}{iii}] glow (cf. Fig.~\ref{f:OIIIim_slit1}, top 
panel). As seen in the lower panel of Fig.~\ref{f:OIIIim_slit1}, high redshifted velocities from
the glow (which is emitted by the halo discussed above) occur close to the position of F1. Despite its projected proximity
to the SNRC, the unresolved spectral components of filament F1 are therefore not part of the inner 
region of the SNR itself (see also below). 

In Sect. 3.2 we connected the maximum recession velocity of the glow at F1 to the pulsar age. For
a maximum velocity of $\gsim +1\,200 \kms$, the pulsar would be older than $1\,200$ years. Fig.~\ref{f:F1_specs} shows that
this could be the case. However, a maximum velocity of $+1\,300 \kms$ at the position of F1 would imply 
$v_{\rm true} \sim 2\,130 \kms$ for a pulsar age of $1\,200$ years, which in turn would correspond to a radius of $\sim 10\farcs8$ 
rather than $\sim 10\farcs0$ in Fig.~\ref{f:O3_diff} for the same expansion as $v_{\rm true}$ in the westward direction orthogonal to 
the line-of sight. 
Fig.~\ref{f:O3_diff} does not really support this, which could indicate aspherical symmetry of the fastest glow even 
on smaller scales. Unfortunately, the VIMOS/IFU field does not reach this far from the pulsar to test the degree of
non-sphericity over a larger scale, but Fig.~\ref{f:Slit_vs_Full} shows that even within the $13\arcsec \times 13\arcsec$ field
of view of VIMOS, there is large asymmetry of the fast ``glow". However, at the position of F1 specifically the broad base is fairly
symmetric and less redshifted than the emission from the SNRC (highlighted in red in Fig.~\ref{f:F1_specs}). Given the results
in Sect. 3.2 and what is shown in Fig.~\ref{f:F1_specs}, the pulsar age is closer to $1\,200$ years than $1\,100$ years. Given that
the data we have analyzed were taken $\sim 20$ years ago, the pulsar age today should be close to $1\,200$ years.

Fig.~\ref{f:F1_specs} also shows the [\ion{O}{ii}]$\wll$3726,3729 doublet along with [\ion{Fe}{ii}]~$\lambda$5159,
[\ion{Fe}{xiv}]~$\lambda$5303, and other lines as marked in Fig.~\ref{f:F1_specs}. [\ion{Fe}{ii}]~$\lambda$5159 and
 [\ion{Fe}{iii}]~$\lambda$5270 both showed hints of broad wings for the SNRC. Nothing like that is seen for the position at F1, 
but this could be due to the limited signal-to-noise. Of greater importance are the two spectral
bumps on each side of the central unresolved blend of the [\ion{O}{ii}] doublet. One possible interpretation is that they
originate in the same part of the ejecta as the [\ion{O}{iii}] halo, and possibly in its outermost part as a result of
a transition of the photoionized state from O$^{2+}$ to  O$^{+}$, at $v_{\rm true} \sim 2\,000 \kms$. 
Another possibility is that the bumps are due to Balmer lines emission from F1. We have marked the positions of them in the middle
panel of Fig.~\ref{f:F1_specs}. On the blue side \ion{H}{i}~$\lambda\lambda$3712,3722 emit at  $\approx -1\,249 \kms$ and 
$\approx -443 \kms$, respectively (when we center on 3727.5 \AA), but on the red side, no \ion{H}{i} line fits. Moreover, the strengths 
of the $\lambda$3712 and $\lambda$3722 lines are expected to be $< 1\%$ of [\ion{O}{ii}]$\wll$3726,3729, using H$\beta$ as template. 
Deeper observations are needed to test the [\ion{O}{ii}] interpretation, along with photoionization modeling like that in 
\citet{Williams08}. 
 
 Closing in on the narrow components of the lines from the filaments, the weighted velocity of individual lines can be used
 to test both the physical velocity of the filaments and to check possible systematics in the wavelength calibration. 
 The calibration is problematic far in the very blue due to warped spectra, which makes the velocities for [\ion{O}{ii}]$\wll$3726,3729
 systematically low. As an example, we show in Fig.~\ref{f:F1_velo} the velocities of the detected lines across the spectrum for F1.
 The velocity of [\ion{O}{ii}]$\wll$3726,3729 was estimated using $3727.5 \AA$ (see below). 
 We note that a group of lines tend to have LMC redshift, while most lines cluster at $\sim +120 \kms$ compared to the LMC. 
 The dashed lines in  Fig.~\ref{f:F1_velo} shows 1$\sigma$ uncertainty for the wavelength solution. Taking the extremes of this fit 
 (i.e., the solid line in the figure) to assign a combined uncertainty, we arrive at $355\pm30 \kms$ for F1. For the other filaments we 
 find $260\pm25 \kms$ for F2, $290\pm30 \kms$ for F3, $255\pm20 \kms$ for F4 and $240\pm20 \kms$ for F5. This means that all of 
them have velocities consistent with LMC velocity, except for F1, which is redshifted by an extra $85\pm30 \kms$.

In Table~\ref{t:Temper4} we list [\ion{O}{iii}] temperatures of all filaments. The temperatures were calculated in the same
way as for the SNRC in Sect. 3.3.2. For the filaments, we have assumed  an electron density of $n_{\rm e} = 10^3 \cm3$ (see below), 
but the difference in estimated temperature is very small for electron densities in the range $10^2 - {\rm few} \times 10^3 \cm3$.
The temperature is highest for F1. If we include [\ion{O}{iii}]$\wll$4959,5007 with its full base we get $(3.6\pm0.4)\EE4$~K.
However, from the Gaussian fit for [\ion{O}{iii}]$\wl$5007 in Fig.~\ref{f:F1_velo} we estimate that only $\sim 60\%$ of the flux is 
in the spectrally unresolved component. If the base originates in a photoionized halo with a presumed temperature of $\lsim 2\EE4$~K, 
no base is expected to be detected from F1 for the weak [\ion{O}{iii}]$\wl$4363 line, which is also confirmed by our data. For the 
unresolved  component, a better estimate of the [\ion{O}{iii}] temperature could therefore be $\gsim 5.5\EE4$~K. For the other 
filaments we estimate $(3.0\pm0.4)\EE4$ K (F2), $(1.8\pm0.2)\EE4$ K (F3) $(1.9\pm0.2)\EE4$ K (F4) and $(2.5\pm0.4)\EE4$ K (F5). 
The temperatures are somewhat higher than reported in our preliminary analysis in \citet{Ser05}. 
For the central peak of  [\ion{O}{ii}]$\wll$3726,3729, we tested possible skewness to reveal the density of F1 (cf. Sect. 3.4.1 for 
how $I_{[\ion{O}{ii}]\wl3729} / I_{[\ion{O}{ii}]\wl3726}$ depends on density). Unfortunately, the spectral resolution is too poor to 
draw any such conclusions.
 
The spectra of all filaments show many forbidden iron lines. Filaments F1 and F3 
have lines of [\ion{Fe}{ii}] and [\ion{Fe}{iii}], whereas F2 also shows [\ion{Fe}{v}]$\wl$4227.
Filaments F1, F4 and F5 show the very highly ionized [\ion{Fe}{xiv}]$\wl$5303 (cf. Fig.~\ref{f:F1_velo}), possibly with some
contamination from [\ion{Ca}{v}]$\wl$5309. There is also a line at $\wl$5537.9 for F1 that
could be [\ion{Ar}{x}]$\wl$5534. Based on these findings, the X-ray morphology of the remnant, the filament dispositions 
(cf. Fig.~\ref{f:OIIIim2}), and the high temperatures of the filaments, it is likely that filaments F1, F2, F4 and F5 are ionized by 
the SNR, most likely through shock activity. Filament F3 sits 1\farcm4 east of the pulsar and has no physical connection
to the remnant. However, it is not a normal \ion{H}{ii} region in the LMC. Typical [\ion{O}{iii}] temperatures in, e.g., the 
30 Doradus region vary rather mildly ($\pm 140$~K) from the average value of 10\,270~K \citep{KC02}. The high [\ion{O}{iii}]
temperature of F3, $18\,000 \pm 2\,000$~K, suggests that it too is affected by the remnant, presumably
through intense photoionization by high-energy photons.

\citet{bra14} studied the radio and X-ray emission from 0540, and we note that one of the regions they studied in detail
includes our filament F4. Their models of the detected X-ray emission from this region indicate a shock temperature of
$T_{\rm s} \sim 7\EE{6}$~K. Three of the five other outer SNR regions they studied have similar temperatures, whereas the two 
others are cooler and hotter by a factor of $\sim 2$. Using the Rankine-Hugoniot relations for fully ionized plasma, this translates 
into a shock speed of $V_{\rm s} \sim 510 \kms$ for the region including F4 \citep[e.g.,][]{bra14}. \citet{bra14} argue for a 
sufficiently short timescale for temperature equilibration between electrons and ions, and we will use this here as well. 
\citet{bra14} also estimated the electron density of the X-ray-emitting regions, and arrive at $n_{\rm e} \sim {\rm a~few} \cm3$, and 
for the region including F4 in particular, $n_{\rm e} \sim 4.0^{+0.6}_{-0.3} \cm3$. 

%-----------------------Table5---------------------------------
%\input{table_4.tex}
\begin{table}[h]
\caption{Temperatures$^1$ in \snr\ and neighboring filaments along the 
VLT slits ``1'' and ``3'' (cf. Fig. 1) estimated from the intensity ratio of 
[\ion{O}{iii}]~$\lambda\lambda$4959,5007 to [\ion{O}{iii}]~$\lambda$4363, and 
assuming $E(B-V) = 0.19$.}%
\label{t:Temper4}
\begin{tabular}{llc}
\hline\hline
 & Source & Temperature (10$^4$~K) \\
\hline
Slit 1 & \snr & $2.35\pm0.18$$^2$ \\
       & F1 (full line)   & $3.6\pm0.4$ \\
       & F1 (unresolved component only)  & $\gsim 5.5$ \\
       & F2   & $3.0\pm0.4$ \\
       & F3   & $1.8\pm0.2$ \\       
\hline
Slit 3 & F4   & $1.9\pm0.2$ \\
       & F5   & $2.5\pm0.4$ \\
\hline
\end{tabular} \\
$^1$~Temperatures were estimated using the method and atomic data described 
in Sect. 3.3.2.\\
$^2$~[\ion{O}{iii}]~$\lambda$4363 was deblended from a contribution of H$\gamma$ (see text). 
\end{table}
%______________________________________________________________

A shock with velocity $V_{\rm s} \sim 510 \kms$ can collisionally ionize iron up to Fe$^{18+}$, with almost no iron in  Fe$^{13+}$ if the 
plasma is in collisional equilibrium \citep[e.g.,][]{AR92}. However, with the low densities in the shocks studied by \citet{bra14}, 
collisional equilibrium is not fully established, and one has instead to resort to non-equilibrium ionization. More specifically, the
collisional ionization time-scale of Fe$^{13+}$ to Fe$^{14+}$ is 
\begin{equation}
t_{\rm ion}({\rm Fe}^{13+}) \approx 3.2~\left(\frac{C_{\rm ion}({\rm Fe}^{13+})}{10^{-9}~{\rm cm}^3~{\rm s}^{-1}}\right)^{-1}~\left(\frac{n_{\rm e}}{10 \cm3}\right)^{-1}~{\rm years},
\end{equation}
where $C_{\rm ion}({\rm Fe}^{13+})$ is the collisional ionization rate coefficient for collisions from Fe$^{13+}$. We use the formalism
of \citet{AR85} which gives $C_{\rm ion}({\rm Fe}^{13+}) \approx 6.8\EE{-10}$~cm$^3$~s$^{-1}$ for $T_{\rm s} \sim 7\EE{6}$~K, 
which means $t_{\rm ion}({\rm Fe}^{13+}) \sim 12$~years for $n_{\rm e} = 4.0 \cm3$. The corresponding time scales 
$t_{\rm ion}({\rm Fe}^{14+})$, $t_{\rm ion}({\rm Fe}^{15+})$ and $t_{\rm ion}({\rm Fe}^{16+})$ are $\sim 15$~years, $\sim 29$~years and
$\sim 230$~years, respectively. For young enough shocks, there will certainly be no iron ionized beyond Fe$^{16+}$, and for
shocks younger than a couple of decades, Fe$^{13+}$ should still be present. \citet{bra14} estimate an ionization time scale of
$\sim 1600$~years for $n_{\rm e} = 4.0 \cm3$, which seems unreasonably long for F4, unless the density of the shocked gas is
substantially higher than $4.0 \cm3$. This could be the case for a low filling factor of the shocked gas.

The ionization time scales should be compared with recombination time scales, and for the recombination 
from Fe$^{13+}$ to Fe$^{12+}$, $t_{\rm rec}({\rm Fe}^{13+})$, we write
\begin{equation}
t_{\rm rec}({\rm Fe}^{13+}) \approx 317~\left(\frac{\alpha_{\rm rec}({\rm Fe}^{13+})}{10^{-11}~{\rm cm}^3~{\rm s}^{-1}}\right)^{-1}~\left(\frac{n_{\rm e}}{10 \cm3}\right)^{-1}~{\rm years},
\end{equation}
where $\alpha_{\rm rec}({\rm Fe}^{13+})$ is the recombination rate coefficient. If we use the results of \citet{Schmidt06}, i.e., 
$\alpha_{\rm rec}({\rm Fe}^{13+}) \approx 1.55\EE{-11}{\rm cm}^3~{\rm s}^{-1}$ at $T_{\rm s} \sim 7\EE{6}$~K, 
we get $t_{\rm rec}({\rm Fe}^{13+}) \sim 510$~years for $n_{\rm e} = 4.0 \cm3$. This means that once ionized to Fe$^{13+}$, 
the ISM shocked by \snr\ would not have time to recombine. Collisional ionization to Fe$^{14+}$ is far superior, which immediately
shows that the low-density shocked ISM with $n_{\rm e} = 4.0 \cm3$ is not the source of the low-ionization lines from filament F4.

This is further strengthened if we study the cooling time of the shocked gas. For this we use an expression for the shocked 
circumstellar matter of SN~1987A, which reads
\citep{Gro06}, valid for $100 \kms \lsim V_{\rm s} \lsim 600 \kms$, 
\begin{equation}
t_{\rm cool} \approx 580~\left(\frac{n_{\rm filament}}{10^3 \cm3}\right)^{-1}~\left(\frac{V_{\rm s}}{500 \kms}\right)^{3.8}~{\rm years}.
\end{equation}
Here $n_{\rm filament}$ is the density of the filament. Even though this expression was tailored for the enriched He and N 
abundances of SN~1987A, it is obvious that the cooling time for 0540 is far too long for the shocked gas to cool down to the 
temperature derived from [\ion{O}{iii}]. 

From our discussion we find that the shocks running into the low-density ISM are adiabatic. Fe$^{13+}$ is most likely in a transitory 
state to becoming further ionized, although full collisional equilibrium will not be attained. This plasma cannot be the same as that 
emitting low-ionization like, e.g., [\ion{O}{iii}] (see below). The situation could therefore be similar to that in the Cygnus Loop 
\citep{Ballet89,FesHur96,Ray15}, N49, N63A, N103b and SN~1987A in the LMC \citep{Denne76,Murdin78,DM79,Gro06,Dop19}, Puppis A 
\citep{Clark79} and SNR S8 in IC 1613 \citep{FW20}, all in which coronal iron lines coexist with lower-ionization iron lines. The situation is 
different from that in, e.g., SNR 1E0102.2-7219 in the SMC \citep{Vogt17}, where only coronal lines are seen, nicely tracing the blast wave.
We return to [\ion{Fe}{xiv}]$\wl$5303 below. 

%-----------------------figure26-----------------------------
\unitlength=1mm
\begin{figure*}[htb]
\begin{center}
\includegraphics[width=180mm, angle=0, clip]{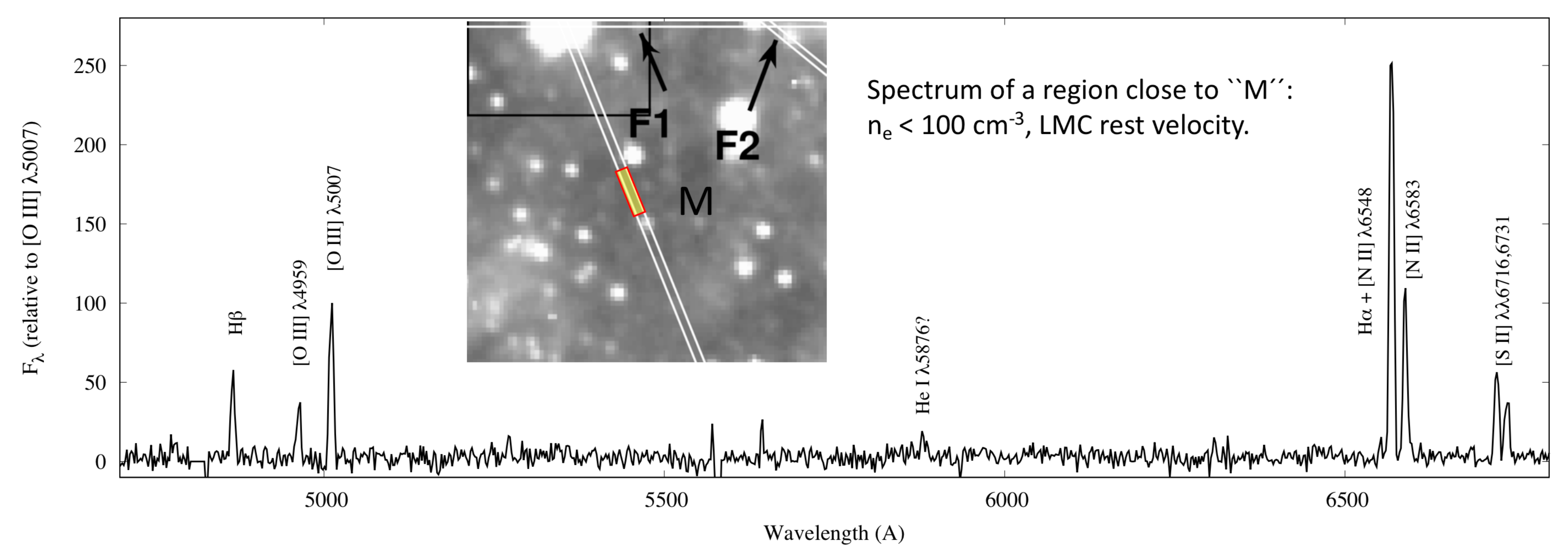}
\end{center}
\caption{Spectrum of a region (marked in yellow) along the NTT/EMMI slit  close to the ``N-rich" region identified by \citet{Mat80}, 
here marked with ``M". The map in the inset is from Fig.~\ref{f:OIIIim2}.  The spectrum was dereddened with $E(B-V) = 0.19$, and
plotted relative to [\ion{O}{iii}]~$\lambda$5007. The electron density is from [\ion{S}{ii}]~$\lambda\lambda$6716,6731, and the velocity 
of the emitting gas is LMC rest velocity. See text for further details.
} \label{f:M_region}
\end{figure*}
%------------------------------------------------------------

The low-ionization lines from F1, F2, F4 and F5 likely originate as results of shocks with lower velocity than the shocks studied 
by \citet{bra14}. From Filament F1 we detect Fe$^{6+}$ lines, and in collisional equilibrium this ion peaks in abundance 
at $\sim (2-3)\EE{5}$~K, corresponding to a shock speed of $\sim 100 \kms$. Since we expect the shock speed in the filaments 
to scale as $\propto n_{\rm filament}^{1/2}$, the density of the filaments emitting the low-ionization lines should be
$\sim 4.0~(V_{\rm X-ray}/V_{\rm opt})^{2} \cm3$, where $V_{\rm X-ray}$ and $V_{\rm opt}$ are the shock speeds in the
X-ray emitting region and low-ionization line emitting region, respectively. This points to $n_{\rm filament} \sim 10^2 \cm3$ for the 
low-ionization line emitting regions. According to Equation 5, this would mean 
$t_{\rm cool} \sim 13$~years (for $V_{\rm opt} = 100 \kms$). These
shocks are therefore radiative (as opposed to the adiabatic shock created by the blast wave). If the postshock
gas is in near pressure equilibrium, the density in the [\ion{O}{iii}]-emitting region should be a factor of $\sim 10$ higher
than that emitting [\ion{Fe}{vii}] lines, i.e., $\sim 10^3 \cm3$. This justifies our choice of $10^3 \cm3$ when we estimated the 
[\ion{O}{iii}] temperature earlier in this subsection. A decisive measurement for this picture would be to measure the
density from [\ion{O}{ii}] and/or [\ion{S}{ii}]. Unfortunately, [\ion{S}{ii}] emits outside the spectral range of our filament spectra, and
the spectral resolution is too low to draw conclusions from [\ion{O}{ii}].

While four of our filaments have velocities consistent with LMC velocity, Figs.~\ref{f:F1_specs} and \ref{f:F1_velo} show that
filament F1 recedes by $85\pm30 \kms$ compared to the others. This is also the only filament which is not situated at, or outside
the projected position of the blast wave. According to the Rankine-Hugoniot shock relations for polytropic index $\gamma = 5/3$,
this means that the shocked gas in the low-ionixzation line-emitting medium is accelerated to $3V_{\rm opt}/4$. For a spherically-
symmetric model with a shock radius corresponding to $\sim 20\arcsec - 25\arcsec$ at the distance of LMC, and $8\farcs5$ for 
the projected distance between the pulsar and F1, the observed  $85\pm30 \kms$ for F1 translates into  $90\pm35 \kms$. 
If we correct for the factor 4/3, we arrive at $V_{\rm opt} = 120\pm45 \kms$, which is fully consistent with what we estimated from
the estimated shock velocity required to ionize up to Fe$^{6+}$ for a radiative shock being responsible for the low-ionization
lines. In reality, there are most likely a range of shock velocities depending on the distribution of densities, much in the same way
as for the well-studied inner ring of SN~1987A \citep{Gro06}. We reiterate that the broad and narrow lines from the
F1 region come from completely different components; the broad [\ion{O}{iii}] lines come from the halo of freely coasting SN ejecta,
while the narrow lines from the shocks at the SNR front on the rear side of the remnant, presumably some $\sim 20\arcsec - 25\arcsec$
from the pulsar. Filaments F2, F4 and F5 are moving mainly orthogonal to the line-of-sight, and do not show any blue- or
redshift compared to LMC velocity. 

In this picture, the [\ion{Fe}{xiv}]$\wl$5303 velocity for F1 should be more redshifted than lines from low-ionization ions, i.e., if it comes
from behind the blast wave. Although [\ion{Fe}{xiv}]$\wl$5303 may be slightly more redshifted than the bulk of emission lines (cf. 
Table~\ref{t:FluxVel_2} and Fig.~\ref{f:F1_velo}), we do not see a several hundred $\kms$ difference in velocity. There could be at least two 
explanations for this. The first is that we have misinterpreted another spectral line for being  [\ion{Fe}{xiv}]$\wl$5303, and the second is that
if [\ion{Fe}{xiv}]$\wl$5303 is indeed the correct identification, it is not emitted immediately behind the inter-cloud fast shock. 
Instead it originates from closer to the clouds. To check the first explanation, we focus on the right panels of Fig.~\ref{f:F1_velo} in
which the stronger iron lines peak around $130-140 \kms$, whereas [\ion{Fe}{xiv}]$\wl$5303 peaks at $\sim 180 \kms$. The Gaussian
fit to  [\ion{Fe}{xiv}]$\wl$5303 is consistent with a single feature. The FWHM is $\approx 335 \kms$, so any thermal broadening
of the line is impossible to trace. That the line may appear somewhat broader than the Gaussian fit is likely due to poor signal-to-noise.
In the line list by \citet{FesHur96}, there are a  few lines close to [\ion{Fe}{xiv}]$\wl$5303, namely [\ion{Fe}{ii}]$\wl$5297 and 
[\ion{Ca}{v}]$\wl$5309. (None of them marked in Fig.~\ref{f:F1_velo}). Their velocities relative to [\ion{Fe}{xiv}]$\wl$5303
are $-340 \kms$ and $+357 \kms$, respectively. Whereas theire is no hint for the [\ion{Fe}{ii}] line, [\ion{Ca}{v}] line could
be responsible for a possible slight extension of the [\ion{Fe}{xiv}]$\wl$5303 main part of the line feature. This is why we have
added the [\ion{Ca}{v}]$\wl$5309 line in Tables~\ref{t:FluxVel_2} and \ref{t:FluxVel_3}. That  [\ion{Fe}{xiv}]$\wl$5303 is present
seems solid.

But why is then [\ion{Fe}{xiv}]$\wl$5303 from F1 not more redshifted than the lower-ionization iron lines in Fig.~\ref{f:F1_velo}? 
The results of \citet{Mic06} may provide a clue. These authors performed 2D axial symmetric simulations of a fast shock interacting
with a cloud in a tenuous medium. The setup was intended to model the X-ray emission from the Vela SNR, but general conclusions 
can be drawn. They tested two geometries, a spherical cloud, and an elliptical cloud with major axis along the shock direction (to
simulate a filament). In the spherical cloud case (which we will concentrate on), the shock interaction results in something
like a two-temperature situation, where soft X-rays are produced in the transmitted shock in the cloud, and a higher-temperature
region in the shocked tenuous medium emits harder X-rays. What is important is that the hard X-ray emission does not mainly come from
region immediately behind the blast wave. It instead comes from a halo around the cloud. Conduction between the cloud and the 
hot shocked tenuous medium drives an evaporation of the cloud and the formation of a hot diffuse halo. We suggest that 
[\ion{Fe}{xiv}]$\wl$5303 originates in this zone along with the hard X-rays. As this zone is tied to the cloud, there should not be
any significant difference in velocity as determined from the line profiles. Future tests of this scenario would be to study 
the 0540 filaments in higher spectral resolution, as well as in the red part of the spectrum to probe, e.g., [\ion{Fe}{x}]$\wl$6374
and [\ion{Fe}{xi}]$\wl$7892, as well as to measure the [\ion{S}{ii}] density.

Moving to regions in the south-western part of \snr, the possibly N-rich filament identified by \citet{Mat80} is outside the reach of
our slits and and the field-of-view of the VIMOS/IFU of \citet{Sand13}. However, our NTT/EMMI slit covers a region
close to it which shows strong nebular emission. In Fig.~\ref{f:M_region} we show this region and its spectrum, dereddened 
using $E_{B-V} = 0.19$, and normalized to [\ion{O}{iii}]~$\lambda$5007. \citep[The region discussed by][is marked with an ``M".]{Mat80} 
The electron density from [\ion{S}{ii}] in the yellow-marked
region along the EMMI slit is $\lsim 10^2 \cm3$, which is fully consistent with an \ion{H}{ii} region in the LMC, as is
also the velocity of the emitting region. However, there is an oddity with the overall spectrum in that it is very red. With
$E_{B-V} = 0.19$ the Balmer decrement H$\alpha$/H$\beta \sim 5.0$., which is almost a factor of $\approx 1.8$ higher than 
expected from an \ion{H}{ii} region. The simplest solution to this is that the emission comes from an \ion{H}{ii} region, but that
$E_{B-V} \sim 0.95$ for this area rather than the value we use for 0540. From a visual inspection of Fig.~\ref{f:OIIIim} it is clear 
that the nebulous emission is less pronounced in an area that includes both the yellow part along the slit in Fig.~\ref{f:M_region} 
and the region M highlighted by \citet{Mat80}. Absence of [\ion{O}{iii}]-emission is also vero clearly seen for this region in 
Figure 1A of \citet{Mat80}. Such patchy structure is expected if there are dusty regions embedded in a large \ion{H}{ii} region
like N~158 \citep{Heinze56} with its superbubble around the OB association LH~104 in the north \citep{LH70,TN98}. which dominates 
this part of the LMC. Finally, we note that the filter used 
by \citet{Mat80} to study [\ion{N}{ii}] was centered on 6584~\AA\ with a FWHM of 16~\AA. With an LMC redshift of $270 \kms$,
H$\alpha$ is redshifted to $\sim 6569$~\AA\ and [\ion{N}{ii}]~$\lambda$6584 to 6589\AA. In the spectrum of Fig.~\ref{f:M_region},
$I_{[\ion{N}{ii}]\lambda6584} / I_{{\rm H}\alpha} \sim 2.2$, and if this is also the case for region ``M", $\sim 20\%$ of the emission
in the 6584~\AA\ filter used by \citet{Mat80} is due to H$\alpha$ (if Gaussian), however, the exact fraction being sensitive to the 
transparency of the blue wing of their 6584~\AA\ filter.

%______________________________________________________________________________ 
\section{Conclusions}
A combination of [\ion{O}{iii}] filters centered on $0 \kms$ and $+3\,000 \kms$ was used to reveal structures in the central
part of \snr\ (SNRC), not previously detected by HST filters, and disproves some structures indicated by the VLT/VIMOS integral-field-unit 
(IFU) observations by \citet{Sand13}. We detect [\ion{O}{iii}] emission out to $\sim 10$\arcsec\ from the pulsar, which corresponds to 
an average ejecta velocity of $\sim 2\,000 \kms$ for an age of 1\,200 years. The filter images complement the IFU images in [\ion{O}{iii}] to 
suggest ring-like structures around a symmetry axis from north-east to south-west, also seen in [\ion{S}{ii}] \citep{Sand13}. In a 
3D-representation, the symmetry axis crosses the projected pulsar position $\sim 500 \kms$ on the receding side of the remnant, which 
if it is due to pulsar activity, could indicate that the pulsar shares the same general redshift as the SNRC. Such pulsar activity could occur 
if the pulsar jet is oriented along this axis rather than as displayed in Fig.~\ref{f:O3_new}. It is emphasized that there is evidence for 
substantial past and present activity along the suggested jet axis \citep[cf.][]{nlun11}, but the rings could also have an origin related to
the supernova explosion.

\snr\ was also studied through two slits roughly orthogonal to each other (PA~$=22\degr$ and PA~$=88\degr$). 
Several new spectral lines are identified for the SNRC, and 2D spectroscopy along the slits reveals fast [\ion{O}{iii}]-emitting ejecta outside 
the SNRC corresponding to a space velocity of $v_{\rm true} \sim 2\,100 \kms$ on the receding western part. This velocity agrees with our 
imaging, as well as a pulsar age of $\approx 1\,200$ years. In the southwest on the receding side of $v_{\rm true}  \gsim +2\,200 \kms$.
On the approaching side, the [\ion{O}{iii}] emission reaches $v_{\rm true} \sim 1\,900 \kms$. Although substantially weaker, [\ion{Ne}{iii}] 
emission has similar structure to that of [\ion{O}{iii}], whereas [\ion{O}{ii}], [\ion{S}{ii}], [\ion{Ar}{iii}] and H$\beta$ are emitted by the more
compact structure of the SNRC, in agreement with the [\ion{S}{ii}] IFU image of \citet{Sand13}. The luminosity-weighted average of the lines 
is $+440\pm80 \kms$ with respect to the host galaxy, which is accord with previous findings, as well as the possible pulsar motion.

The average [\ion{O}{iii}] temperature of the SNRC is $23\,500 \pm 1\,800$~K, which is somewhat lower than previously reported, mainly 
because we have included a correction of the [\ion{O}{iii}]~$\lambda$4363 flux due to the blend with H$\gamma$. The electron density derived  
from [\ion{S}{ii}]~$\lambda\lambda$6716,6731 is found to be $n_{\rm e} \sim 10^3 \cm3$, which is slightly higher than in the density map of 
\citet{Sand13}. 

The VIMOS/IFU data of \citet{Sand13} were used to construct slit correction factors for our spectroscopic data in order to estimate 
overall abundances and masses of various elements in the SNRC. Infrared data from {\sl Spitzer} \citep{Williams08} were also included 
in these estimates. In particular, relative mass ratios  were found to be ${\rm O:Ne:S:Ar} \approx 1:0.07:0.10:0.02$. Those ratios
are consistent with explosion models of $13-20 \Msun$ progenitors \citep[e.g.,][]{CL13,Siever20}, as well as SN~1987A 
\citep[cf.][]{Jerk11}. The argon abundance is a factor of $\sim 2$ lower than used in the models by \citet{Williams08} (which was also
suggested by those authors as a possibility). The fact that we detect H and He mixed in deep into the central region is a clear sign
of explosive mixing, as in SN~1987A. The mass ratio of He/H in the SNRC is estimated to be at least $\sim 0.8$, which is another similarity 
with SN~1987A.

The mass of [\ion{O}{ii}] and [\ion{O}{iii}] emitting oxygen in the SNRC is only $\sim 1.4\EE{-4} \Msun$, but the cooling time of shocked 
gas in the SNRC is so short ($\gsim 0.05$~years) that the mass flux through the low-velocity shocks in the SNRC can be up 
to $\sim 0.003~X_{\rm other} \Msun$~yr$^{-1}$, where $X_{\rm other}$ is the correction factor to include all elements, not only oxygen. 
This is on the low side compared with $\sim 0.01 \Msun$~yr$^{-1}$ estimated by \citet{Williams08}. 
The rapid cooling of the shocked gas in the SNRC could in principle make it possible to detect fluctuations in the flux
of lines other than those from oxygen if the SNRC is not microscopically mixed. We note a possible such fluctuations in
[\ion{S}{ii}]$\lambda\lambda$6716,6731 between 1989--2006. This effect could potentially make abundance estimates for the whole
SNRC vulnerable for elements other than oxygen in the SNRC. IFU images for various spectral lines separated in time be several 
years could resolve this uncertainty.

The fastest emitting [\ion{O}{iii}] ejecta (which have components with velocities in excess of $2\,000 \kms$) come from a freely coasting halo
of photoionized supernova ejecta. The mass of the halo is $\sim 0.12 \Msun$ if spherical symmetry is assumed. However,  the halo 
shows clear asymmetry with no detectable emission on the receding side to the north and only weak emission on the approaching side to 
the south. On the approaching side, there appears to be a ``wall"-like feature centered slightly to the northwest of the projected pulsar position 
and it has velocities more blueward than $\sim -750 \kms$. We have used the VIMOS/IFU results of \citet{Sand13} to constrain that structure.
There is a hint of [\ion{O}{ii}]$\lambda\lambda$3726,3729 emission from fast 
photoionized ejecta (both approaching and receding), which would mark the outer boundary of the photoionized region.
Oxygen-rich ejecta out to $\sim 2\,000-2\,200 \kms$, and a total mass of oxygen which is $\sim 2 \Msun$ are yet other 
similarities with SN~1987A \citep[cf.][and references therein for properties of SN~1987A]{Jerk11}. However, for 0540 most of the oxygen is 
likely to be located in filaments in a pulsar-wind nebula, which for SN~1987A is not the case.

A third slit was placed along the western part of the supernova remnant shock. Two optical filaments, named F4 and F5, are detected
for this slit position, and three further filaments (F1, F2 and F3) are identified for the PA~$=88\degr$ slit. Although having an 
[\ion{O}{iii}] temperature of $18\,000 \pm 2\,000$~K, filament F3 lies too far from \snr\ to be physically connected to it, and is most likely 
an \ion{H}{ii} region in the vicinity of the remnant. The other filaments are hotter, and three of them (F2, F4 and F5) show 
[\ion{Fe}{xiv}]~$\lambda$5303 emission. The hottest (F1), with  an [\ion{O}{iii}] temperature in excess of $55\,000$~K, is redshifted by 
$85\pm30 \kms$ relative to the LMC redshift, and is located only $8\farcs5$ west of the pulsar. We argue that the redshift was
acquired when F1 was shocked by the SNR blast wave. Similar acceleration for F2, F4 and F5 would have occurred orthogonal
to the line of sight since they show no red- or blueshift. Filament F4 overlaps with a region studied in X-rays by \citet{bra14}, and 
including temperatures and densities from the X-ray modeling by these authors, we conclude that most of the optical lines from 
F1, F2, F4 and F5 come from radiative shocks in ISM clouds, overtaken by the supernova blastwave. The electron density of
the [\ion{O}{iii}] emitting  gas is estimated to be $\sim 10^3 \cm3$. Lines from more highly ionized ions, and in particular
[\ion{Fe}{xiv}]~$\lambda$5303, are argued to come from an evaporation zone in connection with the radiatively cooled gas. 
Future observations of these filaments at higher spectral resolution is needed to test these conclusions. 

Finally, a portion of the slit with PA~$=22\degr$ was used to probe a region close to a possibly N-rich filament identified by
\citet{Mat80}. Our spectrum reveals that the emission is highly reddened with a Balmer decrement of H$\alpha$/H$\beta\sim 5.0$, 
and that the emitting gas has a low electron density ($\lsim 10^2 \cm3$). It is most likely emission from an \ion{H}{ii} region unrelated
to the supernova. The severe reddening boosts red lines compared to, e.g., [\ion{O}{iii}] which is weak in this region. The [\ion{N}{ii}] 
filter used by \citet{Mat80} could also have been affected by H$\alpha$ emission. An N-rich supernova filament in this region
therefore seems unlikely.  

\begin{acknowledgements}
We thank R.~J.~Cumming for helping out with the NTT observations and A.~Koptsevich for being part of the
preparations of the VLT observations. We are also grateful to C.~Fransson, J.~Larsson, C.~Sandin and J.~Sollerman for 
discussions. The research of PL is sponsored by the Swedish Research Council. Early on  
in this project, PL was a Research Fellow at the Royal Swedish Academy supported by a grant 
from the Wallenberg Foundation, and NL was supported by the Swedish Institute. 
YuAS is partially supported by the Russian Foundation for Basic Research (RFBR) according the project 19-52-12013.
We made use of the Atomic Line List \citep{VH18}, and the CHIANTI data base. CHIANTI is a collaborative 
project involving George Mason University, the University of Michigan (USA), University of Cambridge (UK) 
and NASA Goddard Space Flight Center (USA). This research has also made use of NASA’s Astrophysics Data System 
Bibliographic Services.
\end{acknowledgements}

\end{document}